\documentclass{article}
\usepackage{e}
\usepackage{epsfig}
\usepackage{latexsym}
\begin{document}
\branch{C}
% the following three lines are contributed by the publisher
\DOI{10.1007/s1010599c0004}                          % do not fill in
\idline{C}{1}{67}                % do not fill in
\editorial{1}{9}{9}{9}              % do not fill in
\title{Correlations between $D$ and $\overline{D}$
       mesons produced in 500 GeV/$c$ $\pi^-$-nucleon
       interactions}
\author{Fermilab E791 Collaboration}
%%%%%%%%%%%%%%%%%%%%   E791 Name list   %%%%%%%%%%%%%%%%%%%%%%%%%%%%%%%
%
\author{Fermilab E791 Collaboration  \\  \\
           E. M. Aitala\inst{9}
\and       S. Amato\inst{1}
\and    J. C. Anjos\inst{1}
\and    J. A. Appel\inst{5}
\and       D. Ashery\inst{13}
\and       S. Banerjee\inst{5}
\and       I. Bediaga\inst{1}
\and       G. Blaylock\inst{8}
\and    S. B. Bracker\inst{14}
\and    P. R. Burchat\inst{12}
\and    R. A. Burnstein\inst{6}
\and       T. Carter\inst{5}
\and    H. S. Carvalho\inst{1}
\and    N. K. Copty\inst{11}
\and    L. M. Cremaldi\inst{9}
\and       C. Darling\inst{17}
\and       K. Denisenko\inst{5}
\and       A. Fernandez\inst{10}
\and       G. F.~Fox\inst{11}
\and       P. Gagnon\inst{2}
\and       C. Gobel\inst{1}
\and       K. Gounder\inst{9}
\and    A. M. Halling\inst{5}
\and       G. Herrera\inst{4}
\and       G. Hurvits\inst{13}
\and       C. James\inst{5}
\and    P. A. Kasper\inst{6}
\and       S. Kwan\inst{5}
\and    D. C. Langs\inst{11}
\and       J. Leslie\inst{2}
\and       B. Lundberg\inst{5}
\and       S. MayTal-Beck\inst{13}
\and       B. Meadows\inst{3}
\and J. R. T. de Mello Neto\inst{1}
\and       D. Mihalcea\inst{7}
\and    R. H. Milburn\inst{15}
\and J. M. de Miranda\inst{1}
\and       A. Napier\inst{15}
\and       A. Nguyen\inst{7}
\and  A. B. d'Oliveira\inst{3,10}
\and       K. O'Shaughnessy\inst{2}
\and    K. C. Peng\inst{6}
\and    L. P. Perera\inst{3}
\and    M. V. Purohit\inst{11}
\and       B. Quinn\inst{9}
\and       S. Radeztsky\inst{16}
\and       A. Rafatian\inst{9}
\and    N. W. Reay\inst{7}
\and    J. J. Reidy\inst{9}
\and    A. C. dos Reis\inst{1}
\and    H. A. Rubin\inst{6}
\and    D. A. Sanders\inst{9}
\and A. K. S. Santha\inst{3}
\and A. F. S. Santoro\inst{1}
\and    A. J. Schwartz\inst{3}
\and       M. Sheaff\inst{4,16}
\and    R. A. Sidwell\inst{7}
\and    A. J. Slaughter\inst{17}
\and    M. D. Sokoloff\inst{3}
\and       J. Solano\inst{1}
\and    N. R. Stanton\inst{7}
\and    R. J. Stefanski\inst{5} 
\and       K. Stenson\inst{16}
\and    D. J. Summers\inst{9}
\and       S. Takach\inst{17}
\and       K. Thorne\inst{5}
\and    A. K. Tripathi\inst{7}
\and       S. Watanabe\inst{16}
\and       R. Weiss-Babai\inst{14}
\and       J. Wiener\inst{11}
\and       N. Witchey\inst{7}
\and       E. Wolin\inst{17}
\and    S. M. Yang\inst{7}
\and       D. Yi\inst{9}
\and       S. Yoshida\inst{7}
\and       R. Zaliznyak\inst{12}
\and       C. Zhang\inst{7}}
\institute{
Centro Brasileiro de Pesquisas F\'\i sicas, Rio de Janeiro, Brazil
\and University of California, Santa Cruz, California 95064
\and University of Cincinnati, Cincinnati, Ohio 45221
\and CINVESTAV, Mexico
\and Fermilab, Batavia, Illinois 60510
\and Illinois Institute of Technology, Chicago, Illinois 60616
\and Kansas State University, Manhattan, Kansas 66506
\and University of Massachusetts, Amherst, Massachusetts 01003
\and University of Mississippi, University, Mississippi 38677
\and Universidad Autonoma de Puebla, Mexico
\and University of South Carolina, Columbia, South Carolina 29208
\and Stanford University, Stanford, California 94305
\and Tel Aviv University, Tel Aviv, Israel
\and Box 1290, Enderby, BC, VOE 1V0, Canada
\and Tufts University, Medford, Massachusetts 02155
\and University of Wisconsin, Madison, Wisconsin 53706
\and Yale University, New Haven, Connecticut 06511
}
\maketitle
\begin{abstract}

We present a study of correlations between $D$ and $\overline{D}$ mesons
produced in 500 GeV/$c$ $\pi^-$-nucleon interactions,
based on data from experiment E791 at
Fermilab.  We have fully reconstructed $791 \pm 44$
charm meson pairs to study correlations between the transverse and
longitudinal momenta of the two $D$ mesons and the relative
production rates for different types of $D$ meson pairs.
We see slight correlations between the longitudinal momenta of the $D$ and
the $\bar D$, and significant correlations between the azimuthal angle
of the $D$ and the $\bar D$.
The experimental distributions are compared to a
next-to-leading-order QCD calculation and to predictions
of the {\sc Pythia/Jetset} Monte Carlo event generator.
We observe less correlation between transverse momenta and different
correlations between longitudinal momenta than these models
predict for the default values of the model parameters.
Better agreement between data and theory might be achieved by tuning
the model parameters or by adding higher order perturbative terms,
thus contributing to a better understanding of charm production.

The relative production rates for the four sets of charm pairs, 
$D^0\overline{D}\!^{\;0}$, $D^0D^-$, $D^+\overline{D}\!^{\;0}$, 
$D^+D^-$, as calculated in the 
{\sc Pythia/Jetset} event generator with the default
parameters,
agree
with data as far as the relative ordering, but predict too many
$D^+\overline{D}\!^{\;0}$ pairs and too few $D^+D^-$ pairs.

\end{abstract}

%%%%%%%%%%%%%%%%%%%%%%%%%%%%%%%%%%%%%%%%%%%%%%%%%%%%%%%%%%%%%%%
%DEFINITIONS
%
\def\cb{\overline{c}}
\def\ccb{c\overline{c}}
\def\bbb{b\overline{b}}
\def\cbcb{\overline{c}\overline{c}}
\def\ddb{D\overline{D}}
\def\db{\overline{D}}
\def\dzb{\overline{D}\!^{\;0}}
\def\dzdb{D^0\overline{D}\!^{\;0}}
\def\dzdm{D^0D^-}
\def\dpdb{D^+\overline{D}\!^{\;0}}
\def\dpdm{D^+D^-}
\def\xf{x_F}
\def\xfd{x_{F,D}}
\def\xfdb{x_{F,\overline{D}}}
\def\dxf{\Delta x_F}
\def\sxf{\Sigma x_F}
\def\yc{y_c}
\def\ycb{y_{\overline{c}}}
\def\yd{y_D}
\def\ydb{y_{\overline{D}}}
\def\yddb{y_{D,\overline{D}}}
\def\dy{\Delta y}
\def\sy{\Sigma y}
\def\dphi{\Delta \phi}
\def\sphi{\Sigma \phi}
\def\pt{p_t}
\def\ptd{p_{t,D}}
\def\ptdb{p_{t,\overline{D}}}
\def\ptt{p_t^2}
\def\pttd{p^2_{t,D}}
\def\pttdb{p^2_{t,\overline{D}}}
\def\pttddb{p^2_{t,D\overline{D}}}
\def\dptt{\Delta p_t^2}
\def\sptt{\Sigma p_t^2}
\def\Mddb{M_{D\overline{D}}}
\def\etal{{\it et al.}}
%
%%%%%%%%%%%%%%%%%%%%%%%%%%%%%%%%%%%%%%%%%%%%%%%%%%%%%%%%%%%%%%%%%%%%%
%\input{offline_doc_290_1_r1.tex} %  Introduction
\section{Introduction}
\label{sec:intro}

Using data from experiment E791 at Fermilab, we
reconstruct pairs of charm mesons produced in 
500 GeV/$c$ $\pi^-$-nucleon interactions, where $\sqrt{s}=30.6$ GeV,
and use correlations between the mesons to probe two aspects of the 
hadroproduction of mesons containing a heavy quark: the dynamics of the
production of heavy quark-antiquark pairs and the subsequent hadronization
of the quarks into hadrons. Correlations between the $D$ and $\db$ momenta
transverse to the beam direction are sensitive to corrections 
to the leading-order calculations of the $\ccb$ cross section. Correlations
between the longitudinal momenta, as well as differences in the production
rates of the four types of $\ddb$ pairs ($\dzdb$, $\dzdm$, $\dpdb$,
and $\dpdm$), provide information regarding the role of the
remnants of the colliding hadrons in the hadronization process that
transforms the charm quarks into charm mesons. 

In most studies of the hadroproduction of charm particles, 
distributions for {\it single}
charm particles are used to probe the underlying production
physics~\cite{ref:bib1,ref:bib2}. 
The variables used to describe the single particle distributions are
the transverse momentum with respect
to the beam direction, $\pt$, and either the rapidity $y$ or the
Feynman scaling variable $\xf$, where
\begin{equation}
\label{eq:rap}
y \equiv \frac{1}{2} \ln \left( \frac{E + p_{z}}{E - p_{z}}
\right) {\rm and}
\end{equation}
\begin{equation}
\label{eq:xf}
x_F \equiv  p_z/{p_{z}^{max}} \approx 2p_z/\sqrt{s}.
\end{equation}
$E$ and $p_{z}$ are the center-of-mass energy and longitudinal 
momentum of the charm particle
and $\sqrt{s}$ is the total center-of-mass energy.  
The center of mass is that of the pion-nucleon system.
Such single charm studies are insensitive to
correlations between the two charm hadrons in a single event.

We have fully reconstructed
$791 \pm 44$
$\ddb$ pairs.
Based on this sample, we present background-subtracted, 
acceptance-corrected distributions for the following variables:
\begin{enumerate}
\item
the invariant mass of the pair of charm mesons, $\Mddb$;
\item
the square of the vector sum of the transverse momenta,
with respect to the beam direction, of the $D$ and $\db$ 
mesons ($p^2_{t,D\overline D} \equiv |\vec p_{t,D} +
\vec p_{t,\overline D}|^2$);
\item
correlations between $\xfd$ and $\xfdb$, as well as
$\yd$ and $\ydb$;
\item
$\dxf \equiv \xfd-\xfdb$ and $\sxf\equiv\xfd+\xfdb$;
\item
$\dy \equiv \yd-\ydb$ and $\sy\equiv\yd+\ydb$;
\item
correlations between the 
squares of the
magnitudes of the transverse momenta of the
$D$ and $\db$ mesons, $\pttd$ and $\pttdb$;
\item
$\dptt \equiv |\pttd-\pttdb|$ and $\sptt \equiv \pttd+\pttdb$;
\item
the azimuthal separation between the momentum vectors of the
$D$ and $\db$ mesons in the plane perpendicular to the beam direction,
$\dphi \equiv$ (minimum of $|\phi_D-\phi_{\db}|$ and
$360^{\circ}-|\phi_D-\phi_{\db}|$);
\item
correlations between the azimuthal separation 
($\dphi$) and the scalar sum and difference of the $D$ and 
$\db$ transverse momenta,
$\dptt$ and $\sptt$ ;
\end{enumerate}

In addition, this paper reports the relative production rates 
for each type of 
$\ddb$ pair ($\dzdb$, $\dzdm$,
$\dpdb$, and $\dpdm$), and compares the rapidity correlations 
for the various $\ddb$ pair combinations.

We also investigate the extent to which the
observed charm-pair correlations can be duplicated by
simply convoluting the observed single charm particle
distributions.
In addition, 
we compare our measured distributions 
to three sets of theoretical predictions:
\begin{enumerate}
\item
the distributions of $\ccb$ pairs from
a next-to-leading-order perturbative 
QCD calculation by
Mangano, Nason and Ridolfi\cite{ref:bib3,ref:bib4};
\item
the distributions of $\ccb$ pairs from the 
{\sc Pythia}/{\sc Jetset} Monte Carlo event 
generator\cite{ref:bib5} which uses a parton-shower 
model to include higher-order perturbative effects\cite{ref:bib6}; and
\item 
the distributions of $\ddb$ pairs from 
{\sc Pythia/Jetset} 
which uses the Lund string model to transform 
$\ccb$ pairs to $\ddb$ pairs\cite{ref:bib7}.
\end{enumerate}

In Table~\ref{tab:tab1}, we compare the E791 charm-pair sample to those  
from other fixed-target experiments (both hadroproduction and photoproduction).
The largest previous sample of fully-reconstructed hadroproduced charm pairs
used to study correlations           
is 20 pairs from                       
the CERN $\pi^-$-nucleon experiment NA32\cite{bib10}.
Some studies have been conducted with 
partially-reconstructed charm hadrons, in which
the direction but not necessarily the magnitude of the charm particle
momentum is determined directly.  
NA32 partially reconstructed 642 such charm pairs\cite{bib11}.
In photoproduction experiments, the largest sample
of charm pairs reconstructed is from the E687 data\cite{bib15}, 
with 325 fully-reconstructed and
4534 partially-reconstructed charm pairs.
In the E687 partially-reconstructed sample, one $D$ meson is
fully reconstructed and the momentum vector of the 
other charm meson is determined by scaling the 
momentum vector of low-momentum charged pions from the decays 
$D^{*\pm}\rightarrow D^0 \pi^{\pm}$.

\begin{table} [htbp]
        \caption{
        Summary of fully-reconstructed and partially-reconstructed
        charm-pair samples from hadroproduction
        and photoproduction fixed-target experiments.
        In the last column, we list the physics variables studied in each
        experiment.  The variables are defined in the text.}
        \label{tab:tab1}

\begin{tabular}{lccc}   \hline  \hline
Experiment                   &Beam             &Number           & Measured Pair \\
                             &Energy(GeV),     &of Pairs         & Variables     \\
                             &Beam Type,       &Reconstructed    &               \\
                             &and Target       &                 &               \\  \hline \hline
E791                         & 500 $\pi^-$     &791 fully        & $\dphi$, $|\Sigma\vec p_t|^2$, $\sptt$,       \\ 
                             &                 &                 & $\dptt$, correlations,                        \\
(this paper)                 &Pt, C            &                 & $\sxf$, $\dxf$, $\sy$,                        \\
                             &                 &                 &$\dy$, $\Mddb$,                                \\
                             &                 &                 &$\sigma_{D^0\overline{D}^0}:\sigma_{D^0D^-}:$  \\
                             &                 &                 &$\sigma_{D^+\overline{D}^0}:\sigma_{D^+D^-}$   \\ \hline
WA92~\cite{bib8}             & 350 $\pi^-$     & 475 partially\footnote{In one of the 475 pairs, both $D$'s are fully reconstructed.} 
                                                                 & $\dphi$, $|\Sigma \vec p_t|^2$, $\Mddb$,      \\
                             & Cu              &                 &$\dxf$, $\sxf$, $\dy$                          \\ \hline
E653~\cite{bib9}             & 800 $p$         & 35 partially    &$\dphi$, $|\Sigma \vec p_t|^2$,                \\
                             &emulsion         &                 &$\dy$, $\Mddb$, $\cos\theta_{cm}$              \\ \hline
NA32~\cite{bib10}\cite{bib11}&  230 $\pi^-$    & 20 fully        &$|\Sigma \vec p_t|^2$, $\dy$, $\Mddb$          \\ \cline{3-4}
(ACCMOR)                     & Cu              & 642 partially   &$|\Sigma \vec p_t|^2$, $\dy$, $\Mddb$, $\dphi$ \\ \hline
WA75~\cite{bib12}            & 350 $\pi^-$     & 177 partially   &$\dphi$, $\dy$                                 \\ \cline{3-4}
                             & emulsion        &120 partially    &$\Mddb$, $\sxf$, $\sptt$                       \\ \hline
NA27~\cite{bib13}            &400 $p$          & 17 fully        &$|\Sigma\vec p_t|^2$, $\sxf$, $\dy$, $\Mddb$   \\ \cline{3-4}
(LEBC)                       & H$_2$           & 107 partially   & $\dphi$, $\sigma_{D^0\bar D^0}:$              \\
                             &                 &                 &$\sigma_{D^0D^-+D^+\overline{D}^0}:\sigma_{D^+D^-}$ \\  \hline
NA27~\cite{bib14}            &360 $\pi^-$      & 12 fully        &$|\Sigma\vec p_t|^2$, $\Sigma x_F$,            \\ 
                             &                 &                 &$\dy$, $\Mddb$                                 \\  \cline{3-4}
(LEBC)                       & H$_2$           & 53 partially    & $\dphi$                                       \\ \hline
E687~\cite{bib15}            &200 $\gamma$     & $325$ fully     &$\dphi$, $|\Sigma\vec p_t|^2$, $\dy$, $\Mddb$  \\ \cline{3-4}
                             & Be              &4534 partially   &$\dphi$, $\dy$, $\Mddb$                        \\ \hline
NA14/2~\cite{bib16}          &100 $\gamma$     & 22 fully        &$\dphi$, $|\Sigma\vec p_t|$, $\Sigma p_{z}$,   \\
                             & Si              &                 &$\dy$, $\Mddb$                                 \\ \hline
\end{tabular} 
\end{table}

In the analysis presented here,
we have completed an
extensive study of acceptance corrections.
Acceptance corrections are made as a function of the eight variables
that describe the $D$ and $\db$ degrees of freedom:
 $((y,p_t,\phi,n)_D,(y,p_t,\phi,n)_{\overline{D}})$.
Here
$n$ is the number of decay tracks from the $D$ meson. 
Corrections are also made for the branching fractions of the reconstructed
$D$ and $\db$ decay modes.

We performed a maximum likelihood fit to the two-dimensional reconstructed
candidate $D$ mass distribution, including terms in the likelihood function
for the true $\ddb$ pairs that are the signal of interest, and also terms for
combinations of a true $D$ with background, combinations of a true $\db$ with 
background, and combinations of two background candidates in the same 
event.  From the full data set, the resulting number of true 
fully reconstructed $\ddb$ pairs
was $791 \pm 44$.  In making the distributions for single-charm and charm-pair
physics variables, a likelihood fit was performed 
for each bin 
of the relevant physics variable.

Since we fully reconstruct both the $D$ and $\db$ meson, our results
have fewer systematic errors than previously published results
based on partially reconstructed pairs.
In particular, we do
not need to correct for missing tracks or possible contamination
from baryons.

In the next section, we
review the current theoretical understanding of the
hadroproduction and hadronization of 
charm quarks.
In the Appendix, we use both theoretical calculations and 
phenomenological models to investigate the 
dependence
of various 
measurable
properties of charm production on higher-order QCD effects,
the charm quark mass, the parton distribution functions, the factorization 
scale and the renormalization scale.
In Sec.~\ref{sec:e791}, we describe the E791 detector and data processing.
In Sec.~\ref{sec:selection}, we describe the optimization of 
selection criteria for charm pairs.
We discuss the extraction of background-subtracted distributions
and  corrections for
acceptance effects in Sec.~\ref{sec:analysis}.
In Sec.~\ref{sec:results}, we present the measured distributions for
the charm pairs 
%%events 
and compare them to the distributions predicted by
(uncorrelated) single-charm distributions and to theoretical predictions.
We summarize our 
%%conclusions 
results
in Sec.~\ref{sec:conclusions}.

%%%%%%%%%%%%%%%%%%%%%%%%%%%%%%%%%%%%%%%%%%%%%%%%%%%%%%%%%%%%%%%%%%%
%\input{offline_doc_290_2_r1.tex} %  theory
\section{Theoretical Overview}
\label{sec:theory}

The charm quark is the lightest of the heavy quarks.  Its relatively 
{\it small} 
mass ensures copious charm particle production at energies typical of 
fixed-target hadroproduction experiments.  Its relatively {\it large} mass 
allows calculation of the large-momentum-transfer processes responsible for 
producing $\ccb$ pairs using perturbative quantum chromodynamics (QCD).
The consequence of the charm quark being the lightest heavy quark --- more 
specifically, having a mass not sufficiently larger than $\Lambda_{QCD}$ --- is 
that there are considerable uncertainties associated with these calculations.
Such large theoretical uncertainties, combined with conflicting experimental 
results from early charm hadroproduction experiments, have made systematic 
comparisons between theory and data difficult to interpret.  
Recent calculations of the full next-to-leading-order (NLO) differential 
cross sections by Mangano, Nason and Ridolfi (MNR)\cite{ref:bib3} and 
others, as well as unprecedented numbers of charm particles reconstructed by 
current fixed-target experiments, have allowed more progress to be made in 
this field.

In this section, we outline the theoretical framework used to describe the 
hadroproduction of charm pairs, focusing on the framework used by the following 
two packages: the FORTRAN program HVQMNR\cite{ref:bib4}, which implements 
the 
MNR NLO perturbative QCD calculation for charm quarks, and the 
{\sc Py\-thia/Jet\-set} Monte Carlo event generator\cite{ref:bib5}, which 
makes predictions for charm particles based on leading order 
parton matrix elements, 
parton showers and the Lund string fragmentation model.
In the Appendix we examine predictions from these two packages for the same 
beam 
type and energy as E791 for a wide range of theoretical assumptions to 
determine 
how sensitive the theoretical predictions are to
\begin{enumerate}
\item
the inclusion of higher order terms
($\alpha_s^3$ or parton shower
contributions); and
\item 
non-perturbative effects, including
\begin{enumerate}
\item
variations in parameters such as the mass of the charm quark and alternative 
parton distribution functions;  
\item
changes in the factorization and renormalization scales; and
\item
other non-perturbative effects (hadronization and 
intrinsic transverse momentum of the colliding partons).
\end{enumerate}
\end{enumerate}
\subsection{Charm Quark Production}
\label{ssec:prod}
Both the  HVQMNR and {\sc Py\-thia/Jet\-set} packages use a 
perturbative QCD framework to obtain the differential cross 
section for producing a $\ccb$ pair:
\begin{equation}
\label{eq:cross}
{\rm d}{\sigma_{c\overline{c}}} = \sum_{i,j}\int{\rm d}x_{b} \,
{\rm d}x_t \, f_i^{b}(x_{b},\, \mu_F) \, f_j^t(x_t, \mu_F) \,
{\rm d}\hat{\sigma}_{ij}(x_{b}P_{b}, x_tP_t,
p_c, p_{\overline{c}}, m_c, \mu_R),
\end{equation}
where
\begin{itemize}
\item $P_b$ ($P_t$) is the momentum of the beam (target) hadron in the center 
of mass of the colliding hadrons;
\item $x_b$ ($x_t$) is the fraction of $P_b$ ($P_t$) 
carried by the hard-scattering parton from the beam (target) hadron;
\item $f_i^b$ ($f_i^t$) is the parton distribution function for the 
beam (target) hadron;
\item $\mu_R$, the renormalization scale, and $\mu_F$, the factorization scale,
 come from the perturbative QCD renormalization procedure which transforms the 
QCD coupling constant $g = \sqrt{4 \pi \alpha_s}$ and the $\pi^-$ and nucleon 
wave functions from ``bare'' (infinite) values to physical ({\it i.e.}, finite 
and measurable) values;
\item ${\rm d}\hat{\sigma}_{ij}$ is the 
differential
cross section for two hard-scattering 
partons to produce a pair of charm quarks, each with mass $m_c$, and with 
four-momenta $p_c$ and $p_{\overline{c}}$.
\end{itemize}

Leading order ($\alpha_s^2$) contributions to the $\ccb$ cross section require 
the charm and anticharm quarks to be produced back-to-back in the center of 
mass of the $\ccb$ pair.  
The (unknown) partonic center of mass is boosted in the beam direction with 
respect to the (known) hadronic center of mass. 
At fixed target energies, this boost smears the longitudinal momentum
correlation while preserving the transverse correlation.
Therefore, leading order calculations predict delta function distributions 
(i.e., maximal correlations) for variables which measure transverse 
correlations, such as $\Delta \phi_{c\overline{c}} = 180^{\circ}$ and 
$p_{t,c\overline{c}}^2 = 0$, but predict small correlations in the 
longitudinal-momentum correlation variables $\dxf$, $\sxf$, $\dy$ and $\sy$.

These leading-order predictions are altered by the inclusion of
higher order effects.  
The HVQMNR program adds the NLO ($\alpha_s^3$) corrections 
to the leading order calculation.  
NLO processes such as $g g \to c \overline{c} g$ produce $\ccb$ pairs that 
are no longer back-to-back, smearing the leading order delta function 
distributions for $\Delta \phi_{c\overline{c}}$ and  $p_{t,c\overline{c}}^2$.

The {\sc Py\-thia/Jet\-set} event generator accounts for higher order 
perturbative 
QCD effects via a 
``parton shower'' approach~\cite{ref:bib17}.
In this approach each of the two incoming and two outgoing 
partons, whose distributions are based on leading-order matrix elements, can 
branch --- backwards and forwards in time respectively --- into two partons, 
each 
of which can branch into two more partons, {\it etc}.  
This evolution continues until a small momentum scale is reached.
In addition, the {\sc Py\-thia/Jet\-set} event generator gives the 
hard-scattering 
partons an intrinsic transverse momentum $k_t$.  
Both the parton showers and the intrinsic transverse momentum tend to 
smear 
the transverse correlations, as shown in the Appendix.

The extent to which transverse-momentum correlations are smeared provides a 
measure of the importance of higher order perturbative effects.
In addition, since the leading order calculation predicts very little
longitudinal-momentum correlation, an enhancement of the longitudinal-momentum
correlation also provides evidence for higher order perturbative effects
or non-perturbative effects such as hadronization, described below.

\subsection{Hadronization}
\label{ssec:hadro}

The process whereby charm quarks are converted to hadrons is known as 
hadronization or fragmentation.  
Since this process occurs at an energy scale too low to be calculable by 
perturbative QCD, fragmentation functions are used to parameterize the 
hadronization of the charm quark.
Such functions have been measured by several $e^+e^-$ experiments.  
The hadroproduction environment in $\pi^-$-N interactions, however, is quite 
different from the $e^+e^-$ environment.  
In $e^+e^-$ interactions, the light quarks in the produced charm hadrons
must come from the vacuum. 
Hadroproduction leaves light quark beam and target remnants which are 
tied by the strong force to the charm quarks.  
Interactions between these remnants and the charm quarks can dramatically 
affect the momentum and flavor of the observed charm hadrons.

The {\sc Py\-thia/Jet\-set} event generator uses the Lund string fragmentation
framework, described in the {\sc Py\-thia/Jet\-set} manual\cite{ref:bib5},
to hadronize the charm quarks.  To illustrate this model we consider an example
from E791 where a gluon from a $\pi^-$ and a gluon from a nucleon in the
target interact to form a $\ccb$ pair.  This accounts for $\sim 90\%$ of the
theoretical cross section for 500 GeV/$c$ $\pi^-$-N interactions.
After the gluon-gluon fusion, the remnant $\pi^-$ and nucleon are no longer 
color-singlet particles.  
The remnant $\pi^-$ is split into two valence quarks, and the remnant nucleon 
into a valence quark plus a diquark.
Given this minimal set of partons --- ($c$, $\overline{c}$), 
$(\overline{u}, d)_{\pi}$, and $(qq, q)_N$ --- 
the two dominant ways to make color-singlet strings, 
and the ones PYTHIA uses are\cite{ref:bib18}:
\begin{eqnarray}
\label{eq:strtop}
(c, \; \overline{u}_{\pi}), &(\overline{c}, \; q_N), &{\rm and} \; \; 
(d_{\pi}, \; qq_N); \; {\rm or} \\
(\overline{c}, \; d_{\pi}),& (c, \; qq_N),& {\rm and} \; \; 
(\overline{u}_{\pi}, \; q_N). \nonumber
\end{eqnarray}
In the center of mass of a particular $q\overline{q}$ system, such as
$\overline{c}d$, the $\overline{c}$ and $d$ are moving apart along the 
string axis.  
As they move apart, energy is transferred to the color field.  
When this energy is great enough, $q\overline{q}$ pairs are created from the 
vacuum with equal and opposite transverse momentum (with respect to the 
string axis) 
according to a Gaussian distribution.  
The transverse momentum relative to the string axis of the resulting 
$\overline{c}q$ 
meson is determined by the $q$ quark since the $\overline{c}$ contributes 
none.  
The longitudinal momentum of the meson is given by a fragmentation function 
which 
describes the probability that a meson will carry off a fraction $z$ of the 
available longitudinal momentum.  
By default, heavy quark fragmentation is performed according to a Lund  
fragmentation function~\cite{ref:bib7} modified by Bowler~\cite{ref:bib19}:
\begin{equation}
\label{eq:flund}
f(z) \; \propto \; \frac{(1-z)^a}{z^{1+b m_Q^2}} \exp
 \left( \frac {-b m_t^2}{z} \right)
\end{equation}
where $m_t^2 \equiv M_h^2 + \, p_t^2$ is the transverse mass of the hadron and
 $m_Q$ is the mass of the heavy quark.
The default {\sc Pythia/Jetset} settings are $a = 0.3$
and $b = 0.58 \; ({\rm GeV}/c^{2})^{-2}$.

When the remaining energy in the string drops below a certain cutoff (dependent 
on the mass of the remaining quarks) a coalescence procedure is followed, which 
collapses the last partons into a hadron while conserving energy.  
The entire string system is then boosted back into the lab frame.  
In the case of a $(\overline{c}, \, d_{\pi})$ or $(c, \,\overline{u}_{\pi})$ 
string, this boost will tend to increase the longitudinal momentum of the charm 
hadron with respect to the charm quark since the $d_{\pi}$ and 
$\overline{u}_{\pi}$ will tend to have large longitudinal momentum.  
The opposite will occur, however, for a $(\overline{c}, \, q_N)$ string.  

In some fraction of events, strings will be formed with too little energy to 
generate $q\overline{q}$ pairs from the vacuum.  
In these cases the $c$ quark (antiquark) will coalesce into a single meson with 
the beam antiquark (quark) or will coalesce into a single baryon (meson) with
the target diquark (``bachelor'' quark).
This will tend to enhance production of charm hadrons with a light quark in 
common with a valence beam quark in the forward direction (beam fragmentation 
region) and production of charm hadrons with a light valence quark or diquark 
in common with the target in the target fragmentation region.
This phenomenon has been used to explain the leading particle effect 
seen in charm hadroproduction experiments\cite{ref:bib20}--\cite{ref:bib25}.

However, in most events, the string has sufficient 
energy\footnote{In contrast, at high $\xf$ most of the particle energy is
taken up by the individual partons, so that the
string has insufficient energy
to produce $q\overline{q}$ pairs, and large asymmetries are
seen by experiments.}
to produce
at least one $q\overline{q}$ pair from the vacuum. 
In this type of beam/target ``dragging,''
the strength of the dragging is
not dependent 
on the light quark content of the produced particle.

\begin{figure} % Figure 1
        \centering
        \centerline{\epsfig{file=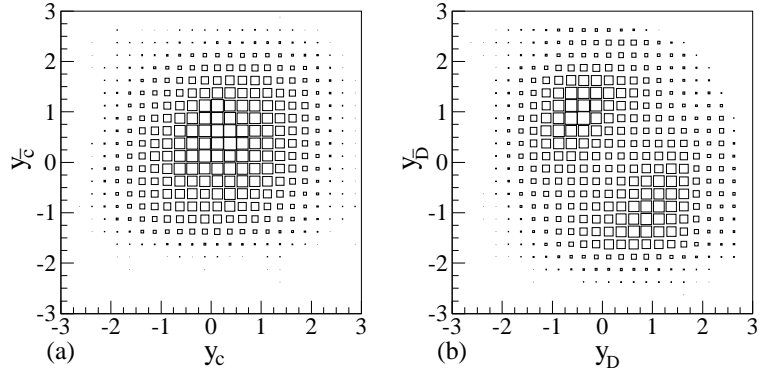,width=4.0in,angle=0}}
        \caption{Scatter plots of $\ycb$ vs.
        $\yc$ and $\ydb$ vs. $\yd$, from
        100,000 {\sc Pythia/Jetset}
        $\ccb$ and $\ddb$ events, showing the correlation introduced by the
        hadronization model.  As discussed in Sec.~\protect\ref{sec:selection},
        we only reconstruct $\ddb$ events in the
        region $-0.5 < \yddb < 2.5$.}
        \label{fig:fig1}
\end{figure}

These effects are evident in Fig.~\ref{fig:fig1}, which shows a 
scatter plot of the charm and anticharm rapidities for {\sc Pythia/Jetset} 
$\ddb$ events.\footnote{Default values are used for all {\sc Pythia/Jetset} 
parameters.}
Comparison of the scatter plot of the charm and anticharm quark rapidities, 
Fig.~\ref{fig:fig1}a, to the scatter plot of the $D$ and $\db$ rapidities, 
Fig.~\ref{fig:fig1}b, clearly demonstrates that significant correlations are 
introduced by hadronization.

Both the degree of correlation between the $D$ and $\db$ longitudinal momenta 
as well as differences in production of the four types of $\ddb$ pairs --- 
$\dzdb$, $\dzdm$, $\dpdb$ and $\dpdm$ --- provide information about the charm 
quark hadronization process in a hadronic environment.

%%%%%%%%%%%%%%%%%%%%%%%%%%%%%%%%%%%%%%%%%%%%%%%%%%%%%%%%%%%%%%%%%%%%%%%%%%%
%\input{offline_doc_290_3_r1.tex} %  E791
\section{Experiment E791} 
\label{sec:e791}

\begin{figure} % Figure 2
        \centering
        \centerline{\epsfig{file=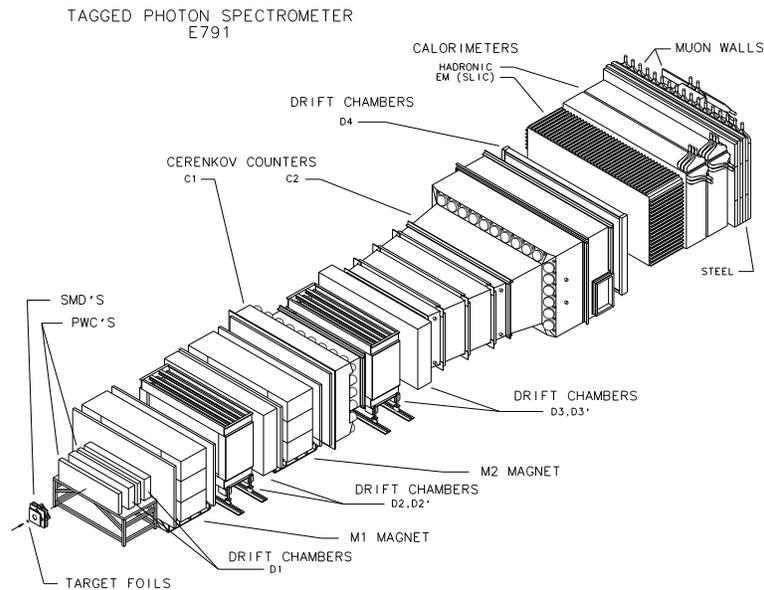,width=4.0in,angle=0}}
        \caption{The E791 spectrometer.}
        \label{fig:fig2}
\end{figure}

The results reported in this paper are based on a data sample recorded
by Fermilab experiment E791 during the 1991/92 fixed-target run.  The
E791 spectrometer is illustrated in Fig.~\ref{fig:fig2}.  A
500~GeV/$c$ $\pi^-$ beam impinged on platinum and carbon targets.
The spectrometer consisted of proportional wire
chambers (PWC's) and silicon microstrip detectors (SMD's) upstream and
downstream of the targets, two magnets, 35 drift chamber (DC) planes,
two \v Cerenkov counters, an
electromagnetic calorimeter, a hadronic calorimeter and a muon detector
composed of an iron shield and two planes of scintillation counters.

The spectrometer was an upgraded version of the apparatus used in
Fermilab experiments E516, E691, and E769\cite{ref:bib26}.
The major differences between the earlier versions and E791 were the
addition of more planes of SMD's, enhancement of
the muon identification system, new front-end detector-signal digitizers and 
a new data acquisition system.
In general, E791 finds that the most important parts of the spectrometer for analysis
are the charged-particle tracking system and the threshold \v Cerenkov
counters; although the threshold \v Cerenkov counters were used minimally in the
analysis reported in this paper.
   
\subsection{Target} 
\label{ssec:targ}

The target consisted of five foils
with center-to-center separations that varied from 14.8 to 15.4 mm.
The most upstream foil was 0.5~mm thick and was made of platinum to provide a 
significant interaction probability in a thin target.
The next four foils were 1.6~mm thick and were made of industrial diamond.
The low Z of these carbon targets minimized multiple scattering, while the
higher density of diamond permitted thinner downstream targets for the same
interaction probability.
The total pion interaction length of all five targets was about 1.9\%.
This target arrangement was chosen so that most of the 
particles with lifetimes
and momenta within the range of interest to this experiment
have a decay vertex in the gaps, where there is less background
from secondary interactions.

\subsection{The Spectrometer}
\label{ssec:spect}

The $\pi^-$ beam particle was tracked with eight PWC planes and six  
SMD planes upstream of the target region. 
Downstream of the targets, the
charged-particle tracking system consisted of 17 SMD planes, 
two PWC planes, and 35 drift chamber planes.  In general several 
planes of tracking chambers with
different angular orientations around the beam
axis were grouped together in each tracking station to provide hit
ambiguity resolution.
The various coordinates ($x$, $y$, $w$, $u$, $v$) measured by the planes 
in the tracking chamber stations
were defined 
relative to a right-handed 
coordinate system $x-y-z$ in which increasing $z$ was in the beam 
direction, $x$ was the horizontal dimension and $y$ increases vertically
upward. 
The $w$, $u$ and $v$ axes were rotated by $+60^\circ$, $+20.5^\circ$,
and $-20.5^\circ$ with respect to the positive $x$ axis.
%The spectrometer was approximately centered on the beam line.

The beam PWC's\cite{ref:bib27} had a wire spacing of 1 mm 
and were arranged in two stations widely separated in $z$ 
to measure the angle of the incoming beam particle with high precision.
The first station was 31 m upstream, and the second was 12 m upstream 
of the last carbon target.
Each station consisted of 4 planes: two staggered $x$ planes, a
$y$ plane and a $w$ plane.

The beam SMD's had a pitch of 25~$\mu$m  and
were also arranged in two stations, each 
with an $x$, $y$ and $w$ plane.
The first SMD station was 80 cm upstream of the most
downstream target, and the second station was 30 cm upstream of
this target.
The system of SMD's downstream of the targets started 2.8~cm 
downstream of the last target and extended for 45~cm.  It had
a maximum angular acceptance of about $\pm$125 mr in both $x$ and $y$. 
Each of the first two planes ($x$ and $y$) had  
an active area of 2.5~cm by 5~cm, a  pitch of 25~$\mu$m in the central  9.6~mm
and 50~$\mu$m in the outer regions, and 
an efficiency of about 84\%.  
The next nine planes were identical to those used in E691\cite{ref:bib28}.
Each plane
had 
a pitch of 50~$\mu$m, 
and an efficiency from 88\% to 98\%. They were instrumented  to give
an acceptance of ~$\pm 100$ mrad with respect to the 
center of the most downstream target.
They measured $x-y-v-y-x-v-x-y-v$ coordinates respectively.
The final six SMD planes 
had active areas of 9~cm by 9~cm. The inner 3~cm had a pitch of 
50~$\mu$m while the outer regions had an effective pitch of 200~$\mu$m.
These measured $v-x-y-x-y-v$ coordinates.
The efficiencies ranged from 96\% to 99\%.

\begin{figure} % Figure 3
        \centering
        \centerline{\epsfig{file=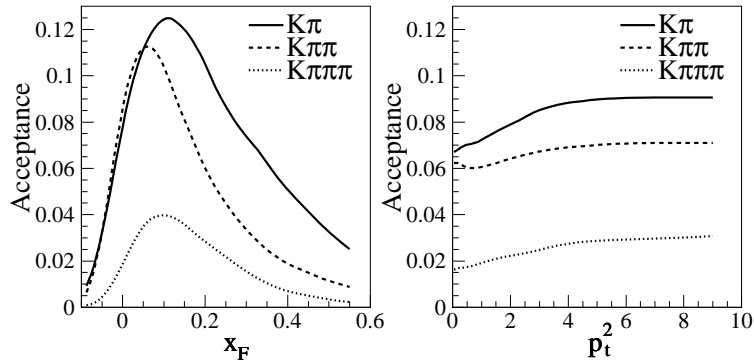,width=4.0in,angle=0}}
%        \caption{E791 acceptance functions vs. $x_F$ and $p_t^2$
%        for $K\pi$, $K\pi\pi$ and $K\pi\pi\pi$ candidates.  The acceptance
%        shown here is for a loose set of single-charm selection criteria.  The
%        $p_t^2$ acceptance is obtained for charm mesons with $-0.1<x_F<0.6$.}
\caption{
E791 acceptance functions vs. $x_F$ and $p_t^2$ for $K\pi$, $K\pi\pi$ and
$K\pi\pi\pi$ candidates. The inclusive charm acceptance shown here was used
to obtain approximately 150,000 reconstructed $D^0$ and $D^+$ charm meson decays
in the E791 spectrometer as described, for example, in
the study of single-charm production~\cite{ref:bib28b}~and~\cite{ref:bib28c}.
The $\ptt$ acceptance is obtained for charm mesons with $-0.1<\xf<0.6$.
The acceptance for events in which both charm particles are detected
is quite different and is documented in detail later in this paper.
(See Section~\ref{ssec:accept}.)
}
        \label{fig:fig3}
\end{figure}

The drift chambers were arranged in four stations as illustrated in 
Fig.~\ref{fig:fig2}.  Each station was subdivided into substations
with plane orientations such that an $x-y-z$ space point could be
reconstructed in each substation. The characteristics of these chambers are 
given in Table~\ref{tab:tab2}.  Since the beam, which operated at about 2 MHz
throughout
the run, passed through the center of the drift chambers in a small region
instrumented with very few wires, each plane
had a central inefficient region in which the efficiency decreased to $<10$\%
and the resolution was degraded by as much as a factor of four. 
The profile of the
efficiency
and resolution degradation region was approximately gaussian 
in an angular region of three to four mrad centered on the beam.
The extent of the inefficient region increased with time during the run
and is the major source of systematic uncertainty associated with 
the acceptance at large $\xf$.
Each substation of the first drift chamber station was augmented
by a PWC which measured the $y$ coordinate. These PWC's had a wire spacing
of 2 mm.  
Typical inclusive single charm acceptances for two, three, and four
particle $D$ decays are shown in Fig.~\ref{fig:fig3}.
 
\begin{table} [htbp]
        \caption{
        Characteristics of the DC tracking chambers
        for E791.
        ``View'' refers to the coordinate measured by that plane,
        with $u$ = $+20.5^\circ$, $v$ = $-20.5^\circ$ and $x'$ staggered by
        one-half a wire spacing relative to the $x$ plane. The efficiency is
        for the region outside the central inefficient area.}
\begin{tabular}{lcccc} \hline \hline
Station & D1 & D2 & D3 & D4 \\ \hline \hline
Approximate size (cm) & 130 $\times$  75 & 280 $\times$ 140 &
                                320 $\times$ 140 & 500 $\times$ 250 \\
Number of substations         &  2 & 4 & 4 & 1 \\
Views per substation          & $x,\ x',\ u,\ v$ & $u,\ x,\ v$
                              & $u,\ x,\ v$       & $u,\ x,\ v$ \\
$u$ and $v$ cell size (cm)    & 0.446  & 0.892  & 1.487  & 2.974 \\
$x$ cell size (cm)            & 0.476  & 0.953  & 1.588  & 3.175 \\
$z$ position of first plane   & 142.4  & 381.4  & 928.1  & 1738. \\
$z$ position of last plane    & 183.7  & 500.8  & 1047.1 & 1749.2 \\
Approx. resolution ($\mu$m)& 430    & 320    & 260    & 500    \\
Typical efficiency            & 92\%   & 93\%   & 93\%   & 85\%   \\  \hline
\end{tabular}
\label{tab:tab2}
\end{table}

Momentum
analysis was provided by two dipole magnets that bent particles in the same
direction in 
the horizontal plane. The transverse momentum kicks 
were 212~MeV/$c$ for the first magnet and 320~MeV/$c$ for the second magnet. 
The centers of the two magnets 
were 2.8 m and 6.2 m downstream of the last target, respectively.
The $x-y-z$ aperture of the pole faces of the first magnet was 
183~cm by 81~cm by 100 cm 
and  that of the second magnet was 183~cm by 86~cm by 100 cm.  

Two segmented, gas-filled, threshold 
\v Cerenkov counters\cite{ref:bib29} provided particle
identification over a large range of momenta.  The threshold momenta
above which a charged particle emits light were 6, 20 and 38~GeV/$c$ for
$\pi$'s, $K$'s, and $p$'s, respectively, for the first counter, and
11, 36, and 69~GeV/$c$ for the second. 
The particle identification algorithm correlates the \v Cerenkov light
observed in a given mirror-phototube segment with the charged particle 
tracking information.  The algorithm indicates the likelihood that
a charged particle of a given mass could have generated the observed
\v Cerenkov light in the segment(s) in question.

The electromagnetic calorimeter,
which we called the Segmented Liquid Ionization Calorimeter (SLIC),  
consisted of 20 radiation
lengths of lead and liquid scintillator and was 19~m from the target. Layers
of scintillator counters 3.17 and 6.24~cm wide were arranged transverse to the
beam and their orientations alternated among horizontal and 
$\pm 20.5^\circ$ 
with respect to the
vertical direction\cite{ref:bib30}. The hadronic calorimeter
consisted of six interaction lengths of steel and acrylic scintillator. There
were 36 layers, each with a 2.5-cm-thick steel plate followed by a
plane of 14.3-cm-wide by 
1-cm-thick scintillator slats; the slats were arranged alternately in the
horizontal and vertical directions, and the upstream and downstream halves
of the calorimeter were 
summed separately\cite{ref:bib31}. The
signals from the hadronic calorimeter as well as those from the
electromagnetic calorimeter were used for electron
identification. Signals from both calorimeters were used to form the
transverse energy requirement in the hardware trigger~\cite{ref:bib32}.
Electron identification was not used in this 
analysis.

Muons were identified by two planes of scintillation counters located
behind a total of 15 interaction lengths of shielding, 
including the calorimeters. The
first plane, 22.4~m from the target, consisted of twelve 
40-cm-wide by 300-cm-long
vertical scintillation counters in the outer region and three counters 60~cm
wide in the central region. 
The second plane, added 
for E791,
consisted of 16 scintillation counters 24.2~m from the target. These counters
were each 14~cm wide and 300~cm long, and measured position in the
vertical plane. These counters were equipped with TDC's which provided 
%%some indication of
information
on the horizontal position of the incident muons~\cite{ref:bib32}. 
Muon identification was not used in this 
analysis.

\subsection{Trigger and Data Acquisition}
\label{ssec:trig}

To minimize biasing the charm data sample,
the trigger requirements were very loose.
The most significant requirements were that the signal in a scintillation
counter downstream of the target be at least four times the 
most likely signal from
one minimum-ionizing particle,
and that 
the sum of the
energy deposited in the electromagnetic and hadronic calorimeters,
weighted by the sine of the angle relative to the beam, be
above a threshold corresponding to 3 GeV of transverse
energy.
The time for the full hardware trigger decision was about 470~ns.
This trigger was fully 100\% efficient for charm decays.

A total of 24,000 channels were digitized and read out in 50~$\mu$s with a
parallel-architecture data acquisition system\cite{ref:bib33}. Events were
accepted at a rate of 9 kHz during the 23-second Tevatron beam
spill.  The typical recorded
event size was 2.5 kbytes. 
Data were written continuously (during both the
23-second spill and the 34-second interspill periods) to 
forty-two 
Exabyte\cite{ref:bib34} model 8200
8-mm-tape drives at a rate of 9.6~Mbytes/s. Over $2\times10^{10}$ hadronic
interactions were recorded on 24,000 tapes.

\subsection{Data Processing} 
\label{ssec:data}  
The $2\times10^{10}$ interactions recorded 
constitute
about
50~Terabytes of data. Event reconstruction and filtering took place over a
period of two and a half years at four locations: the University of 
Mississippi, 
The Ohio State University (moved to Kansas State University in 1993), Fermi
National Accelerator Laboratory, and 
O Centro Brasileiro de Pesquisas F\'\i sicas,
Rio de Janeiro (CBPF). The first three sites used clusters of commercial
UNIX/RISC workstations controlled from a single processor with multiprocessor
management software\cite{ref:bib35}, while CBPF used ACP-II 
custom-built single-board computers\cite{ref:bib36}.

As part of the reconstruction stage, a filter 
was applied which kept $\sim$20\% of
the events. This filter was effectively an offline trigger.  To pass this 
filter, an event was
required to have a reconstructed 
primary 
production vertex whose location coincided with one
of the target foils. The event also had to include at least one of the
following:
\begin{enumerate}
\item At least one reconstructed secondary 
decay
vertex of net charge 0 for an even number of decay tracks and $\pm 1$ for
an odd number of decay tracks. 
The longitudinal
separation of the secondary vertex from the primary had to be at least
four sigma for secondary vertices with three or more tracks and at least six
sigma for vertices with two tracks, where sigma is the error in the separation,
\item At least one reconstructed $K_s\rightarrow \pi^-\pi^+$ 
or $\Lambda\rightarrow p\pi$ candidate whose decay 
was observed  upstream of the first magnet,
\item and for part of the run, 
at least one reconstructed $\phi\rightarrow K^+K^-$ candidate.
\end{enumerate}
For one-third of the data sample, several other classes of events 
were also kept. These are included in analyses not covered in this paper: 
\begin{enumerate} 
\setcounter{enumi}{3}
\item Events in which the net charge of all the reconstructed tracks was
negative and their total momentum was a large fraction of the beam momentum.
\item $K_s\rightarrow \pi^-\pi^+$
or $\Lambda\rightarrow p\pi$ candidates that decayed inside  the
aperture of the first magnet.
\end{enumerate} 
Following the initial reconstruction/filter, which was applied to all events,
additional selections of events were made to further divide the large data
sample into subsets by class of 
physics analysis.

\subsection{Detector Performance}
\label{ssec:perf}

The important detector performance characteristics for this
analysis are the resolution for reconstructing the positions of both
the primary interaction 
and secondary decay vertices, 
the efficiency for reconstructing the trajectories of charged particles,
the resolution for measuring charged track momenta,
and the efficiency and the misidentification rates for identifying 
charged pions and kaons using information from the \v Cerenkov counters.

The  resolution for measuring the position of the primary vertex along the
beam direction varies from about 240~$\mu$m for the most downstream target
foil to 450~$\mu$m for the upstream  foil. 
The variation is due to 
the extrapolation from the SMD system and to
multiple scattering in material downstream of the
interaction. The mean number of reconstructed tracks used to fit the primary
vertex is seven. The measured secondary vertex resolution depends on the 
decay mode, the momentum of the $D$, 
and the selection criteria. 
For example,  the vertex
resolutions along the beam direction for $K^-\pi^+$ and $K^-\pi^+\pi^-\pi^+$
are 320 and 395~$\mu$m, respectively, for a mean $D^0$ momentum of 
65~GeV/$c$, and worsen by 33 and 36~$\mu$m
for every 10~GeV/$c$ increase in $D^0$ momentum.

The total efficiency, including acceptance,
for reconstructing charged tracks is approximately
80\% for particles with a momentum greater than 30~GeV/$c$ and drops to around
60\% for particles of momentum 10~GeV/$c$. (This includes the inefficiency in the
beam region.)
For tracks which pass through both magnets and have
a momentum greater than 10 GeV/$c$, the average resolution 
for measuring charged particle momentum $p$ is 
$\delta p/p=0.6\% \oplus (0.02p)\%$  where $\oplus$ stands for 
the quadratic sum, and $p$ is in GeV/$c$. 
Tracks which pass through 
only the first magnet have a resolution 
$\delta p/p=2\% \oplus (0.1p)\%$.
The mean $D$ mass resolution for hadronic decays to two, three and four
charged particles varies from 13 to 8~MeV/$c^{2}$
as the decay multiplicity increases.
The mass resolution varies by about a factor of 2 between low and high 
momentum $D$ mesons.

In most E791 analyses, the \v Cerenkov counters play a very important role
\cite{ref:bib37}.
%%[ref. for example, DCSD D+ -> K+ pi- pi+ and sigma_c mass difference - Phys.
%%Lett. B404 187 (1997) and Phys. Lett. B379 292 (1996), respectively]
However, in this analysis with the two fully reconstructed D-meson decays, the
\v Cerenkov counters play a minimal role.
We use the \v Cerenkov counters for charged kaon 
identification.
The kaon identification efficiencies and misidentification probabilities
vary with longitudinal and transverse 
momentum and with the signatures required in the \v 
Cerenkov counters.
For typical particle momenta in the range 20~GeV/$c$ to 40~GeV/$c$, 
and for the nonstringent criteria used for some of the final states, 
in this analysis, 
the \v Cerenkov identification efficiency for a kaon ranged from 
64\% to 72\% while
the probability for a pion to be misidentified as a kaon ranged from 6\%
to 12\%. 

A complete Monte Carlo simulation of the apparatus was used in this analysis
to calculate the 
efficiency
and investigate systematic effects.  The 
simulation included all relevant 
physical processes such as multiple interactions and multiple scattering
as well as geometrical apertures and resolution effects. 
It produced
data in the same format as the real experiment.
That Monte Carlo data was
then reconstructed and analyzed with
the same software as 
the real data. 

%%%%%%%%%%%%%%%%%%%%%%%%%%%%%%%%%%%%%%%%%%%%%%%%%%%%%%%%%%%%%%%%%%%%%
%\input{offline_doc_290_4_r1.tex} %  \section{Event Selection}  
\section{Event Selection}  
\label{sec:selection} 
In each E791 event,  
we search for two charm mesons  ($D^0$, $\dzb$, $D^+$ or $D^-$)             
decaying to Cabibbo-favored final states that can be reconstructed with         
relatively high efficiency:   
$D^0\to K^-\pi^+$,
$D^0\to K^-\pi^+\pi^-\pi^+$, 
$D^+\to K^-\pi^+\pi^+$,    and the charge conjugate 
modes.\footnote{Unless noted otherwise, charge conjugate modes are
always implied.}
To optimize the efficiency for reconstructing
charm pairs,
we search for both $D$ candidates simultaneously, rather than
searching for the two candidates consecutively.
In such a simultaneous search, we can require that one candidate
or the other satisfy a fairly stringent selection criterion based on
a particular variable used to discriminate charm decays from background,
or that both candidates satisfy less stringent criteria.

We start with a sample of
events, each
containing two candidate
$K^-\pi^+$, $K^-\pi^+\pi^+$ or $K^-\pi^+\pi^-\pi^+$
combinations 
with invariant mass between 1.7 and 2.0~GeV$/c^2$, and rapidity in the
range $-0.5 < \yddb < 2.5$.  These candidates are found by looping over
all reconstructed tracks.  The primary 
vertex
is then refit after removing
tracks which are associated with either candidate.
No particle identification requirements are applied at this time.
Candidates are rejected if any charged track, the
primary vertex or either of the two secondary vertices
do not meet minimal fit quality criteria.
The sample of candidate pairs that pass just these criteria is
dominated by combinatoric backgrounds.
To choose further selection criteria, we use this sample to 
represent background.
To represent signal, we use reconstructed charm pairs
generated with the Monte Carlo program described
at the end of the previous section.
We then search for selection criteria
that provide optimal discrimination between signal and background.

In order to extract the signal, we use selection criteria defined by
discrimination
variables (properties of the candidate event)
and minimum or maximum allowed values for
each variable.
For candidate pairs with the same final states
({\it i.e.}, both $K\pi$, both $K\pi\pi$, or both $K\pi\pi\pi$),
the same discrimination variables and maximum or minimum values
are used for both candidate $D$'s;
for pairs with different final states, the discrimination variables are
allowed to be different for the two candidate $D$'s.
%The selection criteria can be applied to both candidate $D$'s,
%or to one or the other.

The discrimination variables used
address the following questions.
Is a $D$ candidate
consistent with originating from the
primary interaction vertex?
Is the vertex formed by the decay products of
a $D$ candidate well separated from the primary
interaction vertex and not inside a target foil?
Do any of the decay products of the $D$ candidate
appear to originate from the primary interaction
vertex or from the other $D$ candidate vertex?
Is the scalar sum of the
squares of the transverse momenta of the $D$ candidate
decay products, with respect to the $D$ candidate
trajectory, indicative of a heavy meson decay?
Is the \v Cerenkov information for the kaon candidate
consistent with that for a real kaon? As an example of the cuts used: the 
$p_t$ balance cut was 400 MeV/$c$,
and the secondary vertices were separated from the
primary vertices by 8 times the rms-uncertainty
in the separation.

To optimize the significance of the signal,
we repeatedly choose the selection criterion that maximizes
$N_S/\sqrt{N_S+N_B}$ 
while rejecting no more than 5\% of the (Monte Carlo) 
signal.
$N_S$ is the number of signal
pairs satisfying the selection criterion, 
determined by Monte Carlo simulation, and $N_B$ is 
the number of background pairs, determined from the
data.
This requires properly normalizing
the number of signal pairs in the Monte Carlo to 
the number of signal pairs
in the data.
When the background becomes dominated by pairs with
only one true $D$ decay, we exclude from the background
sample only those pairs in which both $D$ candidates
lie in a narrow range around the $D$ mass.

We iterate the procedure of finding the optimal
selection criterion (always allowing variables to be
reused in subsequent iterations) until the significance of
the signal no longer increases.  
The selection criteria are optimized separately for
each of five decay topologies of $\ddb$ pairs: 
2-2, 3-3, 2-3, 2-4 and 3-4, where each
integer represents the number of charged particles in
the decay.\footnote{We also searched for 4-4 pairs but the efficiency
was too low to add much to the statistical significance of the sample.
We did not use these pairs in the 
final analysis.}
We find that selection criteria are more
often applied to one $D$ candidate or the other,
rather than to both, especially early in the optimization procedure.
In several cases, a criterion will be 
applied to one of the $D$ candidates, and a more stringent
criterion involving the same discrimination variable will be 
applied to the other.

\begin{figure} % Figure 4
        \centering
        \centerline{\epsfig{file=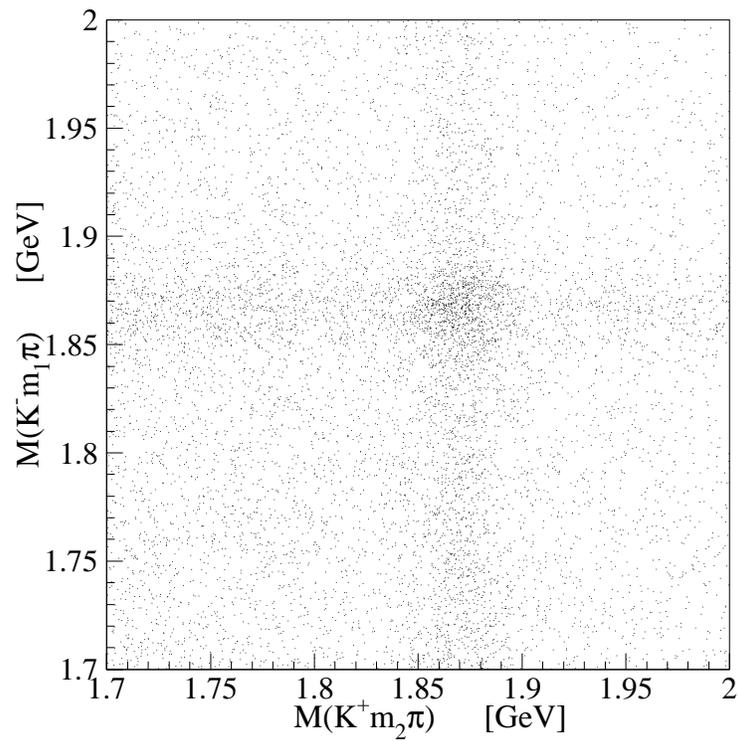,width=4.0in,angle=0}}
        \caption{Scatter plot of the $D$-candidate mass versus
        the $\db$-candidate mass for the final unweighted charm-pair sample.}
        \label{fig:fig4}
\end{figure}

After optimizing our selection criteria, we end up with
a sample of 9254 events in the data with both $D$ candidates in
the mass range 1.7 to 2.0~GeV$/c^2$ and in the rapidity range
$-0.5$ to $2.5$.
Only pairs in which the two $D$ candidates have opposite
charm quantum numbers are included in this sample.
No significant signal for $D D$ or $\db\:\!\db$ pairs
is observed.
In Fig.~\ref{fig:fig4}, we plot the mass of the
$D$ candidate versus the mass of the $\db$ candidate,
for all five types of pairs.
Three types of candidate pairs are evident in 
this scatter plot.
Combinatoric background pairs consisting of a fake
$D$ and a fake $\db$ candidate are spread over 
the entire plot.  
The density of these points decreases linearly with
increasing candidate-$D$ mass.
Pairs containing one real and one fake $D$ candidate
appear as horizontal and vertical bands
(called $D$ and $\db$ ridge events, respectively).
In the center of the plot, we see an enhancement due
to the crossing of the two bands and 
due to real pairs of $D$ mesons.  

%%%%%%%%%%%%%%%%%%%%%%%%%%%%%%%%%%%%%%%%%%%%%%%%%%%%%%%%%%%%%%%%%%%     
%\input{offline_doc_290_5_r1.tex} %  \section{Data Analysis}
\section{Data Analysis}
\label{sec:analysis}
In this section we describe the analysis procedures by which we
determine the number of signal events in the full data sample
shown in Fig.~\ref{fig:fig4}, as well as in each 
bin of the physics variables
used to study the charm-pair production. Acceptance corrections
include geometric acceptance, relative branching
ratios, reconstruction efficiencies, and
event selection efficiencies.

\subsection{Determination of Yields}
\label{ssec:yield}
The experimental resolution for the $D$ mass measurement
in the E791 spectrometer depends on both the $D$ decay mode
and the $\xf$ of the $D$ meson, and the mean reconstructed mass 
depends on the decay mode.
Therefore, we fit to
the normalized $D$ mass defined as
\begin{equation} \label{eq:mn}
M_n \equiv {M - M_D\over \sigma_{M}}, 
\end{equation}
where $M$ is the measured mass, $M_D$ is the mean 
measured mass for the particular decay mode of 
the $D$ candidate, and $\sigma_{M}$ is the experimental
resolution for the particular decay mode and $\xf$
of the $D$ candidate.
%The average resolution varies between 8 and 13~MeV for different
%decay modes. It varies by about a factor of 2
%between low and high $\xf$.

In this analysis we use the maximum likelihood method 
which assumes we have $N$
independent measurements of one or more quantities and
that these quantities $\vec{z}$ are distributed according to some
probability 
density 
function $f(\vec{z};\vec{\alpha})$ where
$\vec{\alpha}$ is a set of unknown parameters to be
determined.  To determine the set of values $\vec{\alpha}$ that maximizes
the joint probability for all events, we numerically solve the set of
equations\cite{ref:bib38}:
\[ \frac{\partial \ln L(\vec{\alpha})}
{\partial \alpha_j} = 0 \; \; {\rm where} \; \;
L(\vec{\alpha}) = \prod_{i=1}^N f(\vec{z}_i;\vec{\alpha}). \]

The quantities that we measure for each event are
the normalized mass of both the
$D$ and $\overline{D}$ candidate; {\it i.e.},
$\vec{z} = (M_n^{K^- m_{1}\pi}$, $M_n^{K^+ m_{2}\pi}$).
The unknown parameters in the maximum likelihood
fit are the number of signal events, $N_S$;
combinatoric events, $N_C$;
events with one real $D$ and one combinatoric background called $D$-ridge
events, $N_D$; 
events with one real $\overline{D}$ and one combinatoric
background called $\overline{D}$-ridge events, $N_{\overline{D}}$;
the slope of the background $K^- m_{1}\pi$ distribution, $S^D$; and
the slope of the background $K^+ m_{2}\pi$ distribution, $S^{\overline{D}}$.
That is, the unknown parameters are
\[ \vec{\alpha} = (N_S,N_C,N_D,N_{\overline{D}},S^D,S^{\overline{D}}).\]
The terms $K^- m_{1}\pi$ and $K^+ m_{2}\pi$ refer to $D$ or $\db$ decays
into a kaon and $m_{i}$ pions.

We construct our probability 
density
function using the following two
assumptions:
(i) the normalized
mass distribution for background $K^- m_{1}\pi$ and $K^+ m_{2}\pi$
is linear in $M_n^{K^- m_{1}\pi}$ and $M_n^{K^+ m_{2}\pi}$,
and (ii) the normalized mass distribution of real $D$'s and
real $\overline{D}$'s is Gaussian with mean of 0 and sigma of 1.
Under these assumptions, which are correct for our data,
the probability density
functions --- normalized to unity in the
two-dimensional window defined by $|M_n^{(K^- m_{1}\pi)}| < 6.5$ and
$|M_n^{(K^+ m_{2}\pi)}| < 6.5$ ---
for each class of events is
\begin{tabbing}
$D$-Ridge background events:. \=     \kill
Combinatoric background: \>  $P_C = 1/169 + S^D M_n^{K^- m_{1}\pi} + S^{\overline{D}} M_n^{K^+ m_{2}\pi},$ \\ 
\>                               \\
$D$-Ridge background events:\>$P_D = (\frac{1}{13 \sqrt{2 \pi}} + \frac{N_C}{N_{D}} S^{\overline{D}}
M_n^{K^+ m_{2}\pi}) \; e^{-(M_n^{K^- m_{1}\pi})^2/2},$ \\ 
\>                               \\
$\overline{D}$-Ridge background events: \>
$P_{\overline{D}} = (\frac{1}{13 \sqrt{2 \pi}} + \frac{N_C}{N_{\overline{D}}} 
S^D M_n^{K^- m_{1}\pi}) \; e^{-(M_n^{K^+ m_{2}\pi})^2/2},$ \\ 
\>                               \\
Signal events: \>
$P_S = \frac{1}{2 \pi} e^{-((M_n^{K^- m_{1}\pi})^2 + 
(M_n^{K^+ m_{2}\pi})^2)/2}.$ \\
\end{tabbing}
The distribution for each
set of events is $N_i P_i$.
The overall probability density
function is then
\[f(\vec{z};\vec{\alpha}) = \frac
{N_C P_C + N_D P_D + N_{\overline{D}} P_{\overline{D}}
+ N_S P_S}{N_C + N_D + N_{\overline{D}} + N_S}.\]

In this analysis, we use the extended maximum likelihood
method\cite[pg. 249]{ref:bib39} in which
the number of $\ddb$~candidates found, $N_{D\overline{D}}$, is
considered 
to be
one more measurement with a Gaussian probability
distribution $G(N_{D\overline{D}},\mu)$ of
mean $\mu = N_C + N_D + N_{\overline{D}} + N_S$
and $\sigma = \sqrt{\mu}$.
Our likelihood function is then
\begin{equation}
\label{eq:likfunc}
L = G(N_{D\overline{D}},\mu) \; \prod_{i=1}^{N_{D\overline{D}}} f(\vec{z}_i;
\vec{\alpha}).
\end{equation}

\begin{figure} % Figure 5
        \centering
        \centerline{\epsfig{file=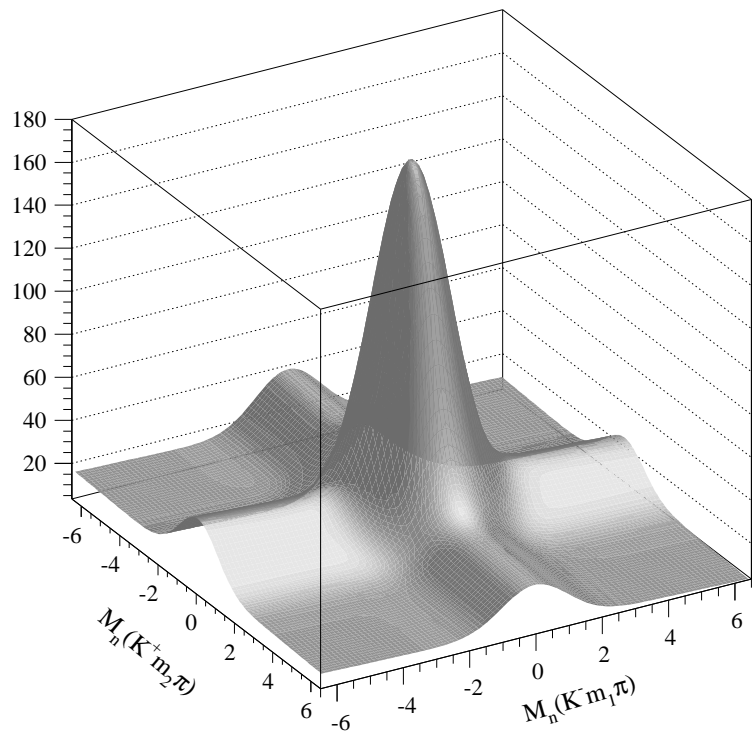,width=4.0in,angle=0}}
        \caption{The function that maximizes the joint
        probability of the unweighted charm-pair
        candidates from Fig.~\protect\ref{fig:fig4} with $|M_n|\leq6.5$.
        The axes are the normalized candidate-$D$
        masses $M_n$ defined in Eq.~\protect\ref{eq:mn} in the text.}
        \label{fig:fig5}
\end{figure}

\begin{figure} % Figure 6
        \centering
        \centerline{\epsfig{file=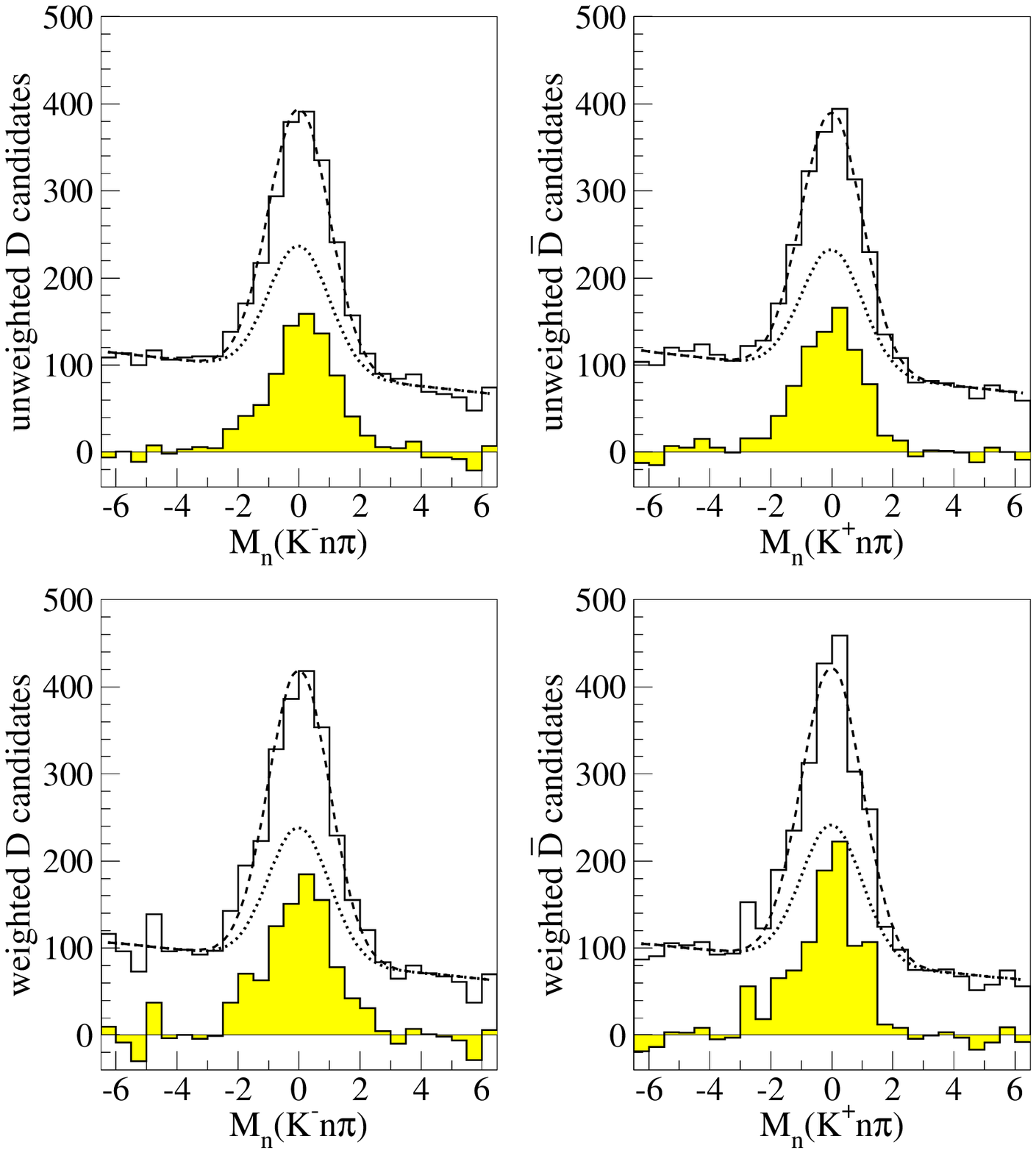,width=4.0in,angle=0}}
        \caption{One-dimensional projections of the charm-pair normalized mass
        distributions for $D$ and $\db$ candidates,
        for unweighted events (top), and weighted events (bottom).
        The dashed curves are the fit projections.  The dotted curves are
        the fit projections for the non-signal part of the fit.  This includes
        background from one real $D$ ($\db$) and one fake $\db$ ($D$).
        The shaded histograms are the background-subtracted signals.
        }
        \label{fig:fig6}
\end{figure}

To maximize the likelihood, we use the function minimization and error
analysis FORTRAN
package MINUIT~\cite{ref:bib38}.
Figure~\ref{fig:fig5} shows the function
\newline
$N_{D\overline{D}}f(\vec{z};\vec{\alpha})$
that maximizes the likelihood function for the final sample of
$\ddb$~candidates from
Fig.~\ref{fig:fig4} with $|M_n| \leq 6.5$, the mass range used
for all fits in this analysis.
The projections of the fit onto the $D$ and $\db$ axes are 
compared to the data in
Fig.~\ref{fig:fig6}.
The projected background contains both ridge events (one real $D$
and one combinatoric background) and events with two combinatoric
background candidates. Therefore, the background under the charm-pair
signal in the projected distribution is a linear distribution
plus a Gaussian distribution, shown as the dotted line in the
figure. The net charm-pair signal is shown as the residual
after background subtraction.

\subsection{Acceptance Corrections}
\label{ssec:accept}
We determine the size of acceptance and smearing
effects with a sample of approximately 7000 
Monte Carlo simulated pairs that pass the same
selection criteria as the real data.
The size of the resolution with which we 
measure each charm-pair physics variable is much 
smaller than the range over which we bin that variable.
Therefore, we ignore smearing effects.

We incorporate acceptance effects 
in the likelihood function for the fit by replacing
the probability $p_i$ for event $i$
by $(p_i)^{w_i}$ where $w_i$ is the weight for
event $i$ \cite{ref:bib40}.
The weight $w_i$ is inversely proportional to the efficiency
and is normalized such
that $\Sigma_{i=1}^{N_{\ddb}} w_i = N_{\ddb}$,
where $N_{\ddb}$ is the number of $\ddb$ candidates in the final sample.
Corrections for relative branching fractions are also included in $w_i$,
as described below.
By construction, the mean of the weights is
equal to 1.
The standard deviation of the weights is 1.3.  
The total number of $\ddb$ events found in the unweighted fit is
$N_s=791\pm44$.  For the weighted likelihood fit, 
we find $N_s=910\pm45$.\footnote{Since the sum of all weights is 
normalized to equal the number of $\ddb$ candidates, the fact that 
the number of signal events is significantly larger for the 
weighted data sample indicates that, on average, the weights for signal
events are larger than for background events.
Since we correct for relative efficiencies, not absolute
efficiencies, the absolute number of weighted signal events has no
significance. It is only of interest in interpreting the 
figures.  
}

The efficiency depends not only on the detector
acceptance but also on the relative branching fractions
for the detected decay modes.
By correcting for branching fractions, the final
efficiency-corrected distributions reflect
the relative production rates of the four types of 
$\ddb$ pairs 
($\dzdb$, $\dzdm$, $\dpdb$, and $\dpdm$)
rather than the relative detected rates.
We use the 
%%1994 Particle Data Group
values
$B(D^+\to K^-\pi^+\pi^+) = (9.1\pm0.6)\%$,
$B(D^0\to K^-\pi^+\pi^-\pi^+) = (8.1\pm0.5)\%$,
and
$B(D^0\to K^-\pi^+) = (4.01\pm0.14)\%$\cite{ref:bib41}.

A minimal independent set of properties that the
acceptance could depend on is
the decay mode of each of the $D$ mesons ($K^-\pi^+$,
$K^-\pi^+\pi^+$, or $K^-\pi^+\pi^-\pi^+$),
the rapidity $y$, the transverse
momentum $p_t$, and the azimuthal angle $\phi$ of each of the
$D$ mesons.
In principle, we can use Monte Carlo simulated events to
determine the acceptance for a particular candidate pair.
The problem is the large number of Monte Carlo events that
is needed to span such a large space
(a 48-dimensional space, six variables for each pair of decay modes used).
However, the efficiency function can be factorized
for each combination of the $D$ decay modes,
greatly increasing
the statistical power of the Monte Carlo.

Using the Monte Carlo simulated events, we find that
the acceptance of the $D$ is independent of the $\db$, and vice versa,
at the level of the statistical
precision of the simulation.
This is true in spite of the correlations in the selection criteria
described in Section~\ref{sec:selection}.
For each one, however,
the shape of the acceptance as a function of $y$ depends on
both
the number of particles in the decay,
$n_D$ or $n_{\db}$,
and, at high $y$, whether the candidate decay is a $D$ or $\db$.
The shape of the acceptance as a function of $p_t$ depends only on the
number of particles in the decay, $n_D$ or $n_{\db}$.
It is also found that the acceptance does not depend on
the azimuthal angle $\phi$ of the $D$ or $\db$.
Therefore, the acceptance function factorizes as follows:
\begin{displaymath}
A(n_D,n_{\db},y_D,y_{\db},p_{t,D},p_{t,{\db}},\phi_D,\phi_{\db})
\end{displaymath}
\begin{equation}
\label{eq:accept}
= b_{n_D} b_{n_{\db}}
  c(y_D,y_{\db})
  d_{n_D}(p_{t,D}) d_{n_{\db}}(p_{t,{\db}}),
\end{equation}
where the subscripts, superscripts, and functional dependences of
the terms b, c, and d are explicit, showing how the
factorization is done.

We next determine which of the variables that describe the candidate
pair, and for which the acceptance is not uniform,
are correlated in the originally generated Monte Carlo.
Such correlations could affect the
apparent acceptance from
the Monte Carlo if we simply integrate over a variable that is
correlated with the variable for which we are determining the acceptance.
We find that the most significant correlations
in the Monte Carlo generator are between the
variables $\yd$ and $\ydb$, where the acceptance is not uniform,
and between the variables $\phi_D$ and $\phi_{\db}$,
where the acceptance is uniform.
Therefore, we cannot simply integrate over $\yd$ when
determining the acceptance as a function of $\ydb$.
Instead, we use the Monte Carlo to
determine the two-dimensional acceptance function
$c(\yd, \ydb)$
---which is independent of the Monte Carlo generated correlations ---
for each of the possible values of
$n_D$ and $ n_{\db}$ and
use this function
in Eq.~\ref{eq:accept}.
(This removes any dependence on the physics assumptions of the
Monte Carlo from this equation.)
Because of the uniform $\phi$ acceptance for the observed events,
the double variable technique is not needed for $\phi$.
Finally, the weight $w_i$ is calculated for each event such that it
is proportional to ${1\over  B(D) B(\db) A}$ and
$\Sigma_{i=1}^{N_{\ddb}} w_i = N_{\ddb}$,
where $N_{\ddb}$ is the number of
unweighted $\ddb$ candidates in the final sample.
Here $B$ is $B(D^+\to K^-\pi^+\pi^+)$
for the charged
$D$ candidates and $B$ is $B(D^0\to K^-\pi^+\pi^-\pi^+) +
B(D^0\to K^-\pi^+)$ for the neutral $D$ candidates except for
$\dzdb$ events where $B(D^0)B(\dzb) = B(D^0\to K^-\pi^+)^2 +
2 B(D^0\to K^-\pi^+) B(D^0\to K^-\pi^+\pi^-\pi^+)$
due to the
exclusion of 4-4 pairs from the final sample.

\subsection{Checks \& Systematic Errors}
\label{ssec:syserrors}

Sources of systematic errors in our
measurements include effects associated with the
fitting procedure used to obtain the yields,
the finite statistics of the Monte Carlo data
sample used to generate the acceptance contributions to the weights,
and imperfections in our modeling of the apparatus in the Monte Carlo.

For all
the measured distributions, we compared the data to the 
two dimensional normalized mass distributions; 
in all cases, the fits 
qualitatively
match the data. (For example, see Fig.~\ref{fig:fig6}.)

We also checked the fitting procedure by
comparing the yields with those given by a simple counting method. 
In this method the normalized mass scatterplot was divided into
regions 
corresponding to different combinations of signal, ridge,
and combinatoric background events.
The number of signal events was
then found by subtracting the properly normalized number of events in the
ridge and background regions from the central signal region.
The results are in agreement, but the fitting technique gave smaller
statistical errors, as expected. 

The effect of the finite
statistics in the Monte Carlo was determined by repeating the fits for the 
yields
in each kinematic bin while varying the weight of each event randomly
according to a Gaussian whose width corresponded to the statistical 
error on the weight.
The systematic errors on the yields generated by this
process were about $20\%$ of the statistical errors from the fit,
which are negligible when added in quadrature.

\begin{figure} % Figure 7
        \centering
        \centerline{\epsfig{file=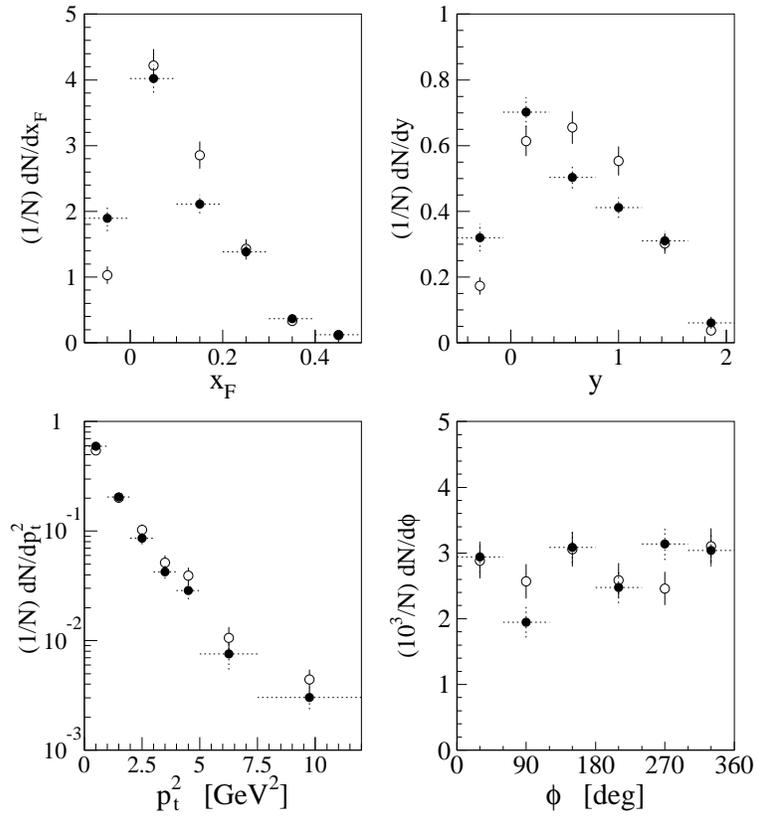,width=4.0in,angle=0}}
        \caption{Single-charm distributions for
        $\xf$, $y$, $\ptt$ and $\phi$, obtained from summing the
        $D$ and $\db$ distributions from our signal $\ddb$ events;
        unweighted ($\circ$) and weighted ($\bullet$)
        data are described in Sec.~\protect\ref{ssec:accept}.}
        \label{fig:fig7}
\end{figure}

\begin{figure} % Figure 8
        \centering
        \centerline{\epsfig{file=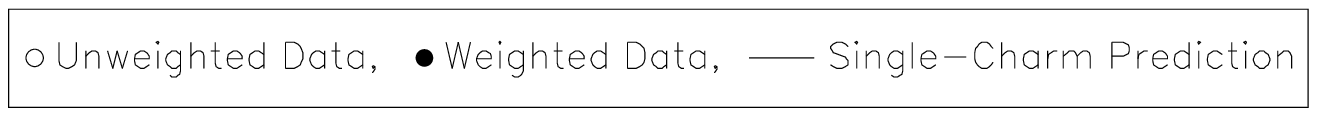,width=4.0in,angle=0}}
        \centerline{\epsfig{file=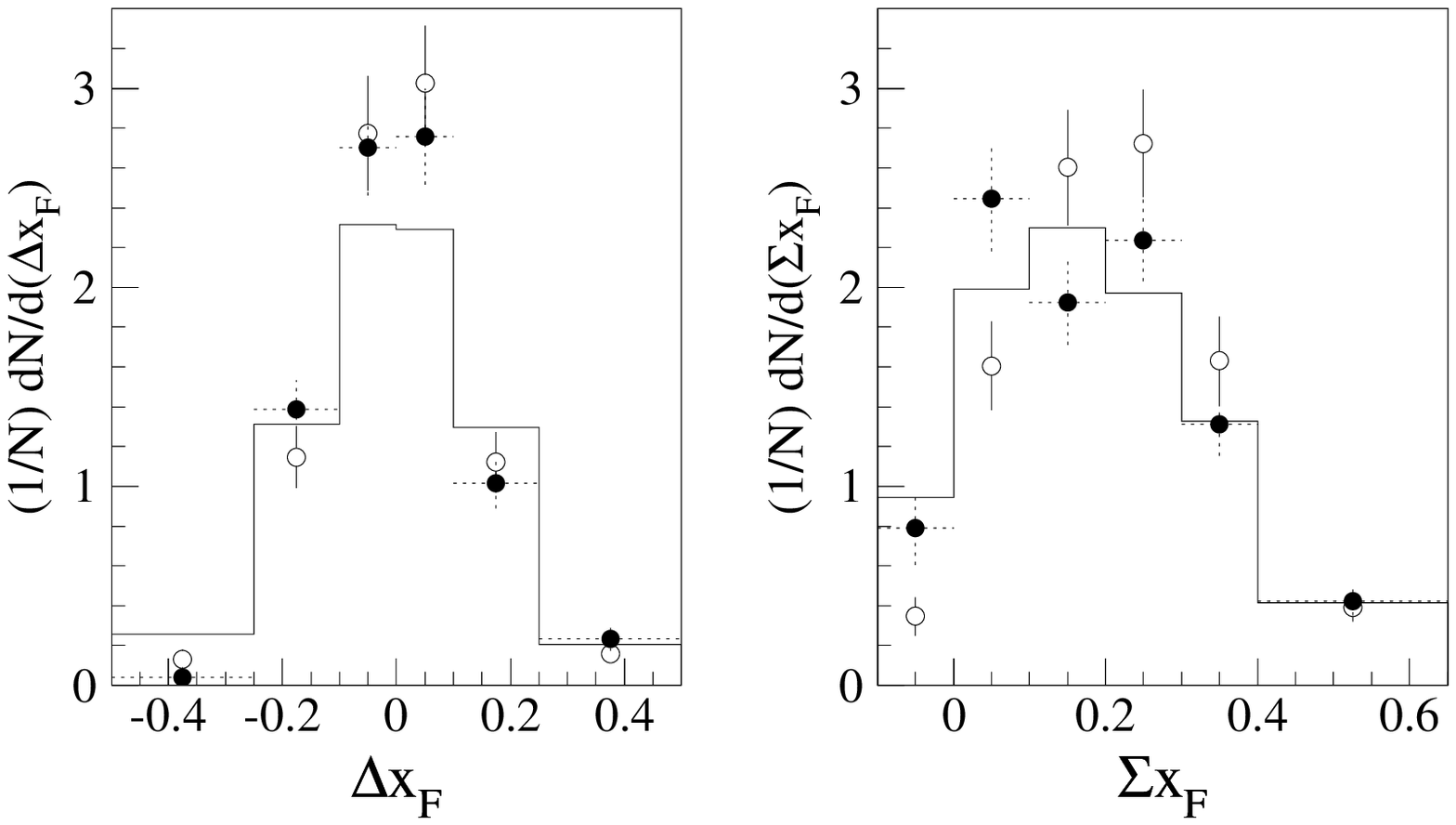,width=4.0in,angle=0}}
        \caption{Charm-pair distributions for $\dxf =
        \xfd - \xfdb$ and
        $\sxf = \xfd + \xfdb$.
        The uncorrelated single-charm predictions (------)
        for $\dxf$ and $\sxf$ are defined in
        Eqs.~\protect\ref{eq:qdv} and~\protect\ref{eq:qsv}.
        Unweighted ($\circ$) and weighted ($\bullet$)
        data are described in Sec.~\protect\ref{ssec:accept}.}
        \label{fig:fig8}
\end{figure}

\begin{figure} % Figure 9
        \centering
        \centerline{\epsfig{file=fig_8to12_label.ps,width=4.0in,angle=0}}
        \centerline{\epsfig{file=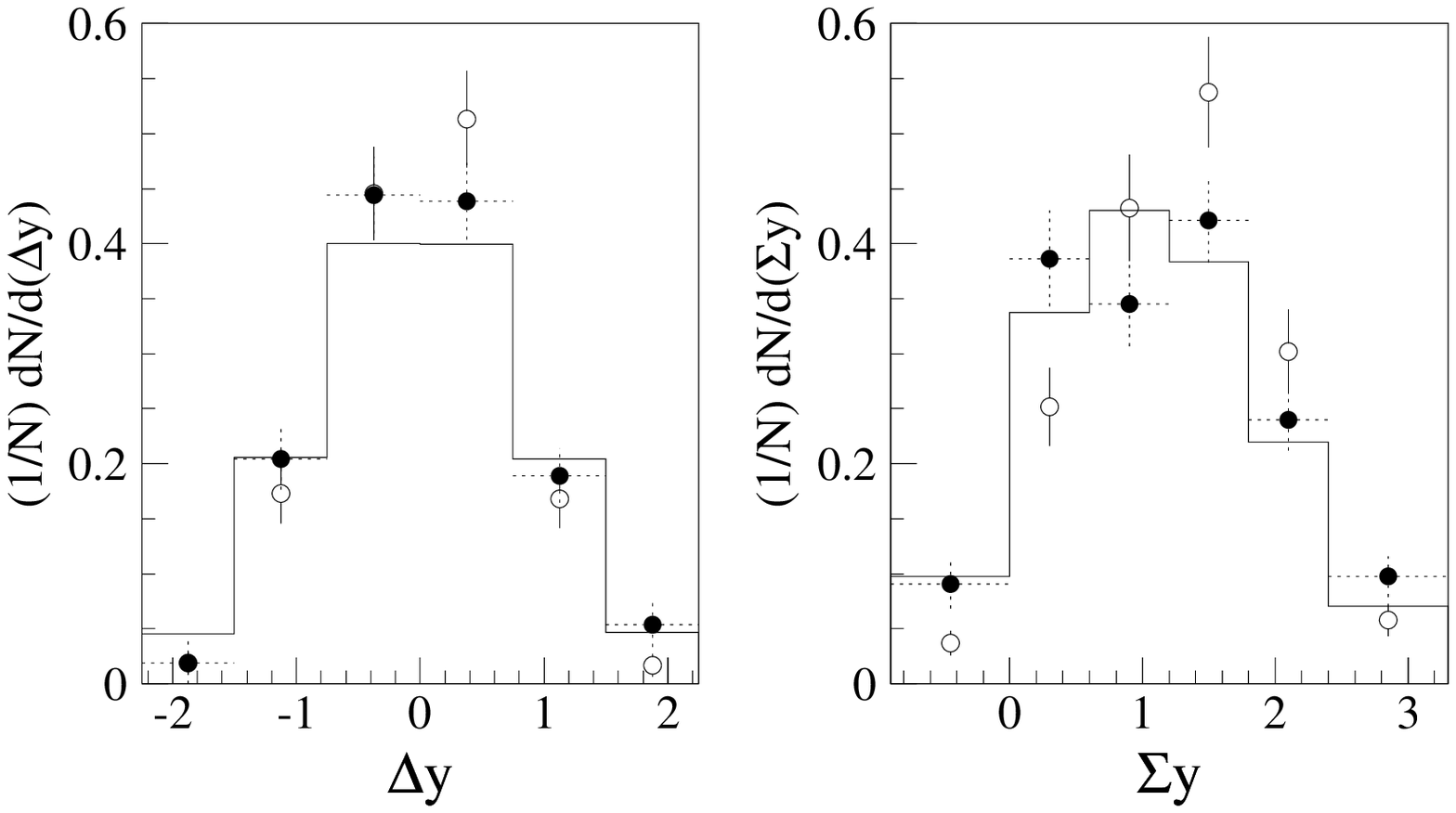,width=4.0in,angle=0}}
        \caption{Charm-pair distributions for $\Delta y =
        \yd - \ydb$ and
        $\Sigma y = \yd+\ydb$.
        The uncorrelated single-charm predictions (------)
        for $\Delta y$ and $\Sigma y$ are defined in
        Eqs.~\protect\ref{eq:qdv} and~\protect\ref{eq:qsv}.
        Unweighted ($\circ$) and weighted ($\bullet$)
        data are described in Sec.~\protect\ref{ssec:accept}.}
        \label{fig:fig9}
\end{figure}

\begin{figure} % Figure 10
        \centering
        \centerline{\epsfig{file=fig_8to12_label.ps,width=4.0in,angle=0}}
        \centerline{\epsfig{file=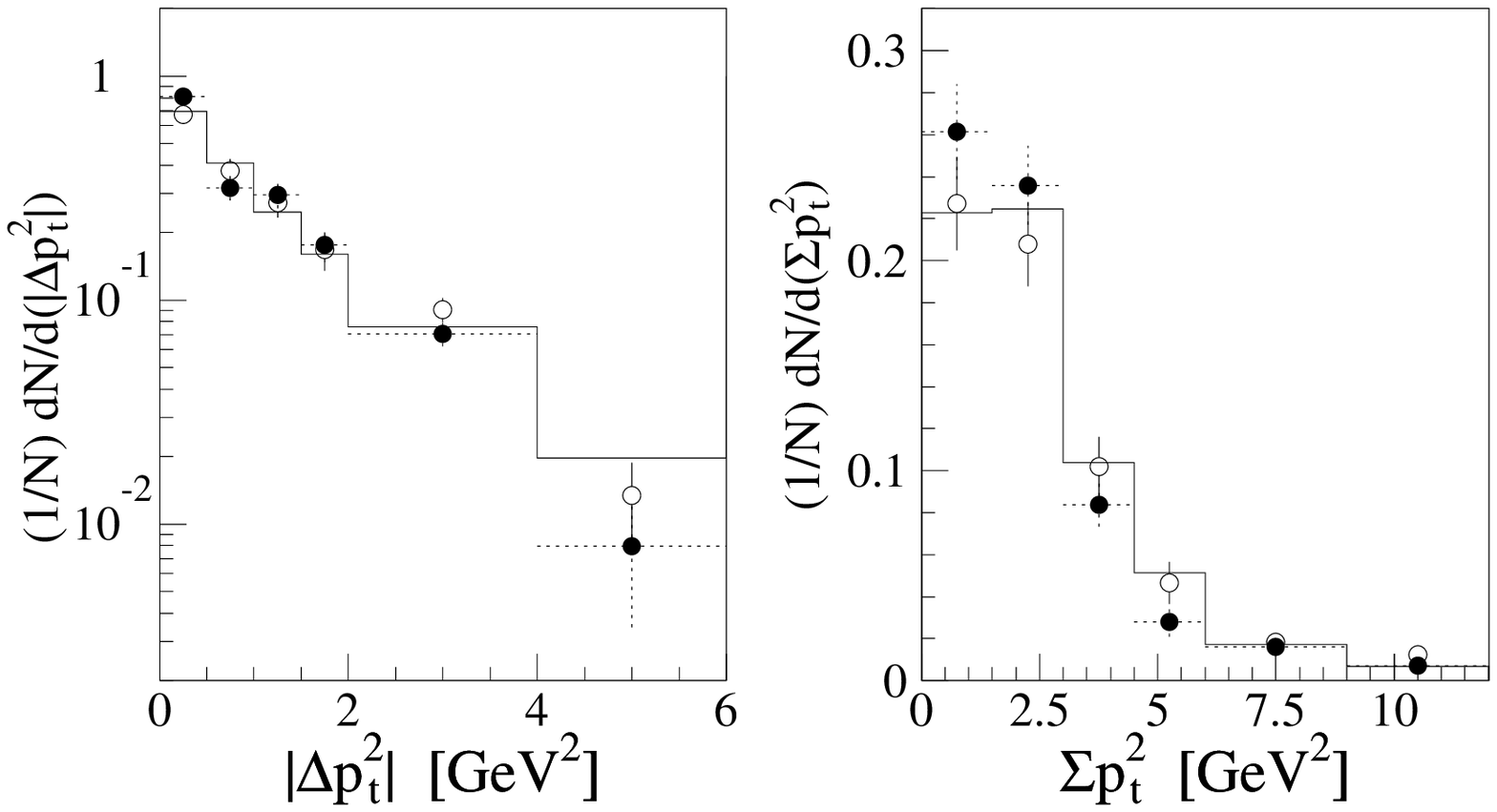,width=4.0in,angle=0}}
        \caption{Charm-pair distributions for $|\dptt| =
        |\pttd- \pttdb|$ and $\sptt= \pttd+ \pttdb$.
        The uncorrelated single-charm predictions (------)
        for $\dptt$ and $\sptt$ are defined in
        Eqs.~\protect\ref{eq:qdv} and~\protect\ref{eq:qsv}.
        Unweighted ($\circ$) and weighted ($\bullet$)
        data are described in Sec.~\protect\ref{ssec:accept}.}
        \label{fig:fig10}
\end{figure}

\begin{figure} % Figure 11
        \centering
        \centerline{\epsfig{file=fig_8to12_label.ps,width=4.0in,angle=0}}
        \centerline{\epsfig{file=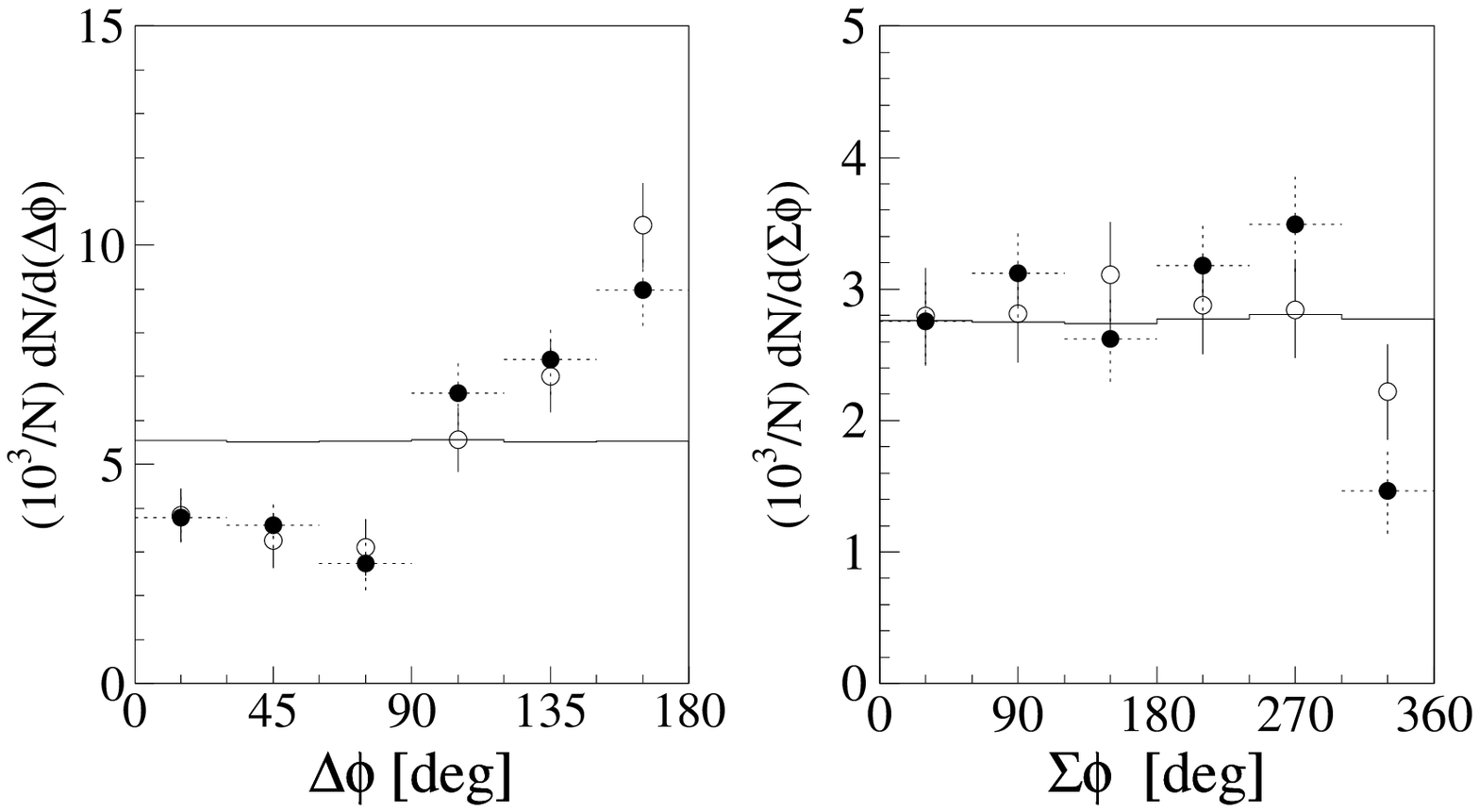,width=4.0in,angle=0}}
        \caption{Charm-pair distributions for $\dphi =$
        (minimum of $|\phi_D-\phi_{\db}|$ and
        $360^{\circ}-|\phi_D-\phi_{\db}|$) and
        $\sphi = (\phi_{D}+\phi_{\db}$ modulo $360^{\circ})$.
        The uncorrelated single-charm predictions (------)
        for $\dphi$ and $\sphi$ are defined in
        Eqs.~\protect\ref{eq:qdv} and~\protect\ref{eq:qsv}.
        Unweighted ($\circ$) and weighted ($\bullet$)
        data are described in Sec.~\protect\ref{ssec:accept}.}
        \label{fig:fig11}
\end{figure}

\begin{figure} % Figure 12
        \centering
        \centerline{\epsfig{file=fig_8to12_label.ps,width=4.0in,angle=0}}
        \centerline{\epsfig{file=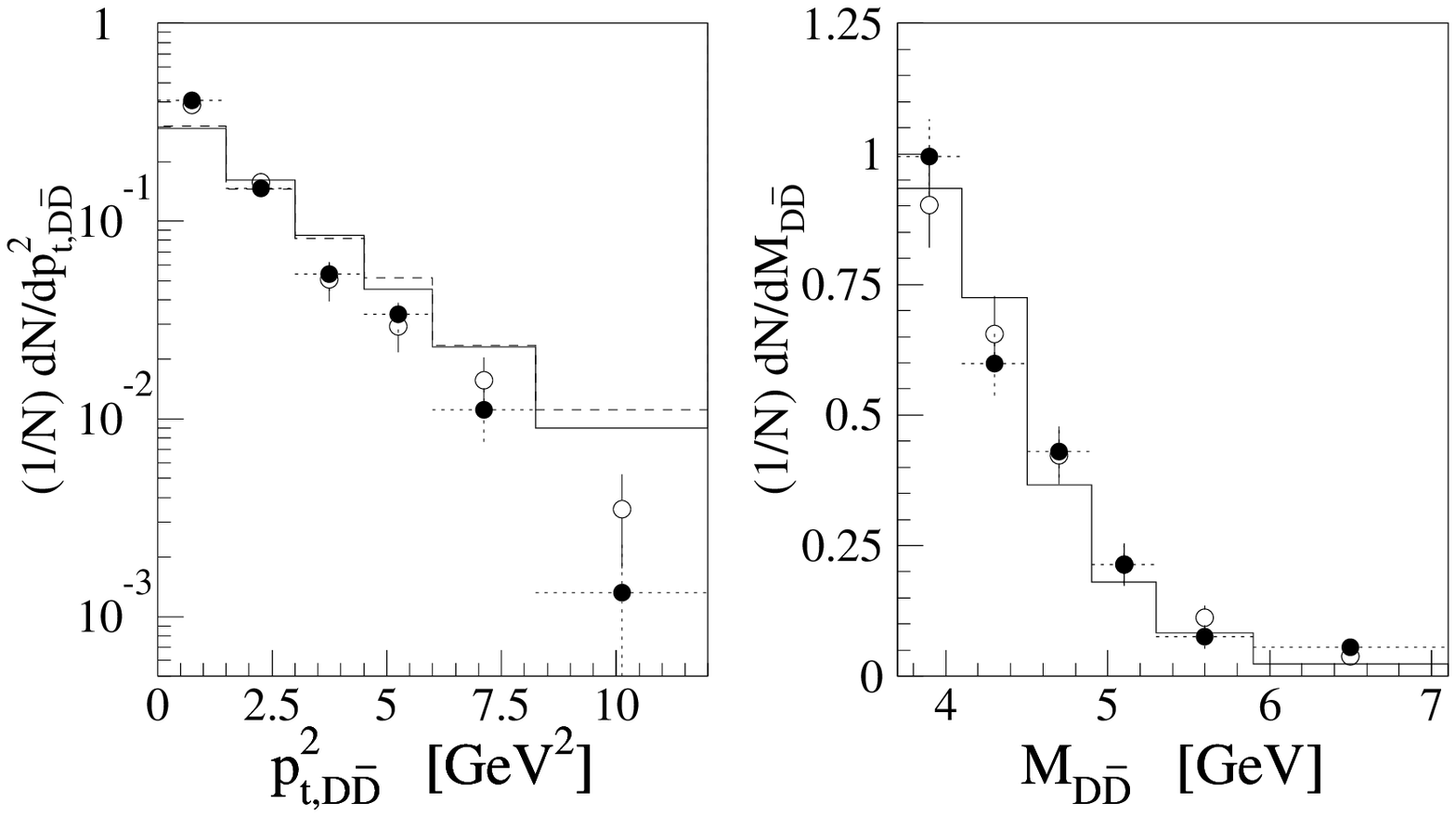,width=4.0in,angle=0}}
        \caption{Charm-pair distributions for
        $\pttddb=|\vec{p}_{t,D}~+ \vec{p}_{t,\db}|^2$~and
        $\Mddb$.  The solid histogram shows the uncorrelated
        single-charm predictions.  The dashed $\pttddb$ histogram
        assumes that $\xfd$ and $\xfdb$ are uncorrelated and
        that $\pttd$ and $\pttdb$ are uncorrelated, but
        that $\phi_{D}$ and $\phi_{\db}$ are correlated as
        shown in Fig.~\protect\ref{fig:fig11}.
        Unweighted ($\circ$) and weighted ($\bullet$)
        data are described in Sec.~\protect\ref{ssec:accept}.}
        \label{fig:fig12}
\end{figure}

As demonstrated in
Figs.~\ref{fig:fig7} to \ref{fig:fig12}, in
most cases the weighted and unweighted distributions are very similar.
Statistical errors associated with modeling the acceptance are most
important for events with large weights, but
the number of events with large weights is small. We checked the effect of
large weights by generating distributions
without the large weight events, with no significant change.
The distributions were also examined with all $K\pi\pi\pi$ 
candidates eliminated,
the
source of most of the 
events with a large acceptance correction. Again, the change did not
significantly affect the distributions. 

Another potential source of
systematic error is 
uncertainty in the modeling of the beam-induced
inefficiency in the centers of the drift chambers.
The inefficiencies increased as the run progressed,
primarily affecting $D$'s at large $\xf$.  
Since most $D$'s in the $\ddb$ events are at low to modest $\xf$,
the drift chamber inefficiencies 
did not have a significant effect on the relative efficiencies used
in this analysis.
We checked this by comparing the experimental results that had been
efficiency-corrected with Monte Carlo simulations corresponding to
different parts of the run,
and found no significant differences.

In summary, systematic errors were found to be small relative to statistical
errors.

%%%%%%%%%%%%%%%%%%%%%%%%%%%%%%%%%%%%%%%%%%%%%%%%%%%%%%%%%%%%%%%%%%%%%%%%
%\input{offline_doc_290_6_r1.tex} %  \section{Results}                       
\section{Results}
\label{sec:results}

In this section, we present the 
background-subtracted, acceptance-corrected charm-pair 
distributions from the data and compare them to theoretical predictions.
The Appendix contains an extensive discussion of the theoretical predictions for
charm-pair distributions.
As discussed in Sec.~\ref{sec:selection}, all 
distributions --- experimental and theoretical --- are obtained after 
excluding any events in which the center-of-mass rapidity 
of either $D$ meson or either charm quark is less than $-0.5$ or 
greater than $2.5$.

For the experimental results, the acceptance-corrected distributions
are obtained from maximizing the likelihood function with weighted events 
as discussed in Sec.~\ref{ssec:accept}. 
The uncorrected distributions
%%used to determine the fractional statistical error,
are obtained 
from maximizing the unweighted likelihood function;
the total number of signal $\ddb$ events found in the
unweighted fit is $N_s=791\pm44$. 

If the two charm mesons in each $\ddb$ event are completely
uncorrelated, then the charm-pair distributions contain
no more information than the single-charm distributions.
Before comparing 
the observed distributions to 
theoretical predictions,  
we use two methods to 
determine whether there exist
correlations
in the data. 
In the first method, described in Section~\ref{ssec:single}, 
we convolute acceptance-corrected single-charm 
distributions to predict what the charm-pair 
distributions would be if the $D$ and $\db$ were 
uncorrelated.  
Comparing 
these single-charm predictions,
the measured charm-pair 
distributions 
provide
one measure of the degree of correlation between the $D$ and $\db$.
In the second method, described in Section~\ref{ssec:twod},
we look directly for correlations 
by
examining several two-dimensional distributions.  For example, by finding the 
number of signal $\ddb$ events per $\yd$ interval in several $\ydb$
intervals, 
we can determine whether the shape of the $\yd$ distribution 
depends on the value of $\ydb$.
In Section~\ref{ssec:theory}, we compare our experimental distributions to
the theoretical predictions discussed in Sec.~\ref{sec:theory} and in
the Appendix.
In Section~\ref{ssec:asym}, we examine integrated production asymmetries
among the four types of $\ddb$ pairs --- 
$\dzdb$, $\dzdm$, $\dpdb$, {\rm and} $\dpdm$ ---
and compare our experimental results to the predictions from 
the {\sc Pythia/Jetset} event generator.    
                          
\subsection{Single-Charm Predictions}
\label{ssec:single}

In Fig.~\ref{fig:fig7} we show the measured single-charm 
distributions for 
$x_F$, $y$, $p_t^2$ and $\phi$, as defined in Sec.~\ref{sec:intro}.  
The single-charm distributions
are obtained by fitting the two-dimensional
normalized mass distributions for only those $\ddb$ pairs in
which the value of the single-charm variable for the candidate $D$ 
(or $\db$) is in the appropriate interval for each bin.
In this way, the contribution to the single-charm signal from the
$D$ and $\db$ ridge events is excluded.
The distributions shown in
Fig.~\ref{fig:fig7} 
correspond to single-charm mesons from $\ddb$ pairs in which the
center-of-mass rapidity of {\it both} charm mesons lies 
between $-0.5$ and $+2.5$.
Each distribution shown in Fig.~\ref{fig:fig7} is obtained by summing
the $D$ and $\db$ distributions.
We have checked and found that the $D$ and $\db$ distributions are the
same
within statistical errors. 
\footnote{This might not be the case if the experiment had
greater statistics, or if the data sample extended to a higher
region in $x_F$.}

The vertical axis of each distribution gives the fraction of signal
mesons per variable $v$ interval, 
$P(v) = \frac{1}{N_D}\frac{{\rm d}N_D}{{\rm d}v}$, 
where the total number of signal $D$ mesons $N_D$ is simply
twice the number of signal $\ddb$ events. 
Only a very small fraction (0\% -- 3\%) of the signal events lie 
outside any of the ranges used in 
Figs.~\ref{fig:fig7}--\ref{fig:fig11}.

For each single-charm variable $v = \xf$, $y$, $\ptt$, and $\phi$ 
we obtain two measured charm-pair 
distributions: the difference in $v$ for the two $D$'s,
$\Delta v = v_D-v_{\overline{D}}$, and the sum of the $v$'s for the two $D$'s, 
$\Sigma v = v_D+v_{\overline{D}}$.  
($\Delta \phi$ is defined to be the minimum 
of $|\phi_D-\phi_{\overline{D}}|$ and 
$360^{\circ}-|\phi_D-\phi_{\overline{D}}|$, and $\Sigma \phi$ is 
defined to be $\phi_D + \phi_{\overline{D}}$ modulo $360^{\circ}$.)
In Figs.~\ref{fig:fig8}--\ref{fig:fig11}, 
we compare these measured charm-pair distributions to the 
charm-pair distributions
one would generate from the measured single-charm distributions assuming the
$D$ and $\db$ are completely uncorrelated,
calculated as follows:
{\renewcommand{\arraystretch}{1.5}
\begin{eqnarray}
\label{eq:qdv}
Q(\Delta v) & = 
& \int \int \delta(\Delta v - v_D + v_{\overline{D}}) P(v_D) 
P(v_{\overline{D}}) dv_D dv_{\overline{D}}\,, 
\end{eqnarray}
}
and 
{\renewcommand{\arraystretch}{1.5}
\begin{eqnarray}
\label{eq:qsv}
Q(\Sigma v) & = 
& \int \int \delta(\Sigma v - v_D - v_{\overline{D}}) P(v_D) 
P(v_{\overline{D}}) dv_D dv_{\overline{D}}\,,
\end{eqnarray}
}where $P(v)$ refers to the single-charm distributions shown in 
Fig.~\ref{fig:fig7}.  

This convolution cannot be done with previously reported inclusive
single charm distributions~\cite{ref:bib25}
since the inclusive distributions
contain events which are excluded from the charm-pair
sample, for example, events in which the $\xf$ of the unobserved
$D$ is outside the acceptance 
or in which
the unobserved charm
particle is a charm baryon.

If the $D$ and $\db$ in the
signal $\ddb$ events are completely uncorrelated,
then the measured charm-pair distributions for $\Delta v$ and $\Sigma v$
should agree
with the single-charm predictions, because both the charm-pair
and single-charm distributions are for $D$ mesons with exactly the same
restrictions on the rapidity of both $D$ mesons in the event.
With the exception of the $\dphi$ distribution 
(Fig.~\ref{fig:fig11}), the measured distributions
are quite similar to the uncorrelated single-charm predictions, indicating 
both that the correlation between the $D$ and $\overline{D}$ longitudinal
momenta is small and that the correlation between the 
amplitudes of the $D$ and $\overline{D}$ transverse momenta is small.  
The measured $\dxf$ and $\dy$ distributions,
however, are somewhat more peaked near zero than the single-charm predictions, 
possibly indicating slight longitudinal correlations.

Two other commonly used charm-pair variables are
the square of the net transverse momentum of the charm pair,
$\pttddb = |\vec{p}_{t,D} + \vec{p}_{t,\db}|^2$,
and the invariant mass of the charm pair, $\Mddb$. 
The 
measured distributions and the uncorrelated single-charm predictions
for these two variables are shown in Fig.~\ref{fig:fig12}. 
Obtaining these single-charm predictions for $\pttddb$ and 
$\Mddb$
is slightly more involved than for the $\Delta v$ and $\Sigma v$ variables
because $\pttddb$
and $\Mddb$ are not linear functions of 
$\xfd$, $\xfdb$, $\phi_{D}$, $\phi_{\overline{D}}$,
$\pttd$, and $\pttdb$.
Rather, in terms of these single-charm variables,
\[ \pttddb = \pttd + \pttdb + 
2\sqrt{\pttd \pttdb}\cos(\phi_{D}-\phi_{\overline{D}}), \; {\rm and} \]
\[ \Mddb = \sqrt{2M_D^2 + 
2 E_D E_{\overline{D}}
-2\sqrt{\pttd \pttdb}\cos(\phi_{D}-\phi_{\overline{D}})
-\frac{s \; \xfd \; \xfdb}{2}}, \]
where the $D$ meson energy
$E$ is $\sqrt{M_D^2 + p^2_t + \frac{s \; x^2_F}{4}}$, and $s$ is the 
square of the center-of-mass energy of the colliding hadrons.
We obtain single-charm predictions by randomly generating
$10^8$ $\ddb$ events in which all three variables 
($\xf$, $\phi$, and $\ptt$) 
for both $D$ mesons from each $\ddb$ event
are selected independently and randomly
from a probability density
function 
that is flat within the 
bins
shown in Fig.~\ref{fig:fig7},
and zero elsewhere.  Each event is weighted by 
\[ \frac{1}{|J|} P(x_{F,D}) P(\phi_{D}) P(p^2_{t,D}) P(x_{F,\overline{D}}) 
P(\phi_{\overline{D}}) P(p^2_{t,\overline{D}}), \]
where $P(v)$ refers to the single-charm distributions shown in
Fig.~\ref{fig:fig7} and $|J|$ is
the Jacobian determinant of the transformation from 
the complete and independent set of variables 
($x_{F,D}$, $x_{F,\overline{D}}$, $\phi_{D}$, $\phi_{\overline{D}}$,
$p^2_{t,D}$, and $p^2_{t,\overline{D}}$) to the
set ($x_{F,D}$, $x_{F,\overline{D}}$, $\phi_{D}$, 
$\phi_{\overline{D}}$, $p^2_{t,D\overline{D}}$, and $M_{D\overline{D}}$).
Specifically, $|J| =$
\begin{displaymath}
\label{eq:jac}
       \left| \begin{array}{cc}
       1 + \sqrt{\frac{p^2_{t,\overline{D}}}{p^2_{t,D}}}\cos(\phi_{D}-
       \phi_{\overline{D}}) 
     & 1 + \sqrt{\frac{p^2_{t,D}}{p^2_{t,\overline{D}}}}
       \cos(\phi_{D}-\phi_{\overline{D}})                              \\
       \frac{1}{2 \sqrt{2} M_{D\overline{D}}} \left( \frac{E_{\overline{D}}}{E_D} 
       - \sqrt{\frac{p^2_{t,\overline{D}}}{p^2_{t,D}}}\cos(\phi_{D}-
       \phi_{\overline{D}}) \right) 
     &\frac{1}{2 \sqrt{2} M_{D\overline{D}}} \left( \frac{E_D}{E_{\overline{D}}} 
       - \sqrt{\frac{p^2_{t,D}}{p^2_{t,\overline{D}}}}\cos(\phi_{D}-
      \phi_{\overline{D}}) \right) \end{array} \right|.       \nonumber
\end{displaymath}

The measured distribution for $\Mddb$ agrees
quite well with the uncorrelated single-charm prediction.  The
measured distribution for $\pttddb$, however, 
is steeper than the uncorrelated single-charm prediction, indicating
the presence of correlations between 
$\vec{p}_{t,D}$ and $\vec{p}_{t,\overline{D}}$.  
The dashed histogram in Fig.~\ref{fig:fig12} 
demonstrates that this lack of agreement is not due to the 
correlations between $\phi_D$ and $\phi_{\overline{D}}$ evident 
in the $\dphi$ distribution in 
Fig.~\ref{fig:fig11}.  
This latter prediction is obtained
by assuming that $\xfd$ and $\xfdb$ are uncorrelated and
that $\pttd$ and $\pttdb$ are uncorrelated, but
that $\phi_{D}$ and $\phi_{\overline{D}}$ are correlated as
shown in Fig.~\protect\ref{fig:fig11}.  
The correlations in $\pttddb$ should reflect similar correlations in
$\dphi$, $\pttd$, and $\pttdb$, since $\pttddb$ is a
function of these variables. The fact that the disagreement in
Fig.~\ref{fig:fig12} is not so readily explained is a sign that
the correlations can be subtle, and that there are 
additional correlations among the variables.
In the following section we investigate
correlations between $\dphi$,
$\ptd$, and $\ptdb$ in
more detail.

\subsection{Two-Dimensional Distributions}
\label{ssec:twod}
\begin{figure} % Figure 13
        \centering
        \centerline{\epsfig{file=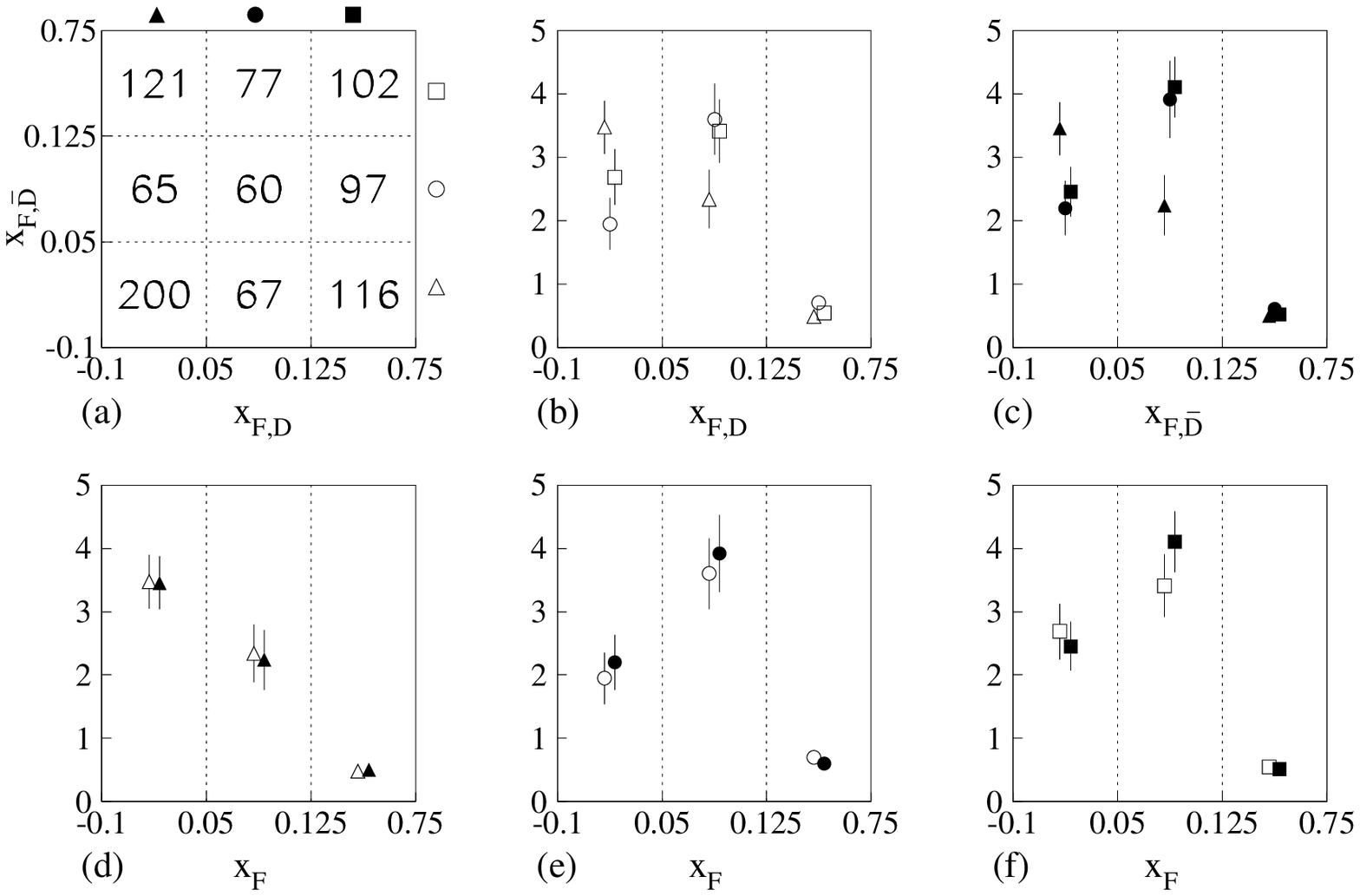,width=4.0in,angle=0}}
        \caption{(a) Number of weighted $\ddb$ signal events d$N_s$
        found in nine ($\xfd$, $\xfdb$) bins.
        \,(b) $\xfd$ distribution
        $({\rm d}N_s/{\rm d}x_{F,D})/N_i$
        for each $\xfdb$ bin,
        where $N_i$ is the total number of events in the three
        $\xfd$ bins.
        \,(c) $\xfdb$ distribution for each $\xfd$ bin.
        \,(d)-(f) $\xfd$ ($\xfdb$) distribution for the
        first, second and third $\xfdb$ ($\xfd$) bins,
        respectively.
        Open symbols show the $\xfd$ distributions;
        closed symbols show the $\xfdb$ distributions.
        The weighting procedure is described in Sec.~\protect\ref{ssec:accept}.}
        \label{fig:fig13}
\end{figure}

\begin{figure} % Figure 14
        \centering
        \centerline{\epsfig{file=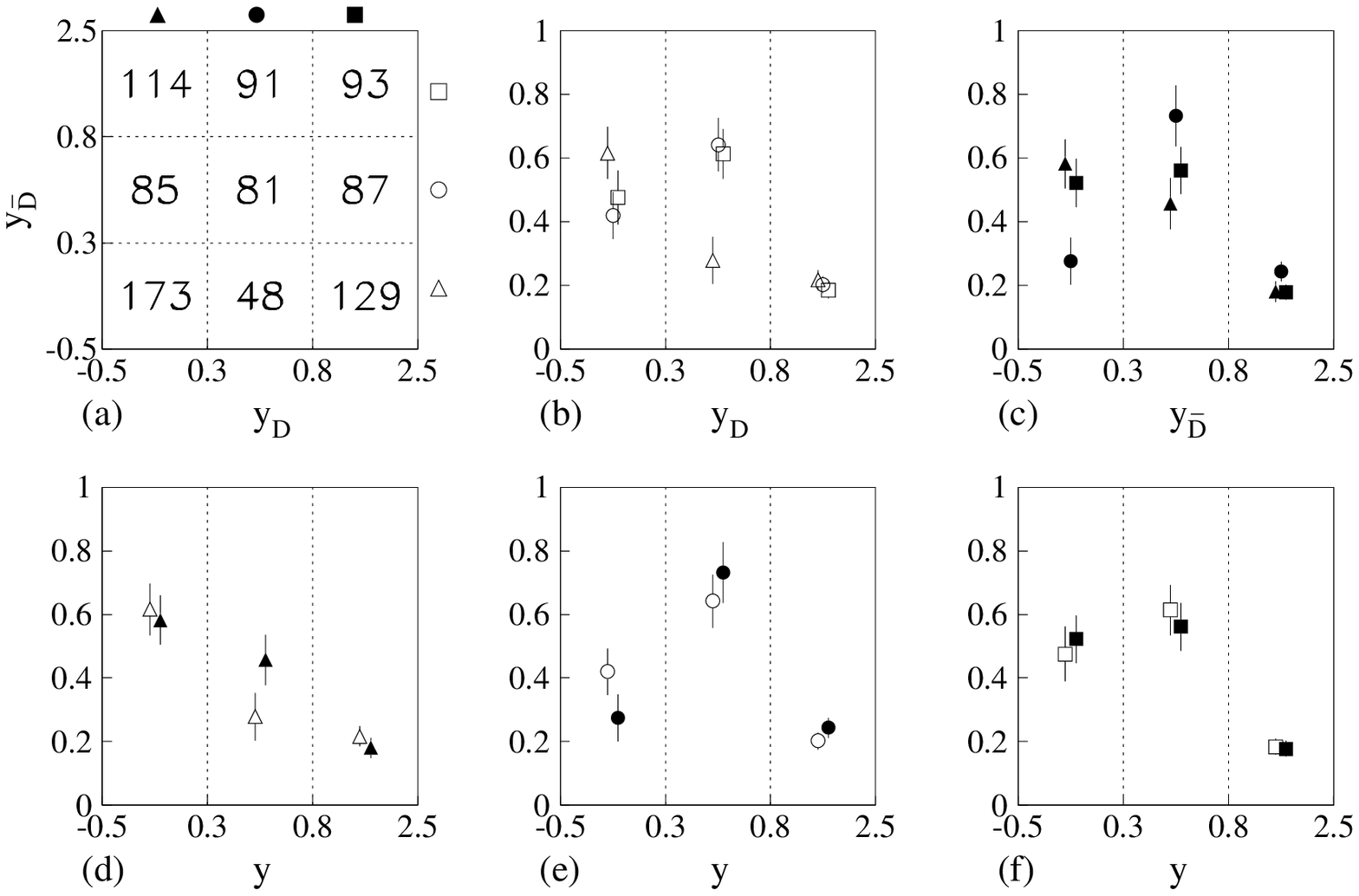,width=4.0in,angle=0}}
        \caption{(a) Number
        of weighted $\ddb$ signal events d$N_s$ found in nine
        ($\yd$, $\ydb$) bins.
        (b) $\yd$ distribution
        $({\rm d}N_s/{\rm d}y_D)/N_i$
        for each $\ydb$ bin,
        where $N_i$ is the total number of events in the three
        $\yd$ bins.
        (c) $\ydb$ distribution
        for each $\yd$ bin.  (d)-(f) $\yd$ ($\ydb$)
        distribution for the first, second and third $\ydb$
        ($\yd$) bins, respectively.
        Open symbols show the $\yd$ distributions;
        closed symbols show the $\ydb$ distributions.
        The weighting procedure is described in Sec.~\protect\ref{ssec:accept}.}
        \label{fig:fig14}
\end{figure}

\begin{figure} % Figure 15
        \centering
        \centerline{\epsfig{file=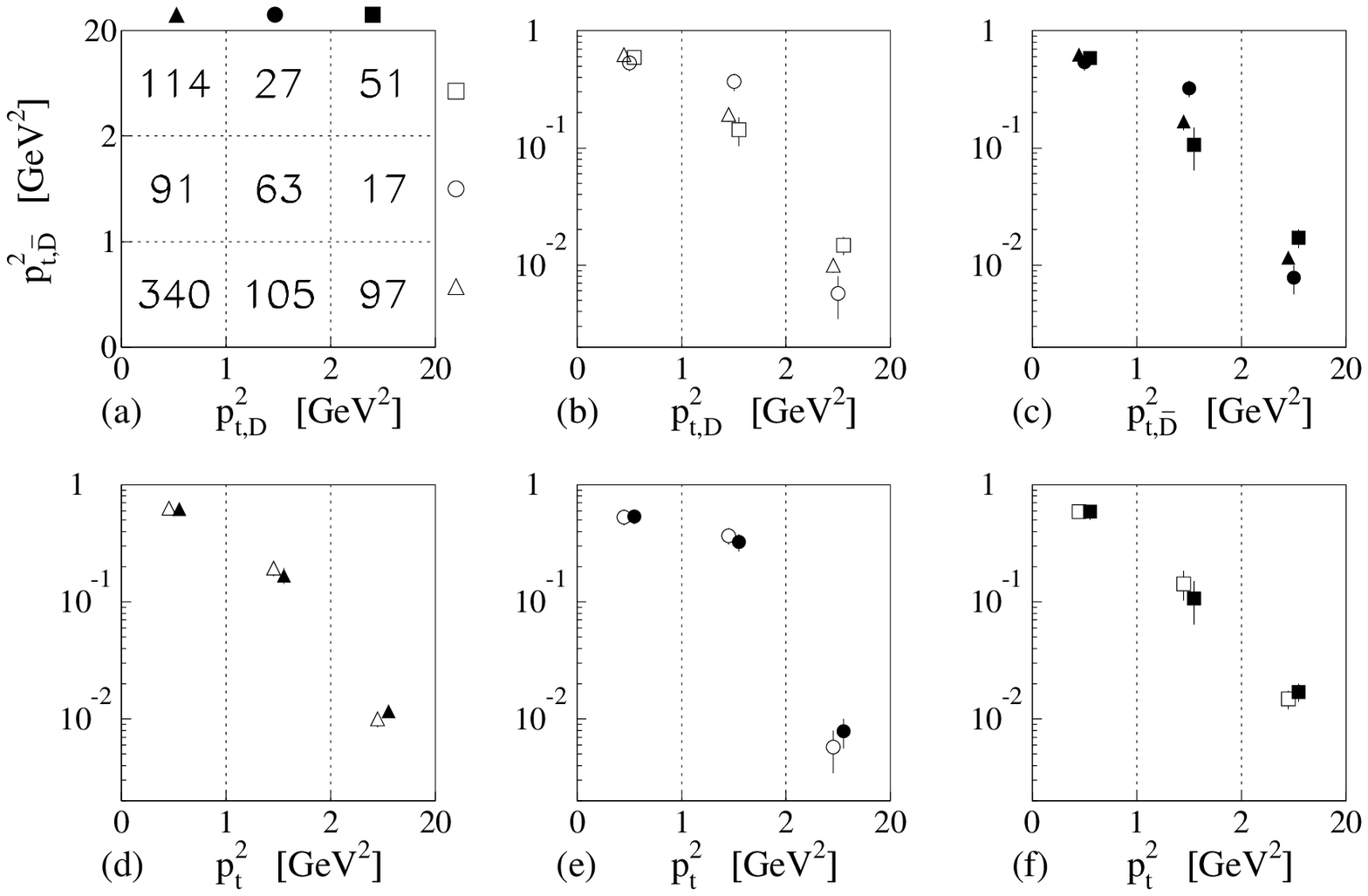,width=4.0in,angle=0}}
        \caption{(a) Number
        of weighted $\ddb$ signal events d$N_s$ found in nine
        ($\pttd$, $\pttdb$) bins.
        (b) $\pttd$ distribution
       $({\rm d}N_s/{\rm d}p^2_{t,D}/N_i)$
       for each $\pttdb$ bin,
        where $N_i$ is the total number of events in the three
        $\pttd$ bins.
        (c) $\pttdb$ distribution
        for each $\pttd$ bin.  (d)-(f) $\pttd$
        ($\pttdb$)
        distribution for the first, second and third $\pttdb$
        ($\pttd$) bins, respectively.
        Open symbols show the $\pttd$ distributions;
        closed symbols show the $\pttdb$ distributions.
        The weighting procedure is described in Sec.~\protect\ref{ssec:accept}.}
        \label{fig:fig15}
\end{figure}

A direct method for investigating whether the
variables $v_D$ and $v_{\overline{D}}$ are correlated
is to determine the number of $\ddb$ signal events per $v_D$ interval 
for a series of
$v_{\overline{D}}$ intervals. Such two-dimensional
distributions show
whether the $v_D$ distribution depends
on the value of $v_{\overline{D}}$, and vice-versa.
Given the limited number of $\ddb$ pairs, we 
use coarse binning
to see statistically meaningful effects.
In Figs.~\ref{fig:fig13}, \ref{fig:fig14} and \ref{fig:fig15}, 
we show the results
for $v = x_F$, $y$ and $p^2_t$, respectively.  In part (a) of each 
figure, we show the number of 
acceptance-corrected $\ddb$ signal events reconstructed in
nine $(v_D, \; v_{\overline{D}})$ bins --- 
three $v_D$ bins times three $v_{\overline{D}}$ bins. 
Note that the three bin sizes are not equal.
From the
information in this two-dimensional plot, several normalized 
one-dimensional plots are created, facilitating our ability to 
detect differences in the shapes of the distributions.
In particular, plot (b) in each figure shows the $v_D$ distribution
for each $v_{\overline{D}}$ bin, $\frac {{\rm d}N_s}{{\rm d}v_D}/N_i$,
where $N_s$ is the number of events in the relevant bin, and
$N_i$ is the total number of events in the three $v_D$ bins.
This normalization is chosen so that the integral over each $v_D$ distribution
equals one.
The symbols are defined in plot (a).
 Similarly, plot (c) in each figure shows the 
normalized $v_{\db}$ distribution for
each $v_D$ bin.  
If the three sets of points in the figures (b) and (c) are statistically
consistent, there are no significant correlations.
Lastly, plots (d)--(f) simply rearrange
the information shown in (b) and (c).  Plot (d) shows the normalized
$v_D$ and $v_{\overline{D}}$
distributions for the first $v_{\overline{D}}$ and $v_D$ bin, respectively; 
plot (e) shows results for the second bins; and plot (f) shows results
for the third bins.  
In (d)-(f), agreement of the two sets of points implies 
that correlations in $v_{\overline{D}}$
are the same as correlations in $v_D$.
The two-dimensional plots show the
actual number of acceptance-corrected $\ddb$ signal events in each
bin, whereas the one-dimensional plots, proportional to 
$\frac {{\rm d}N_s}{{\rm d}v_D}$, take into account the variation 
in bin size.

\begin{figure} % Figure 16
        \centering
        \centerline{\epsfig{file=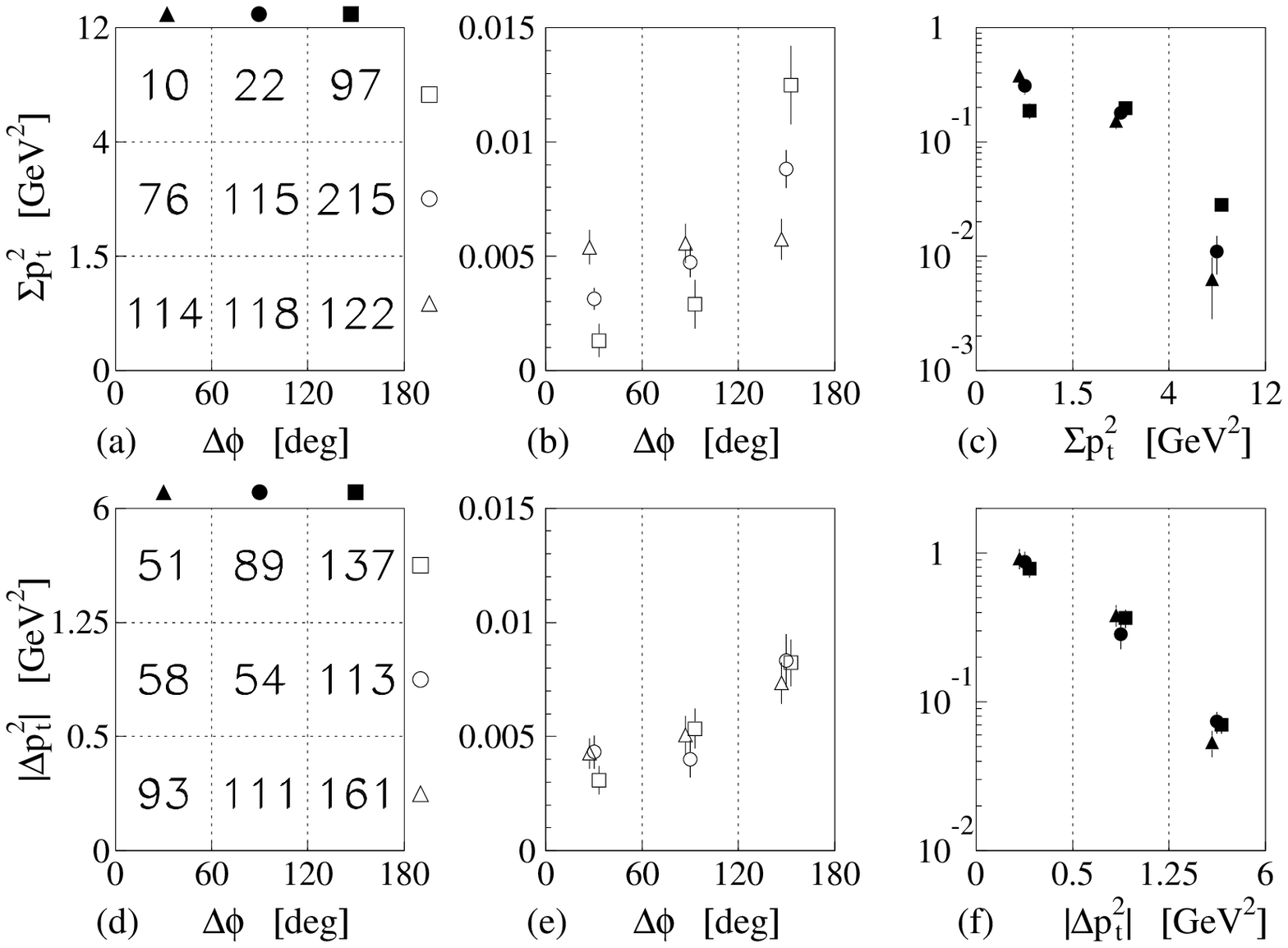,width=4.0in,angle=0}}
        \caption{(a) Number
        of weighted $\ddb$ signal events d$N_s$ found in nine
        ($\dphi$, $\sptt$) bins.
        (b) $\dphi$ distribution
        $({\rm d}N_s/{\rm d}\Delta \phi)/N_i$
        for each $\sptt$ bin,
        where $N_i$ is the total number of events in the three
        $\dphi$ bins.
        (c) $\sptt$ distribution
        for each $\dphi$ bin.  (d) Number
        of weighted $\ddb$ signal events d$N_s$ found in nine
        ($\dphi$, $|\Delta p_t^2|$) bins.
        (e) $\dphi$ distribution for each $|\Delta p_t^2|$ bin.
        (f) $|\Delta p_t^2|$ distribution
        for each $\dphi$ bin.
        The weighting procedure is described in Sec.~\protect\ref{ssec:accept}.}
        \label{fig:fig16}
\end{figure}

Figure~\ref{fig:fig13} indicates some correlation between
$\xfd$ and $\xfdb$.
In particular, the first-bin distributions are
more peaked at low $\xf$ than the second- and
third-bin distributions
of figures (b) and (c).  
This result is consistent with 
Fig.~\ref{fig:fig8}, discussed above, which shows that the 
measured $\dxf$ distribution is somewhat steeper than 
the single-charm prediction.  Because $\xf$ and $y$ are 
highly correlated, Fig.~\ref{fig:fig14} shows the same trends
as Fig.~\ref{fig:fig13}.  
Figure~\ref{fig:fig15} indicates that 
$\pttd$ and $\pttdb$ are also slightly correlated ---
the second-bin $\pttd$ and $\pttdb$ distributions
are enhanced in the second bin; and the third-bin 
$\pttd$ and $\pttdb$ distributions
are enhanced in the third bin
of figures (b) and (c).  
This result should be compared with
Fig.~\ref{fig:fig10}, which is consistent with no correlation.
Correlations are also seen in Fig.~\ref{fig:fig12}
which shows that the
measured $\pttddb$ distribution is somewhat steeper than
the uncorrelated single-charm prediction.  In all three 
figures (\ref{fig:fig13}--\ref{fig:fig15}), the shapes of the three $v_D$ 
distributions are remarkably 
similar to the shapes of the respective $v_{\overline{D}}$ distributions
as seen in figures (d)--(f).
In Fig.~\ref{fig:fig16} we investigate whether the 
separation in azimuthal angle between the $D$ and $\overline{D}$
is correlated to the amplitude of the transverse momenta of 
the $D$ and $\overline{D}$.  In particular, we determine
the number of signal $\ddb$ events per $\Delta \phi$ interval in
$\Sigma p^2_t$ intervals and the number 
of signal $\ddb$ events per $\Delta \phi$ interval in
$|\dptt|$ intervals.
Although we find no significant correlation between $\Delta \phi$ and 
$|\dptt|$, we find that $\dphi$ and $\sptt$ 
are quite correlated.  The $\dphi$ distribution is
more peaked at large $\sptt$ and the $\sptt$ distribution
is flatter at large $\dphi$.  A theoretical explanation for
these correlations is discussed in the following section.

\subsection{Comparisons with Theory}
\label{ssec:theory}

In this section, we compare all the acceptance-corrected 
distributions discussed in the previous two sections 
(Figs.~\ref{fig:fig7}--\ref{fig:fig16}) to three sets of 
theoretical predictions:
the distributions of $\ccb$ pairs from
a next-to-leading-order (NLO) perturbative 
QCD calculation by
Mangano, Nason and Ridolfi (MNR)\cite{ref:bib3,ref:bib4};
the distributions of $\ccb$ pairs from the 
{\sc Pythia}/{\sc Jetset} Monte Carlo event 
generator\cite{ref:bib5} which uses a parton-shower 
model to include higher-order perturbative effects\cite{ref:bib6}; and
the distributions of $\ddb$ pairs from the
{\sc Pythia} (Version 5.7)/{\sc Jetset} (Version 7.4)
Monte Carlo event generator\cite{ref:bib5}
which uses the Lund string model to transform 
$\ccb$ pairs to $\ddb$ pairs\cite{ref:bib7}.
For all theoretical predictions, we use the default parameters
suggested by the respective authors, which are discussed in 
the Appendix.
All distributions are obtained  after
excluding any candidates in which the center-of-mass rapidity
of either the $D$ or $\overline{D}$ 
(or, for the MNR calculation, the $c$ or $\cb$) is less than $-0.5$ or
greater than 2.5.

\subsubsection{Single-Charm Distributions}

\begin{figure} % Figure 17
        \centering
        \centerline{\epsfig{file=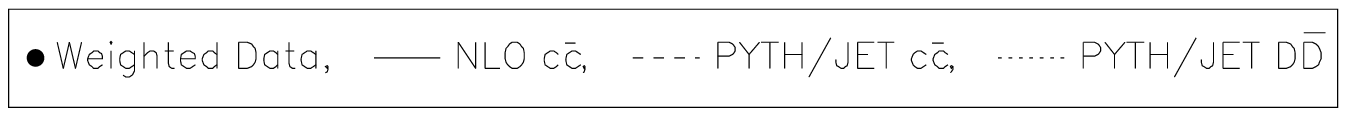,width=4.0in,angle=0}}
        \centerline{\epsfig{file=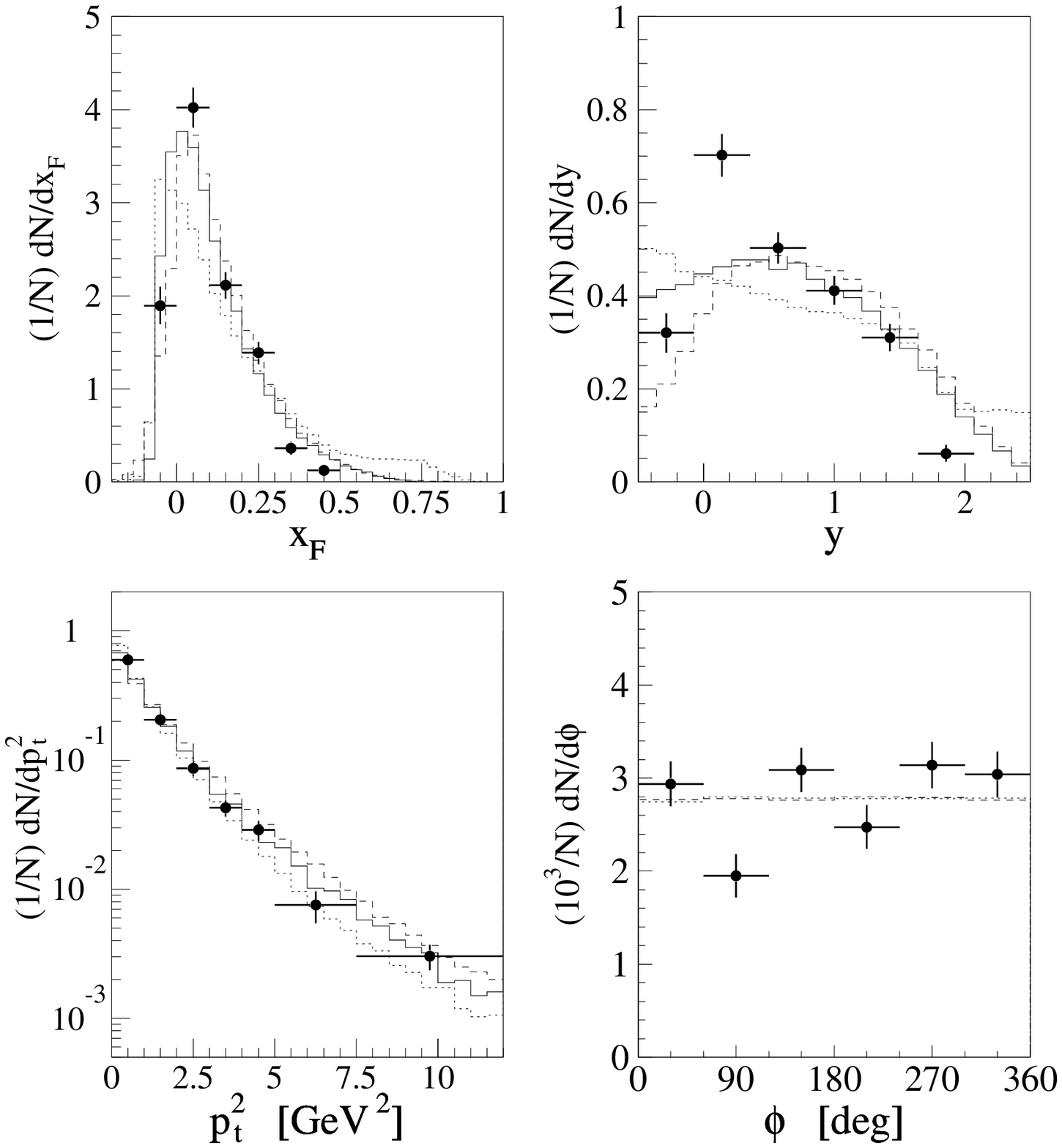,width=4.0in,angle=0}}
        \caption{Single-charm distributions for
        $\xf$, $y$, $\ptt$ and $\phi$:
        weighted data ($\bullet$) as described in
        Sec.~\protect\ref{ssec:accept};
        NLO QCD prediction (------);
        {\sc Pythia/Jetset}
        charm quark prediction (${\scriptscriptstyle -\;\!-\;\!-\;\!-}$);
        and {\sc Pythia/Jetset} $D$ meson prediction
        (${\scriptstyle \cdots\cdots\cdots}$).
        All distributions are obtained by summing the charm and anticharm
        distributions from charm-pair events.}
        \label{fig:fig17}
\end{figure}

Lack of agreement between an experimental charm-pair distribution
and a theoretical prediction can arise if the theory does not 
model the correlations between the two charm particles correctly.
However, it can also arise if the theory models the
correlations correctly but does not
correctly model the single-charm distributions.
Hence, before comparing the experimental charm-pair distributions
to theory, we first compare the 
acceptance-corrected single-charm distributions 
($x_F$, $y$, $p_t^2$ and $\phi$) to theory in Fig.~\ref{fig:fig17}.

For the longitudinal momentum distributions ---
$x_F$ and $y$ --- the experimental results and theoretical predictions
based on the default parameters
do not agree.  The experimental distributions are most 
similar to the NLO and {\sc Pythia/Jetset} $\ccb$ distributions, but are 
narrower than all three.  The difference between the 
{\sc Pythia/Jetset} $\ccb$ and the {\sc Pythia/Jetset} $\ddb$ longitudinal 
distributions 
shows the effect of the hadronization scheme that color-attaches one
charm quark to the remnant beam and the other to the remnant target, 
broadening the longitudinal distributions.

The experimental $\ptt$ distribution agrees
quite well with all three theoretical distributions.  
However, the {\sc Pythia/Jetset}
$\ccb$ distribution is somewhat flatter; the {\sc Pythia/Jetset}
$\ddb$ distribution is somewhat steeper.  As expected, both the 
theoretical and experimental $\phi$ distributions are consistent with being
flat.

\subsubsection{Longitudinal Distributions for Pairs}

\begin{figure} %Figure 18
        \centering
        \centerline{\epsfig{file=fig_17to23_label.ps,width=4.0in,angle=0}}
        \centerline{\epsfig{file=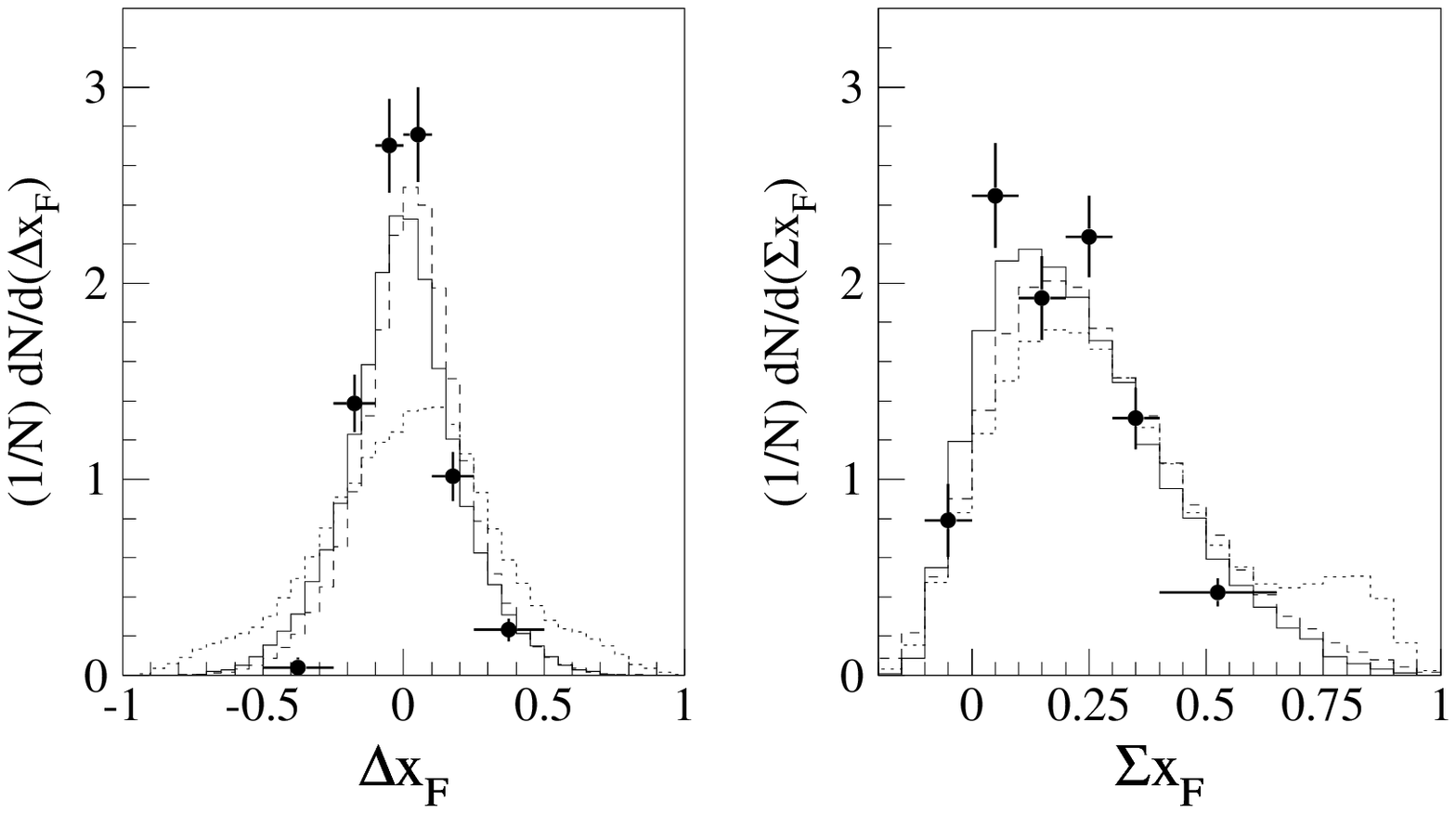,width=4.0in,angle=0}}
        \caption{Charm-pair $\dxf$ and $\sxf$ distributions;
        weighted data ($\bullet$) as described in
        Sec.~\protect\ref{ssec:accept};
        NLO QCD prediction (------);
        {\sc Pythia/Jetset}
        charm quark prediction (${\scriptscriptstyle -\;\!-\;\!-\;\!-}$);
        and {\sc Pythia/Jetset} $D$ meson prediction
        (${\scriptstyle \cdots\cdots\cdots}$).
        }
        \label{fig:fig18}
\end{figure}

\begin{figure} %Figure 19
        \centering
        \centerline{\epsfig{file=fig_17to23_label.ps,width=4.0in,angle=0}}
        \centerline{\epsfig{file=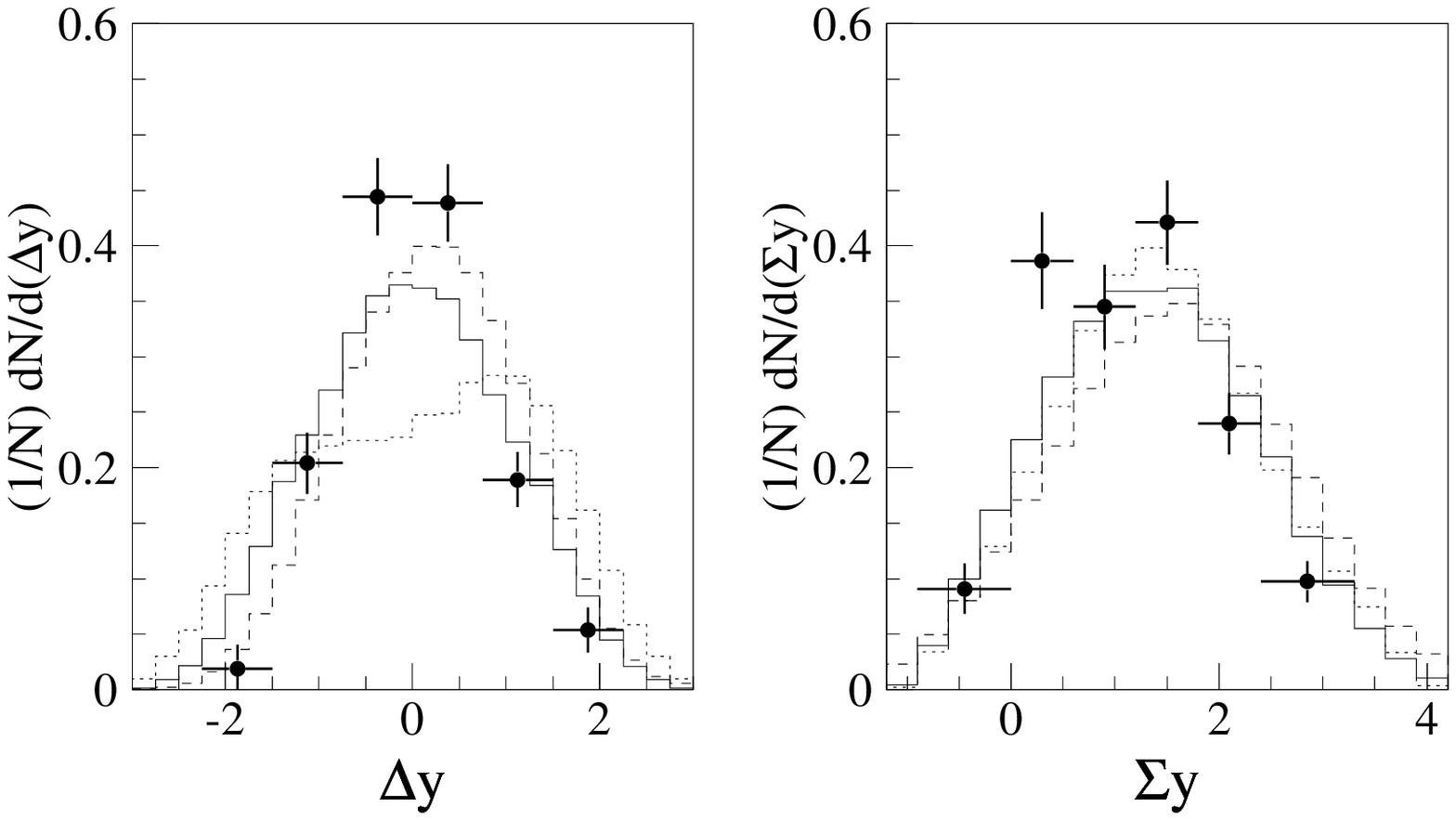,width=4.0in,angle=0}}
        \caption{Charm-pair $\dy$ and $\sy$ distributions;
        weighted data ($\bullet$) as described in
        Sec.~\protect\ref{ssec:accept};
        NLO QCD prediction (------);
        {\sc Pythia/Jetset}
        charm quark prediction (${\scriptscriptstyle -\;\!-\;\!-\;\!-}$);
        and {\sc Pythia/Jetset} $D$ meson prediction
        (${\scriptstyle \cdots\cdots\cdots}$).
        }
        \label{fig:fig19}
\end{figure}

\begin{figure} %Figure 20
        \centering
        \centerline{\epsfig{file=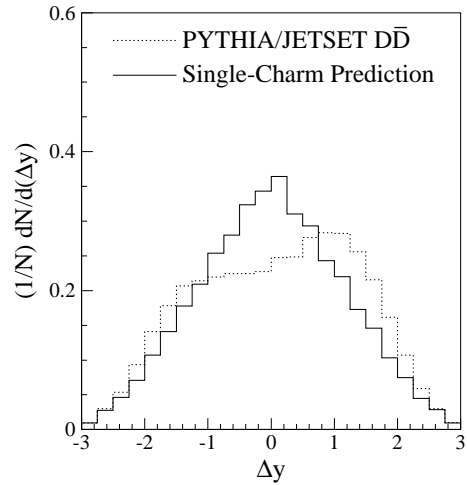,width=2.5in,angle=0}}
        \caption{
        {\sc Pythia/Jetset} $\ddb$ prediction for $\dy$ compared to the
        {\sc Pythia/Jetset} single-charm prediction that is obtained by assuming
        that the $D$ and $\db$ mesons are completely uncorrelated
        (see Eq.~\protect\ref{eq:qdv}).}
        \label{fig:fig20}
\end{figure}

The experimental and theoretical
$\dxf$ and $\sxf$ distributions (Fig.~\ref{fig:fig18})  and $\dy$ and
$\sy$ distributions (Fig.~\ref{fig:fig19}) do not agree with theoretical
predictions derived from default parameters. This
may be a reflection of the disagreement between the measured single-charm
longitudinal distributions and theoretical models.
As with the single-charm
distributions, the experimental results are much closer to the 
two $\ccb$ predictions than to the {\sc Pythia/Jetset} $\ddb$ prediction, but 
narrower than all three predictions. 
The {\sc Pythia/Jetset} hadronization
scheme introduces a strong correlation between the $D$ and $\overline{D}$ 
which significantly broadens the $\dy$ distribution.
That is, hadronization tends to pull the $\ddb$ apart, due to
color string attachment to the incident hadronic remnants.
As we show in Fig.~\ref{fig:fig20}, 
the {\sc Pythia/Jetset} $\ddb$ $\dy$ distribution is broader than
the prediction we 
obtain by using the predicted
single-charm distributions
and assuming they are uncorrelated, as calculated using 
Eqn.~\ref{eq:qdv}.
In 
contrast,
the experimental $\dy$ distribution is slightly
narrower than its uncorrelated single-charm prediction (Fig.~\ref{fig:fig9}).

\subsubsection{Transverse Distributions for Pairs}
\renewcommand{\textfraction}{0.}

\begin{figure} %Figure 21
        \centering
        \centerline{\epsfig{file=fig_17to23_label.ps,width=4.0in,angle=0}}
        \centerline{\epsfig{file=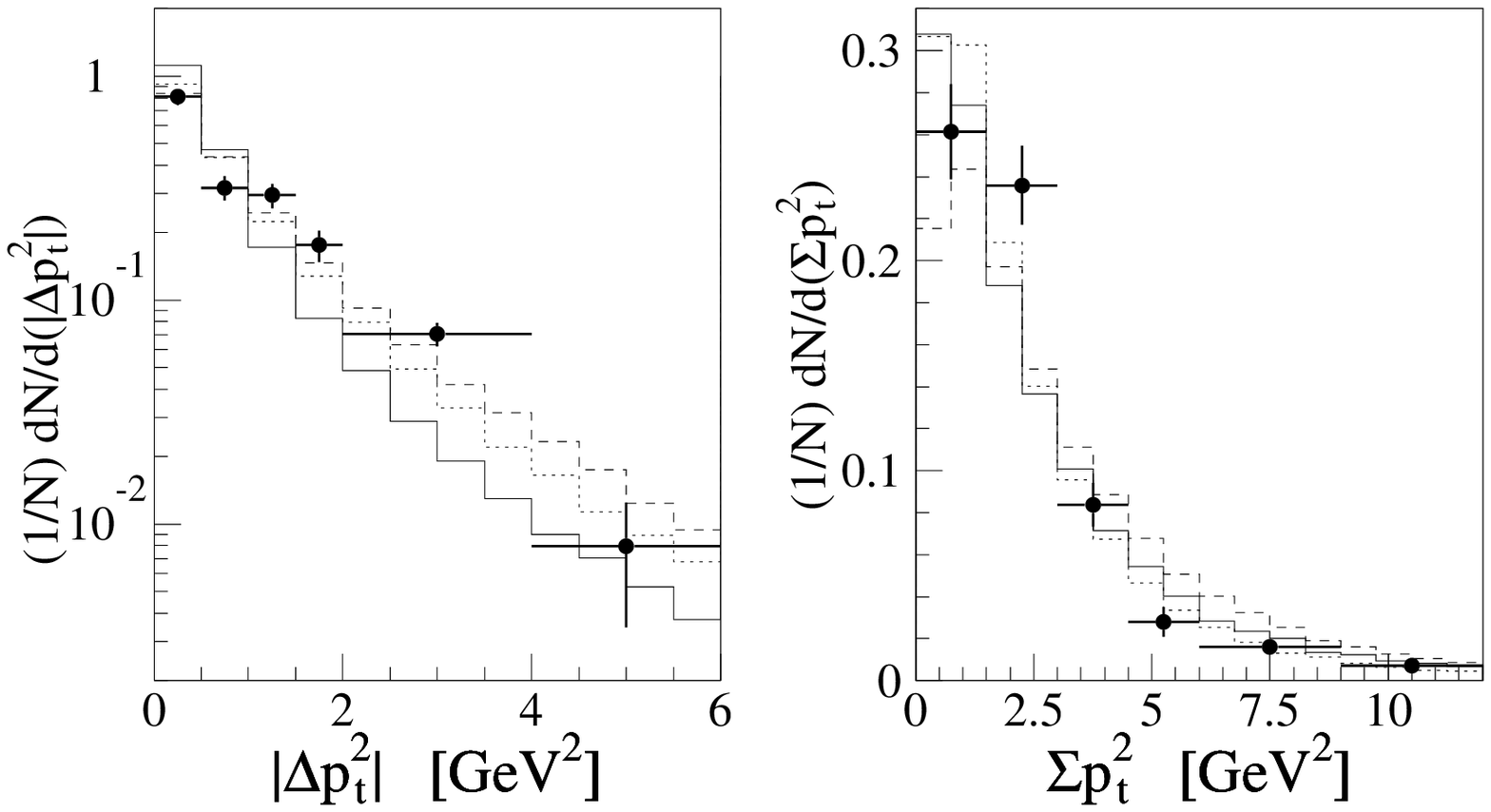,width=4.0in,angle=0}}
        \caption{Charm-pair $|\Delta p^2_t|$ and $\sptt$ distributions;
        weighted data ($\bullet$) as described in
        Sec.~\protect\ref{ssec:accept};
        NLO QCD prediction (------);
        {\sc Pythia/Jetset}
        charm quark prediction (${\scriptscriptstyle -\;\!-\;\!-\;\!-}$);
        and {\sc Pythia/Jetset} $D$ meson prediction
        (${\scriptstyle \cdots\cdots\cdots}$).
        }
        \label{fig:fig21}
\end{figure}

\begin{figure} %Figure 22
        \centering
        \centerline{\epsfig{file=fig_17to23_label.ps,width=4.0in,angle=0}}
        \centerline{\epsfig{file=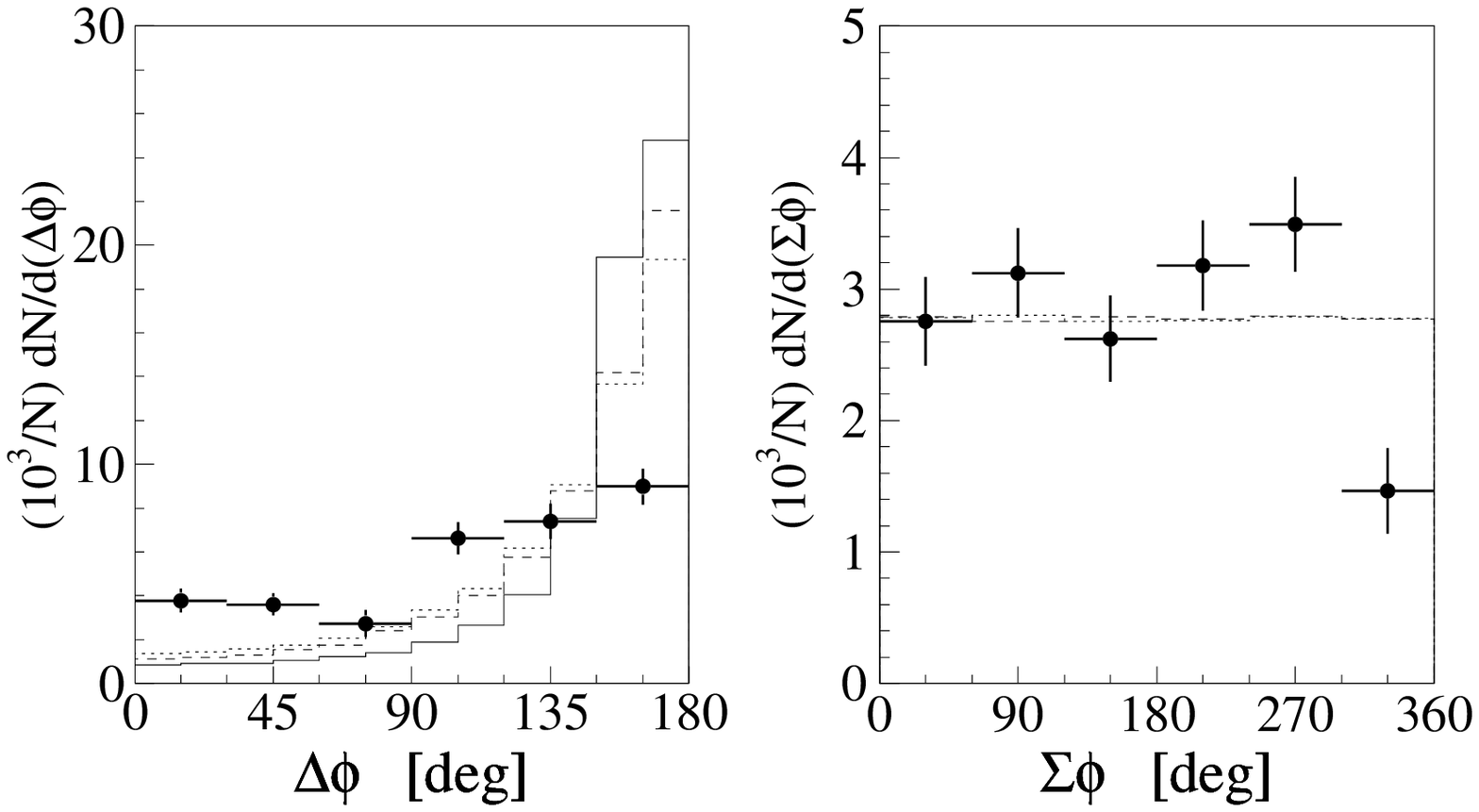,width=4.0in,angle=0}}
        \caption{Charm-pair $\dphi$ and $\sphi$ distributions;
        weighted data ($\bullet$) as described in
        Sec.~\protect\ref{ssec:accept};
        NLO QCD prediction (------);
        {\sc Pythia/Jetset}
        charm quark prediction (${\scriptscriptstyle -\;\!-\;\!-\;\!-}$);
        and {\sc Pythia/Jetset} $D$ meson prediction
        (${\scriptstyle \cdots\cdots\cdots}$).
        }
        \label{fig:fig22}
\end{figure}

\begin{figure} %Figure 23
        \centering
        \centerline{\epsfig{file=fig_17to23_label.ps,width=4.0in,angle=0}}
        \centerline{\epsfig{file=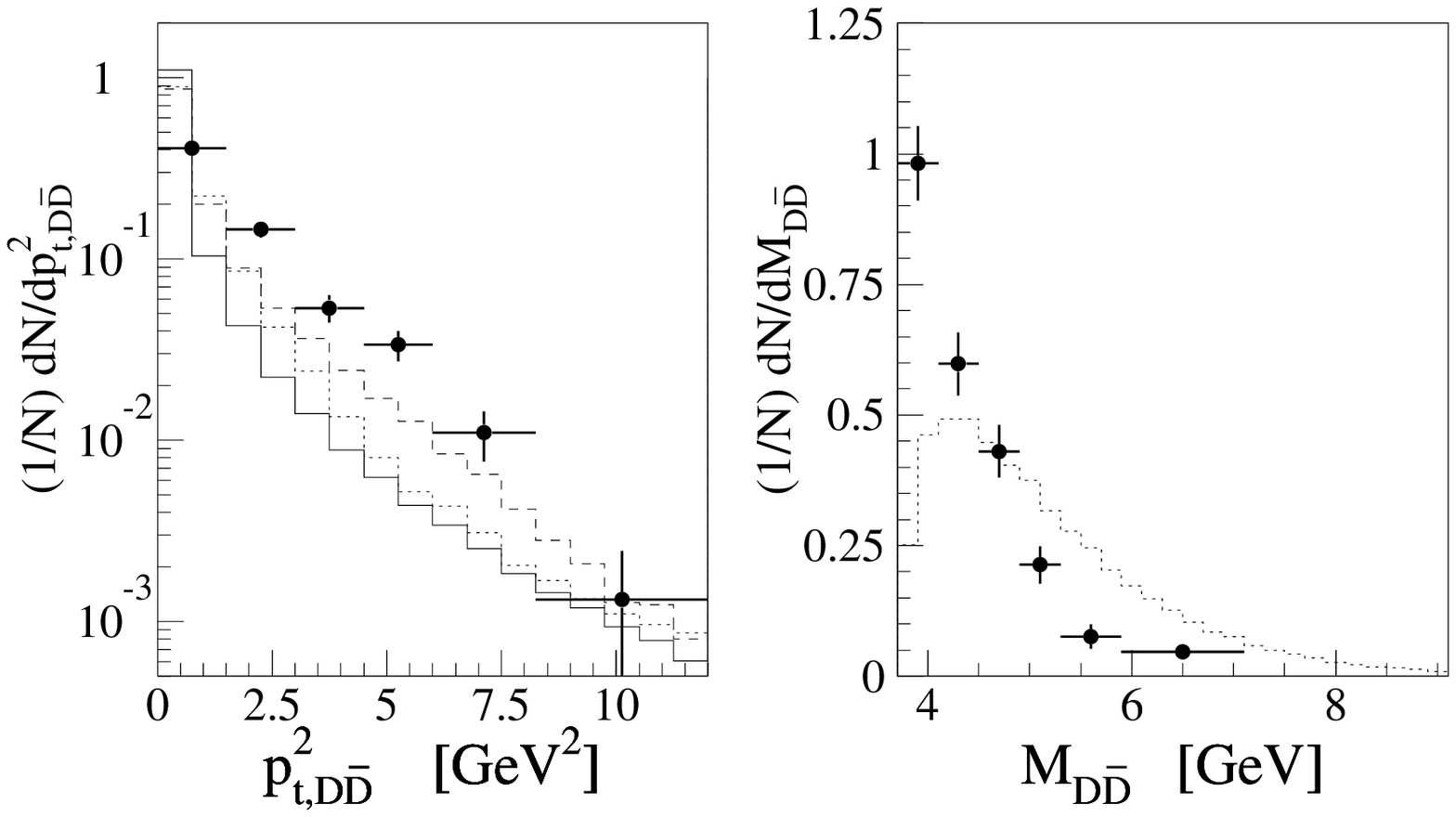,width=4.0in,angle=0}}
        \caption{Charm-pair $\pttddb$ and $\Mddb$ distributions:
        weighted data ($\bullet$) as described in
        Sec.~\protect\ref{ssec:accept};
        NLO QCD prediction (------);
        {\sc Pythia/Jetset}
        charm quark prediction (${\scriptscriptstyle -\;\!-\;\!-\;\!-}$);
        and {\sc Pythia/Jetset} $D$ meson prediction
        (${\scriptstyle \cdots\cdots\cdots}$).
        }
        \label{fig:fig23}
\end{figure}

In Figs.~\ref{fig:fig21}--\ref{fig:fig23}, we compare 
experimental distributions to theoretical predictions for the 
following transverse variables: $|\dptt|$, 
$\sptt$, $\dphi$, $\sphi$, and $\pttddb$.
Any observed discrepancy between theory and 
data for the $|\Delta p^2_t|$, $\Delta \phi$, and 
$p^2_{t,D\overline{D}}$ 
distributions is 
noteworthy because the single-charm $p^2_t$ and $\phi$ experimental
distributions agree 
well with theory.
An observed discrepancy, therefore, must derive from the theory
modeling the correlation between $\vec{p}_{t,D}$ and 
$\vec{p}_{t,\overline{D}}$ incorrectly.

If $\vec{p}_{t,D}$ and $\vec{p}_{t,\overline{D}}$ were 
completely uncorrelated, then the single-charm predictions
(Figs.~\ref{fig:fig10}--\ref{fig:fig12}) would provide
good estimates for these three distributions.
At the opposite extreme, if $\vec{p}_{t,D}$ and $\vec{p}_{t,\overline{D}}$
were completely anticorrelated --- 
as in the leading-order perturbative
QCD prediction --- then
the $\Delta p^2_t$ distribution would be a delta function 
at $\Delta p^2_t = 0\; {\rm GeV}^2$; 
the $p^2_{t,D\overline{D}}$ distribution a 
delta function at $p^2_{t,D\overline{D}}=0 \; {\rm GeV}^2$; and 
the $\Delta \phi$ distribution
a delta function at $\Delta \phi=180^{\circ}$.
Both the experimental distributions and the three sets of 
theoretical predictions lie between these extremes.
None of the
three experimental distributions, however, is as steep as any of the 
theoretical predictions.  The next-to-leading-order predictions
are the steepest --- that is, the next-to-leading-order calculation
predicts the most correlation between $\vec{p}_{t,c}$ and 
$\vec{p}_{t,\overline{c}}$ as a model for 
$\vec{p}_{t,D}$ and $\vec{p}_{t,\overline{D}}$.
Thus hadronization and higher-order perturbative effects
smear out the $\ccb$ correlations.
The {\sc Pythia/Jetset} hadronization scheme
broadens the $\Delta \phi$ distribution, bringing it 
closer to
the experimental results.  The same hadronization scheme also
narrows the $\pttddb$ and $\dptt$ distributions, which
makes them disagree even more with the experimental results.
One mechanism which would broaden the $\pttddb$ and $\dptt$
distributions as well as the $\dphi$
distribution (bringing all
into better agreement with the experimental results) is
to increase the intrinsic transverse momentum of the 
colliding partons in the 
beam and target hadrons 
(see the Appendix, Secs.~\ref{ssec:npeffects} and~\ref{ssec:itm}).
An improved theoretical understanding may involve 
adding terms higher than NLO to
calculations,
although other authors
find good agreement by choosing appropriate values for
nonperturbative parameters~\cite{ref:bib42}.

\subsubsection{Charm-Pair Invariant Mass}
In Fig.~\ref{fig:fig23}, we compare the experimental 
charm-pair invariant mass distribution to the {\sc Pythia/Jetset} 
prediction.
The experimental $\Mddb$ distribution is
steeper than the theoretical predictions. This is similar to
the experimental single-charm $x_F$ (or $y$) distributions, which
are also steeper than the theoretical predictions. (See Fig.~\ref{fig:fig17}.)

In addition, the correlations introduced by the 
{\sc Pythia/Jetset} hadronization scheme broaden the invariant mass 
distribution.

\subsubsection{Two-Dimensional Distributions}
\label{sssec:twodd}
\begin{figure} %Figure 24
        \centering
        \centerline{\epsfig{file=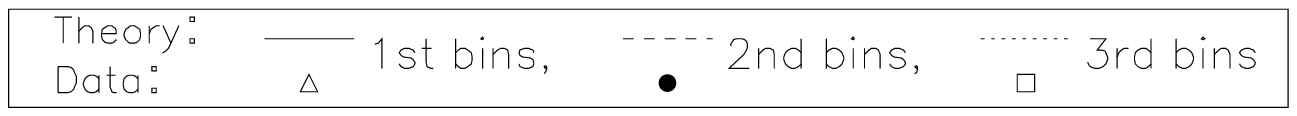,width=4.0in,angle=0}}
        \centerline{\epsfig{file=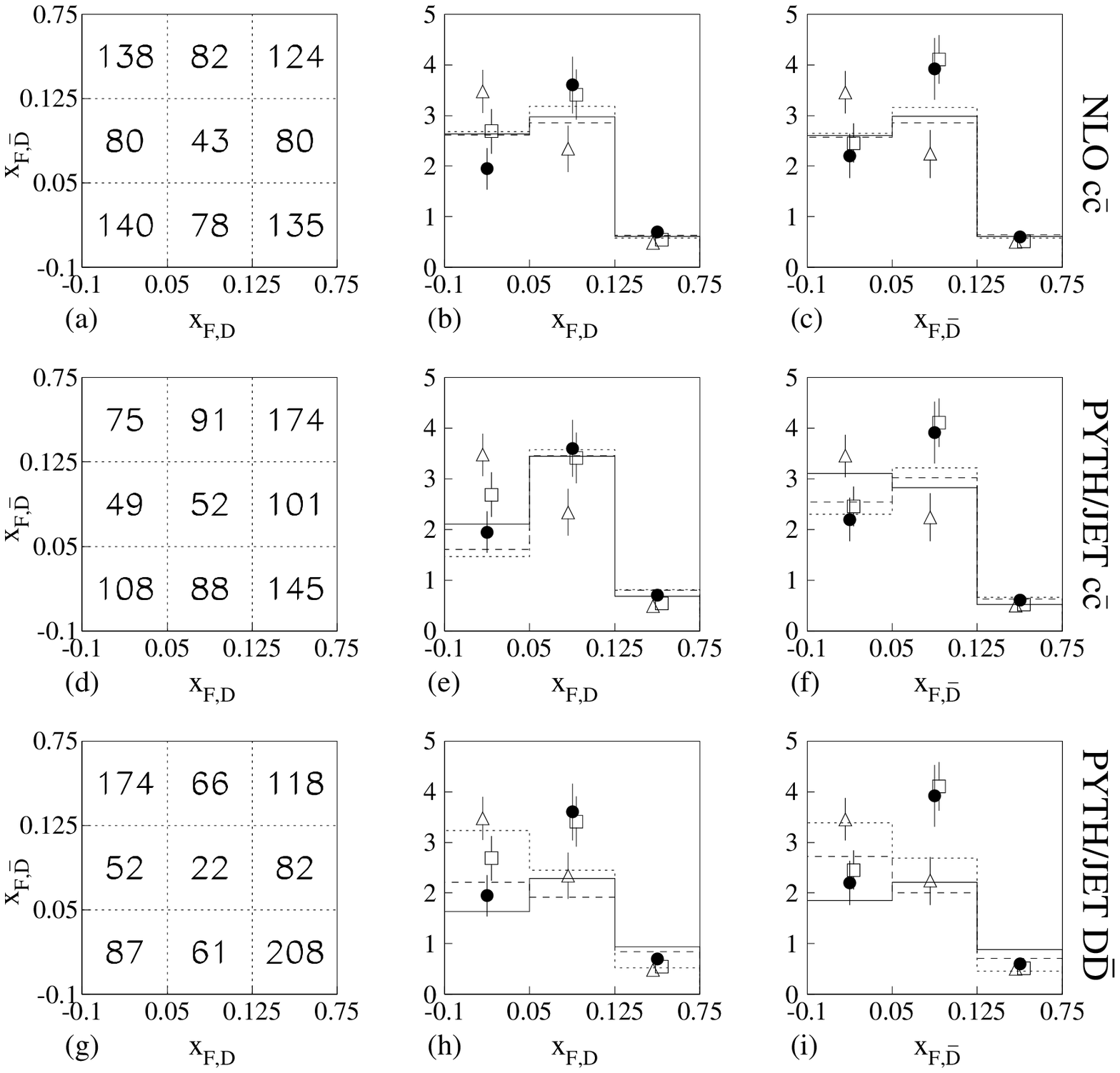,width=4.0in,angle=0}}
        \caption{
        (a) The NLO QCD prediction for the number
        of $\ccb$ events in nine
        ($\xfd$, $\xfdb$) bins, normalized such that the
        number of NLO events equals the number of weighted
        $\ddb$ signal events.
        (b) Experimental $\xfd$ distribution
        for each $\xfdb$ bin
        compared to the NLO QCD predictions.  Each $\xfd$
        distribution
        is normalized such that the integral over $\xfd$ equals one.
        (c) Same as (b) for the $\xfdb$ distributions.
        (d)-(f)~Same as (a)-(c) for the {\sc Pythia/Jetset} $\ccb$ prediction.
        (g)-(i)~Same as (a)-(c) for the {\sc Pythia/Jetset} $\ddb$ prediction.
        Symbols represent weighted data; histograms represent
        theoretical predictions.
        $\triangle$ and ------ correspond to the low bin; $\bullet$
        and ${\scriptscriptstyle -\;\!-\;\!-\;\!-}$ to the middle bin;
        $\Box$ and ${\scriptstyle \cdots\cdots\cdots}$ to the high bin.}
        \label{fig:fig24}
\end{figure}

\begin{figure} %Figure 25
        \centering
        \centerline{\epsfig{file=fig_24to27_label.ps,width=4.0in,angle=0}}
        \centerline{\epsfig{file=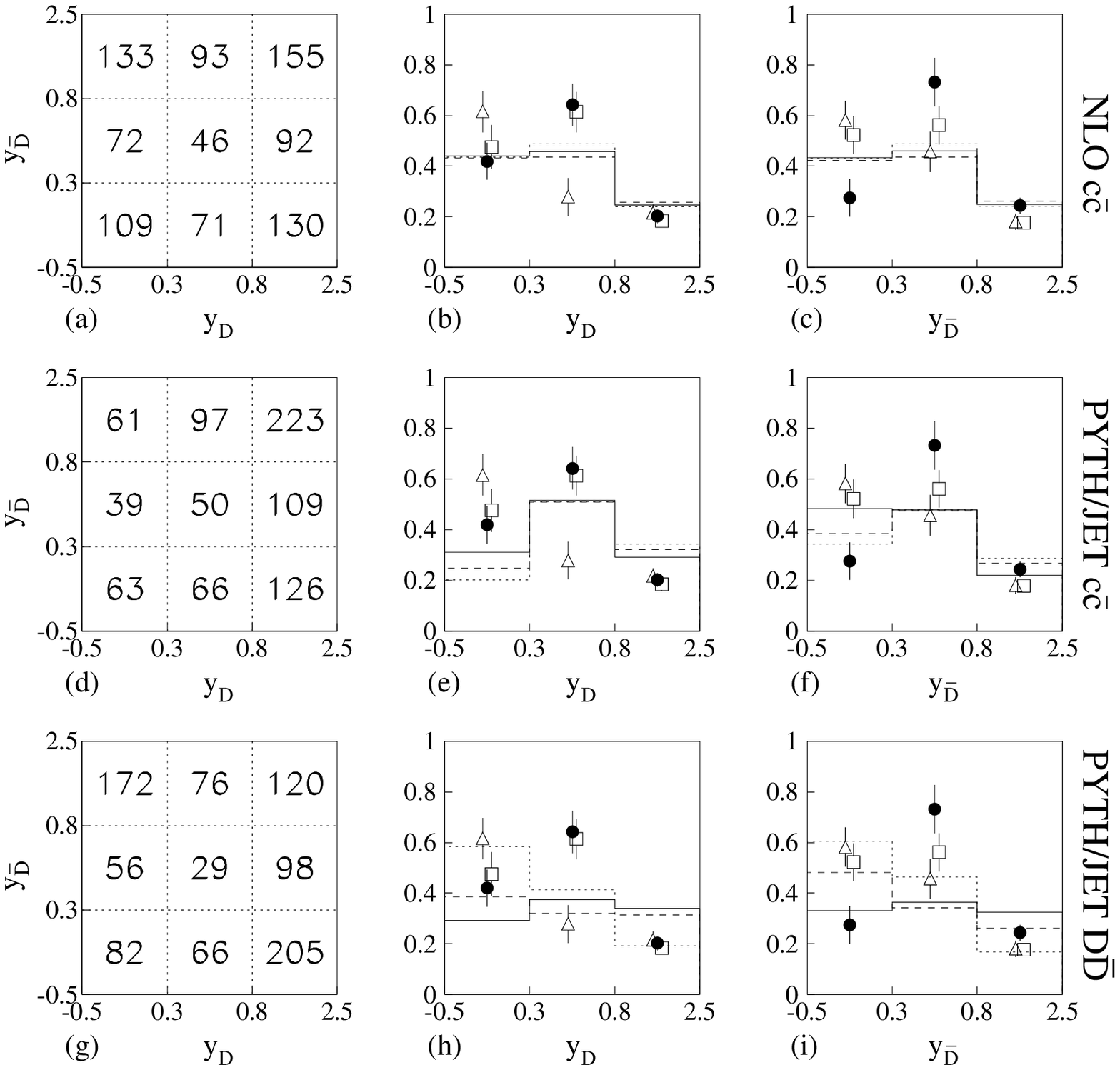,width=4.0in,angle=0}}
        \caption{(a) The NLO QCD prediction for the number
        of $\ccb$ events in nine
        ($\yd$, $\ydb$) bins, normalized such that the
        number of NLO events equals the number of weighted
        $\ddb$ signal events.
        (b) Experimental $\yd$ distribution
        for each $\ydb$ bin compared to the NLO QCD
        predictions. Each $\yd$ distribution
        is normalized such that the integral over $\yd$ equals one.
        (c) Same as (b) for the $\ydb$ distributions.
        (d)-(f)~Same as (a)-(c) for the {\sc Pythia/Jetset} $\ccb$ prediction.
        (g)-(i)~Same as (a)-(c) for the {\sc Pythia/Jetset} $\ddb$ prediction.
        Symbols represent weighted data; histograms represent
        theoretical predictions.
        $\triangle$ and ------ correspond to the low bin; $\bullet$
        and ${\scriptscriptstyle -\;\!-\;\!-\;\!-}$ to the middle bin;
        $\Box$ and ${\scriptstyle \cdots\cdots\cdots}$ to the high bin.
        }
        \label{fig:fig25}
\end{figure}

\begin{figure} %Figure 26
        \centering
        \centerline{\epsfig{file=fig_24to27_label.ps,width=4.0in,angle=0}}
        \centerline{\epsfig{file=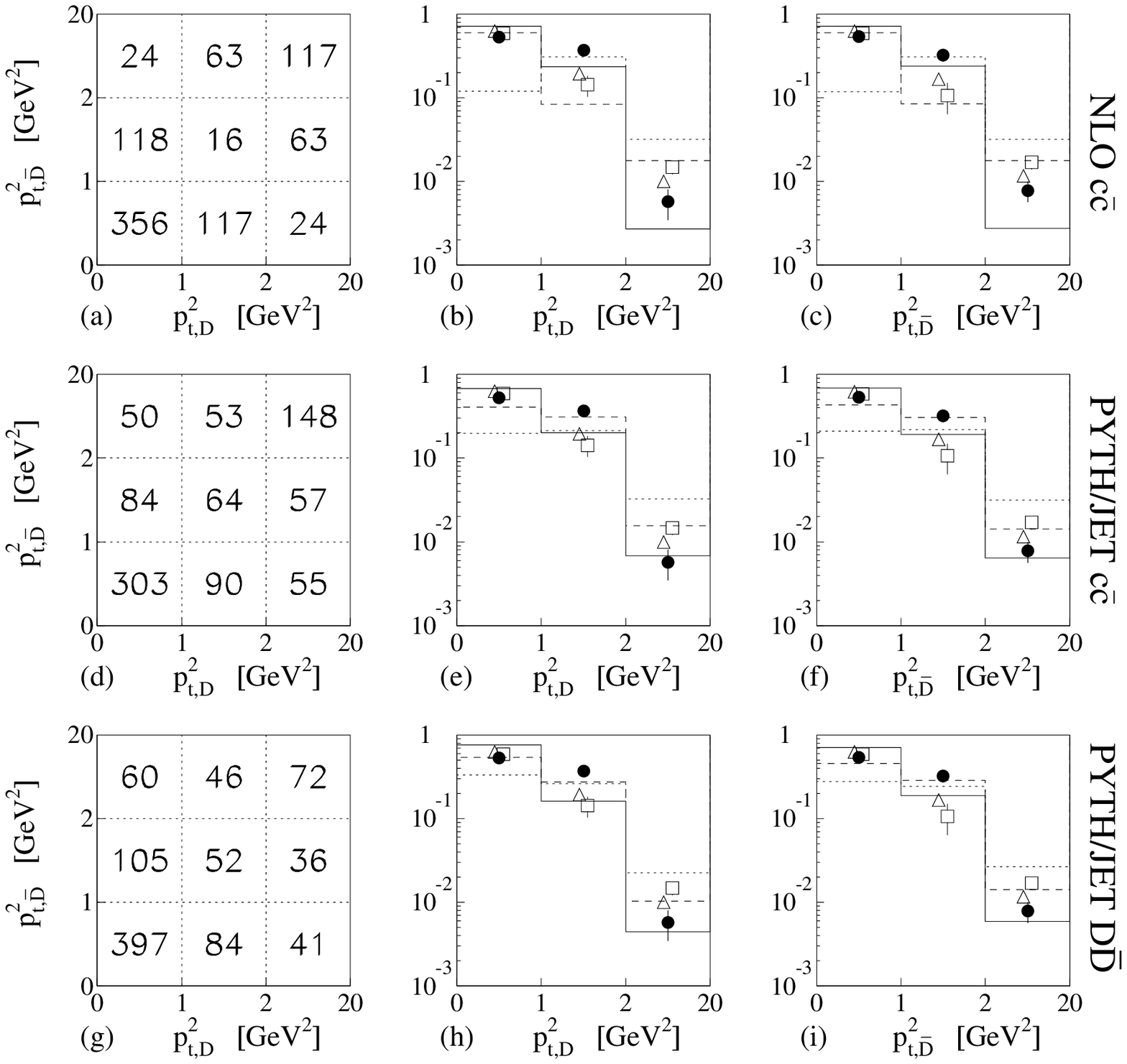,width=4.0in,angle=0}}
        \caption{(a) The NLO QCD prediction for the number
        of $\ccb$ events in nine
        ($\pttd$, $\pttdb$) bins, normalized such that the
        number of NLO events equals the number of weighted
        $\ddb$ signal events.
        (b) Experimental $\pttd$ distribution
        for each $\pttdb$ bin compared to the NLO
        QCD predictions.  Each $p^2_{t,D}$ distribution
        is normalized such that the integral over $\pttd$ equals one.
        (c) Same as (b) for the $\pttdb$ distributions.
        (d)-(f)~Same as (a)-(c) for the {\sc Pythia/Jetset} $\ccb$ prediction.
        (g)-(i)~Same as (a)-(c) for the {\sc Pythia/Jetset} $\ddb$ prediction.
        Symbols represent weighted data; histograms represent
        theoretical predictions.
        $\triangle$ and ------ correspond to the low bin; $\bullet$
        and ${\scriptscriptstyle -\;\!-\;\!-\;\!-}$ to the middle bin;
        $\Box$ and ${\scriptstyle \cdots\cdots\cdots}$ to the high bin.
        }
        \label{fig:fig26}
\end{figure}

\begin{figure} %Figure 27
        \centering
        \centerline{\epsfig{file=fig_24to27_label.ps,width=4.0in,angle=0}}
        \centerline{\epsfig{file=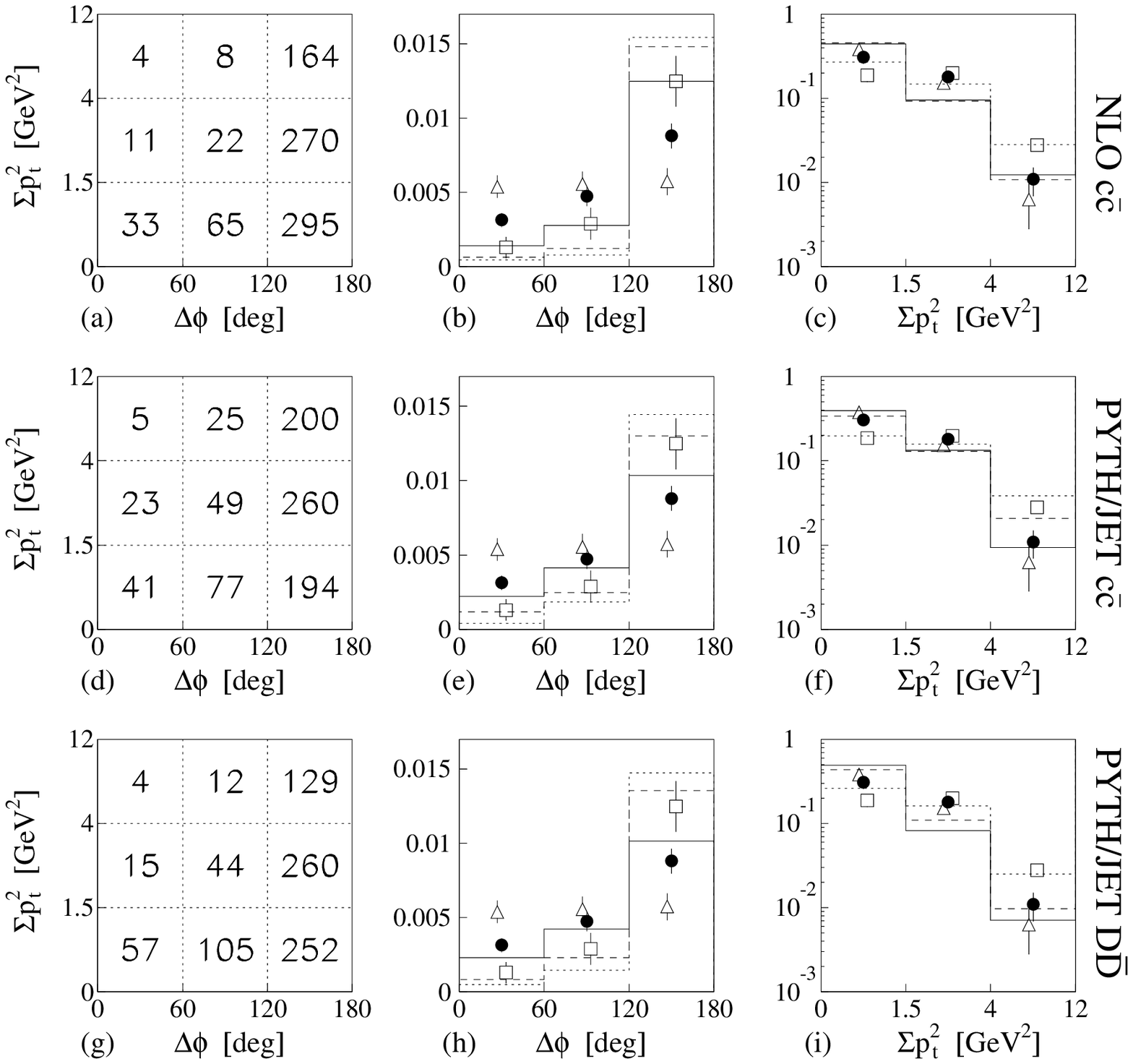,width=4.0in,angle=0}}
        \caption{(a) The NLO QCD prediction for the number
        of $\ccb$ events in nine
        ($\dphi$, $\Sigma p^2_t$) bins, normalized such that the
        number of NLO events equals the number of weighted
        $\ddb$ signal events.
        (b) Experimental $\dphi$ distribution
        for each $\Sigma p^2_t$ bin
        compared to the NLO QCD predictions.  Each $\Delta \phi$
        distribution
        is normalized such that the integral over $\Delta \phi$ equals one.
        (c) Same as (b) for the $\Sigma p^2_t$ distributions.
        (d)-(f)~Same as (a)-(c) for the {\sc Pythia/Jetset} $\ccb$ prediction.
        (g)-(i)~Same as (a)-(c) for the {\sc Pythia/Jetset} $\ddb$ prediction.
        Symbols represent weighted data; histograms represent
        theoretical predictions.
        $\triangle$ and ------ correspond to the low bin; $\bullet$
        and ${\scriptscriptstyle -\;\!-\;\!-\;\!-}$ to the middle bin;
        $\Box$ and ${\scriptstyle \cdots\cdots\cdots}$ to the high bin.
        }
        \label{fig:fig27}
\end{figure}

In Figs.~\ref{fig:fig24}--\ref{fig:fig27}, we examine the 
same two-dimensional experimental 
distributions discussed in Section~\ref{ssec:twod}.
We now compare these experimental results 
to the three sets of theoretical predictions.  In each figure, the top
row shows the NLO perturbative QCD $\ccb$ prediction; the
middle row, the {\sc Pythia/Jetset} $\ccb$ prediction; and the bottom
row, the {\sc Pythia/Jetset} $\ddb$ prediction.  The experimental
data points and errors
are repeated in each row.

The longitudinal distributions, $x_F$ and $y$, are shown in 
Figs.~\ref{fig:fig24} and~\ref{fig:fig25}.
The three theoretical predictions are quite different.  The
NLO $\ccb$ predictions show no significant correlation 
between $\xfd$ and $\xfdb$ 
(or between $\yd$ and $\ydb$) and
the $\xfd$ and $\xfdb$ distributions are quite 
similar.  The {\sc Pythia/Jetset} $\ccb$ predictions show a slight correlation
and
the $\xfd$ and $\xfdb$ distributions
are somewhat different.  Due to the {\sc Pythia/Jetset} hadronization scheme,
the {\sc Pythia/Jetset} $\ddb$ prediction shows the 
strongest correlation between $\xfd$ and $\xfdb$.
Interestingly, in the {\sc Pythia/Jetset} $\ddb$ prediction, $\xfd$ and 
$\xfdb$ are anticorrelated; in the
{\sc Pythia/Jetset} $\ccb$ prediction they are positively correlated.
The correlation patterns in the experimental results, although 
inconsistent with any of the theoretical predictions,
are closest to the {\sc Pythia/Jetset} $\ccb$ predictions.

In Fig.~\ref{fig:fig26},  we investigate the correlations
between $p^2_{t,D}$ and $p^2_{t,\db}$.  
The three theoretical
predictions and the experimental results all show similar
trends.  Although 
all the distributions are broader than the leading-order perturbative
QCD prediction --- a delta function at $p^2_{t,D} = p^2_{t,\db}$ ---
they all shows signs of an enhancement in the
$p^2_{t,D}=p^2_{t,\db}$ bins.  
The {\sc Pythia/Jetset} $\ccb$ distributions and the {\sc Pythia/Jetset} 
$\ddb$ distributions are very similar and resemble the experimental 
results more so than the NLO $\ccb$ distributions.  All of the 
theoretical third-bin distributions are significantly flatter
than the experimental third-bin distributions.  In contrast to the
longitudinal distributions, all the $p^2_{t,D}$ distributions 
are very similar
to the respective $p^2_{t,\overline{D}}$ distributions.

In Fig.~\ref{fig:fig27}, we investigate correlations
between $\dphi$ and $\sptt$.
For the $\dphi$ dependence, 
a leading-order perturbative QCD calculation predicts a 
delta function at $\dphi = 180^{\circ}$.  We expect
perturbative predictions to be more accurate as the 
energy scale $Q$ of the partonic hard scattering increases:
\begin{equation}
\label{eq:q}
Q \equiv \sqrt{m_c^2 + \frac{p^2_{t,c} + p^2_{t,\overline{c}}}{2}}.
\end{equation}
That is, we expect the $\dphi$ distribution to be more 
peaked at $180^\circ$ for $\ddb$ events with larger 
$\sptt$.  This behavior 
is clearly evident in our experimental distributions as well as in 
all three theoretical predictions.  The theoretical
$\dphi$ distributions, however, for all three $\sptt$ bins, 
are significantly steeper than the respective experimental distributions.
The NLO $\ccb$ $\dphi$ distributions are the steepest.
The experimental and theoretical $\sptt$ distributions are 
in fairly good agreement, with the $\sptt$ distribution 
broadening as $\dphi$ increases.
No significant correlation between $\dphi$ and $|\dptt|$ 
or between $\dphi$ and $\dy$
is seen
in the data or theory.

\subsection{Dependence of Yields and Longitudinal Correlations on
Type of $\ddb$ Pair}

\label{ssec:asym}

\subsubsection{Relative Yields}

In Table~\ref{tab:tab3}, we compare the
experimental yields for
each type of $\ddb$ pair to the predictions from the {\sc Pythia/Jetset} event 
generator and to a naive spin-counting model.
The experimental results are obtained by maximizing the 
weighted likelihood function where
the weights account for both acceptance effects and the relative 
branching fractions of the reconstructed decay modes
(Sec.~\ref{ssec:accept}).
Again, for both data and {\sc Pythia/Jetset} predictions,
the results are for pairs in which the rapidities of both the
$D$ and $\db$ lie between $-0.5$ and $2.5$.
The {\sc Pythia/Jetset} and naive spin-counting models both
assume that vector $D^*$ production is three times more likely than 
pseudoscalar $D$ production due to the number of spin states and that
contributions from
higher spin states are negligible.  They also use the
known $D^{*\pm}$ branching fractions, $B(D^{*+}\to D^0\pi^+) = 68.3\%$ and 
$B(D^{*+}\to D^+X) = 31.7\%$, to determine $D$ production.
The differences between the {\sc Pythia/Jetset} and naive spin-counting model
come from the {\sc Pythia/Jetset} hadronization scheme --- in particular,
the rate of coalescence.  As discussed in Sec.~\ref{ssec:hadro}, a charm
quark tied to a 
valence quark by a
low-mass string can coalesce 
with that valence quark
into a meson.  This will
tend to increase the rate of $D^-(\overline{c}d)$ and $D^0(c\overline{u})$ 
production
in the forward region for the E791 $\pi^-(\overline{u}d)$ beam.
Since $D^{*-}$ decays to $\dzb$, production of $\dzb$ will also 
be enhanced.
This effect is seen in Table~\ref{tab:tab3} where 
the number of pairs that contain a 
$D^+$ is
reduced while the number of pairs that contain a $D^-, D^0,$ or $\dzb$ is
increased
in the {\sc Pythia/Jetset} model relative to the naive model.  Both models 
agree
with data as far as the relative ordering but predict too many 
$\dzdb$
pairs and too few $\dpdm$ pairs.

\begin{table} [htbp]
        \caption{Normalized acceptance-corrected
        experimental yields for the four types of
        $\ddb$ pairs compared to predictions from a naive spin-counting
        model and
        the {\sc Pythia/Jetset} event generator.
        Experimental and {\sc Pythia/Jetset} yields are for
        $-0.5 < y_{D,\overline{D}} < 2.5$. 
Statistical and systematic errors are given for the data.}
        \label{tab:tab3}
\begin{tabular}{lccc}  \hline \hline
& \multicolumn{1}{c}{Data} & \multicolumn{1}{c}{Spin-counting model} &
\multicolumn{1}{c}{{\sc Pythia/Jetset}} \\  \\ \hline \hline
$\dzdb$             & $0.50\pm0.04\pm0.01$ & 0.572 & $0.604\pm0.002$ \\
$\dzdm$             & $0.20\pm0.02\pm0.01$ & 0.184 & $0.208\pm0.002$ \\
$\dpdb$             & $0.18\pm0.02\pm0.01$ & 0.184 & $0.138\pm0.002$ \\
$\dpdm$             & $0.12\pm0.02\pm0.01$ & 0.060 & $0.051\pm0.001$ \\ \hline
\end{tabular}
\end{table}

\subsubsection{Correlations Between the $D$ and $\db$ Longitudinal 
Momenta}
\begin{figure} %Figure 28
        \centering
        \centerline{\epsfig{file=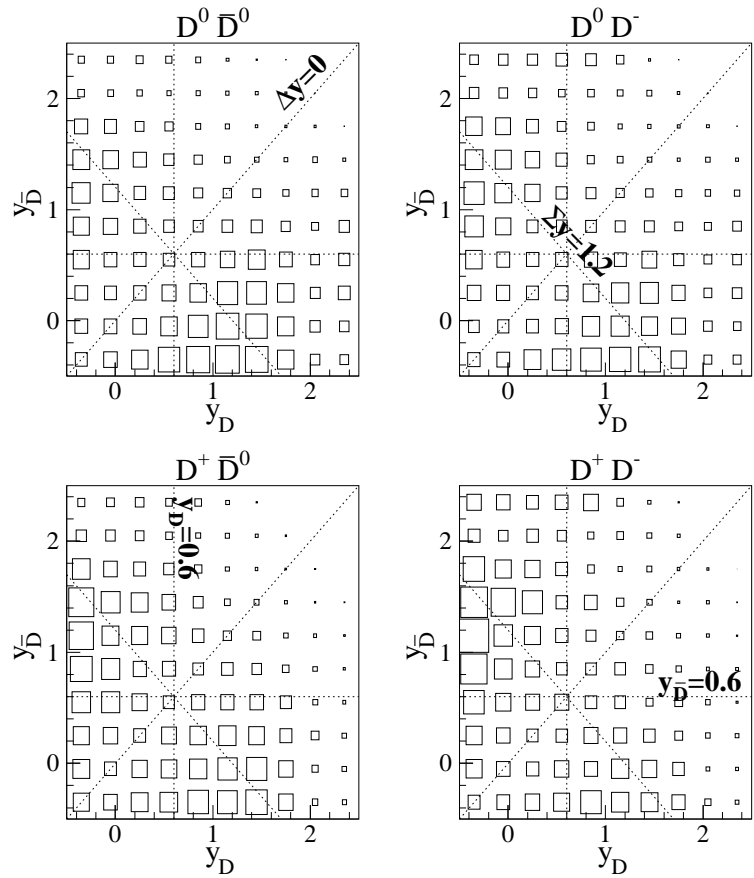,width=4.0in,angle=0}}
        \caption{{\sc Pythia/Jetset} prediction for the
        ($\yd$, $\ydb$)
        distribution for each of the four types of $\ddb$ pairs.  The dashed
        lines help define the asymmetry functions $A_v(i)$
        (Eq.~\protect\ref{eq:av}) shown in Fig.~\protect\ref{fig:fig29}.}
        \label{fig:fig28}
\end{figure}

\begin{figure} %Figure 29
        \centering
        \centerline{\epsfig{file=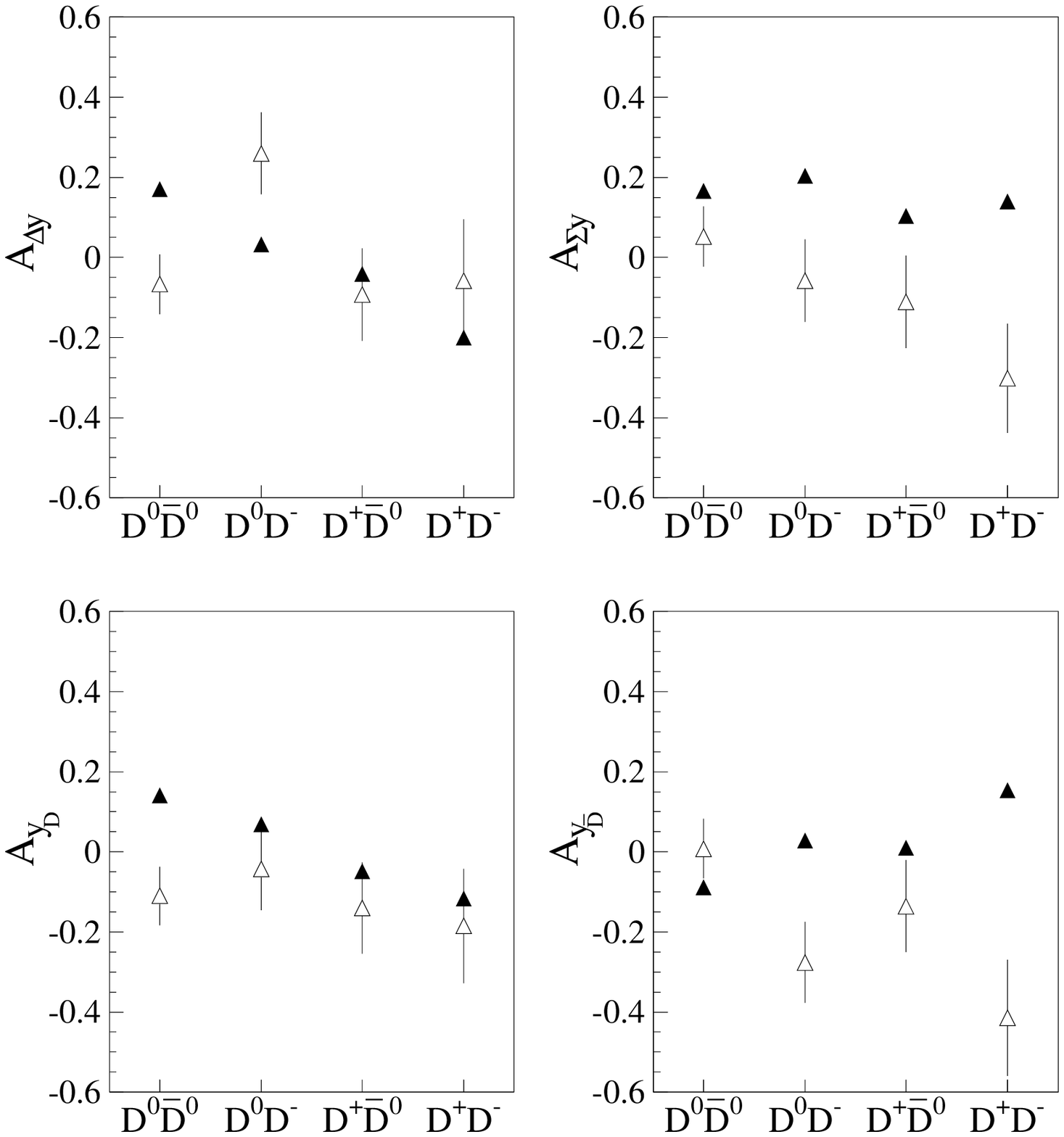,width=4.0in,angle=0}}
        \caption{The asymmetry function
        $A_v(i) = \frac{N_i(v>a) - N_i(v<a)}{N_i(v>a) + N_i(v<a)}$, where
        $i=$($\dzdb$, $\dzdm$, $\dpdb$, $\dpdm$) and
        $N_i$ is the number of signal $\ddb$ events of type $i$, for
        ($v$, $a$) = ($\dy$, $0$), ($\sy$, $1.2$), ($\yd$, $0.6$), and
        ($\ydb$, $0.6$).  Both the weighted data (open symbols) and the
        {\sc Pythia/Jetset} (closed symbols) distributions
        correspond to events in which $-0.5 < \yd,\ydb < 2.5$.}
        \label{fig:fig29}
\end{figure}

As shown in Fig.~\ref{fig:fig28}, in the {\sc Pythia/Jetset} hadronization
scheme, the correlation between 
$\yd$ and $\ydb$ is quite different for 
each of the four types of $\ddb$ pairs.
In Fig.~\ref{fig:fig29}, we investigate whether this is also
true for data.  Given the limited size of our data
sample, we can only search for gross asymmetries in the 
($\yd$, $\ydb$) distributions.
We obtain the four plots in Fig.~\ref{fig:fig29} by bisecting 
the two-dimensional ($\yd$, $\ydb$)
distributions along the following four lines ($v=a$):
$\dy = 0$, $\sy = 1.2$, $y_D = 0.6$, and
$\ydb=0.6$.  
These four lines are indicated by dashed lines in Fig.~\ref{fig:fig28}.

To search for possible differences in asymmetries, 
we determine 
whether the fraction of signal events on one side of a given line 
depends on the type of $\ddb$ pair.
Specifically, 
for both theory and data, we show in Fig.~\ref{fig:fig29}
\begin{equation}
\label{eq:av}
A_v(i) = \frac{N_i(v>a) - N_i(v<a)}{N_i(v>a) + N_i(v<a)} 
\end{equation}
where $i=(\dzdb, \dzdm, \dpdb, \dpdm)$ and
$N_i$ is the number of signal $\ddb$ events of type $i$.
The {\sc Pythia/Jetset} $A_{\Sigma y}$ distribution is fairly flat, 
indicating no significant differences among the four $\ddb$ types for 
the $\sy$ distribution. The 
{\sc Pythia/Jetset} $A_{\dy}$, $A_{\yd}$, and $A_{y_{\db}}$ 
distributions, 
however, indicate significant differences, all of which are easily 
interpreted in terms
of the {\sc Pythia/Jetset} coalescence mechanism discussed in 
Sec.~\ref{sec:theory}.  
Unfortunately, our experimental errors are
of the same order as the degree of differences in the 
{\sc Pythia/Jetset} predictions.
The experimental $A_{\yd}$ distribution, for example,
is consistent with the {\sc Pythia/Jetset} prediction, but it is also consistent
with being flat.  Similarly, the experimental $A_{\sy}$ distribution 
is fairly consistent with the flat {\sc Pythia/Jetset} prediction, but 
it also shows some indication of a difference between $\dzdb$ and
$\dpdm$.
The most significant difference between the experimental
results and the {\sc Pythia/Jetset} predictions occurs for the $\ddb$ types
$\dzdb$ and $\dpdm$ in the 
$A_{y_{\db}}$ distribution.  
Both theory and data indicate a difference between $\dzdb$ 
and $\dpdm$; however, we find experimentally that
$A_{y_{\db}}(\dzdb) > A_{y_{\db}}(\dpdm)$,
whereas the {\sc Pythia/Jetset} model finds 
$A_{y_{\db}}(\dpdm) > A_{y_{\db}}(\dzdb)$.

%%%%%%%%%%%%%%%%%%%%%%%%%%%%%%%%%%%%%%%%%%%%%%%%%%%%%%%%%%%%%%%%%%%%%%  
%\input{offline_doc_290_7_r1.tex} %  \section{Conclusions}                  
                  
\section{Conclusions}                                                           
\label{sec:conclusions} 

We fully reconstructed $791\pm44$ true $\ddb$ pairs 
after all background subtractions.   This is the largest such
sample of charm pairs used in an analysis of the hadroproduction 
of $c\overline c$ to date.
The full reconstruction of the final states of both $D$ mesons
offers several advantages over some of the previous studies that have used 
partially reconstructed $D$ candidates.
We are able to correct the data for both detector inefficiencies and for
the branching fractions of the observed decays so that the 
acceptance-corrected distributions represent the produced mixture of 
$D$ mesons, rather than the detected mixture.
Because the final states are fully reconstructed, we are able to calculate
both the magnitude and direction of the $D$ momenta.
Therefore, we are able to thoroughly investigate
the degree of correlation between both the transverse and longitudinal 
components of the momenta, with respect to the beam direction, 
of the $D$ and $\db$.

We have compared all the measured acceptance-corrected distributions to 
predictions of the fully differential next-to-leading-order
calculation for $\ccb$  production by Mangano, Nason and 
Ridolfi~\cite{ref:bib3,ref:bib4}, as well as to predictions from the 
{\sc Pythia/Jetset} Monte Carlo event generator~\cite{ref:bib5}
for $\ccb$~\cite{ref:bib6} 
and $\ddb$ production~\cite{ref:bib7}.

\subsection{Transverse Correlations}
Our measurements indicate that the transverse momenta of the $D$ and 
$\overline{D}$ in charm-pair events are correlated in several ways.  
(See Secs.~\ref{ssec:twod} and~\ref{sssec:twodd}.)
The square of the amplitudes of the $D$ and $\overline{D}$ 
transverse momenta are slightly correlated (Fig.~\ref{fig:fig15}).
The directions of the $D$ and $\overline{D}$ in the plane 
transverse to the beam axis are significantly correlated (Fig.~\ref{fig:fig11}).
The separation in azimuthal angle, $\dphi$,
is significantly correlated with the sum of the squares of the 
$D$ and $\db$
transverse momenta, $\Sigma p^2_t$ (Fig.~\ref{fig:fig16}).
These features have been observed by several other
experiments~\cite{bib8,bib9,bib10,bib11,bib12,bib13,bib14,bib15,bib16}.
Using the default parameters, the three models yield the same
trend in correlations as we find in data. The models also predict
that the relative transverse angles of the $D$ and $\db$
are more correlated than we find in data (Fig.~\ref{fig:fig22}).
These results provide an opportunity to tune the default parameters,
or add higher order terms
in the models, to obtain better agreement with data~\cite{ref:bib42}.

\subsection{Longitudinal Correlations}
Our measurements indicate that the longitudinal momenta of the $D$ and
$\db$ from charm-pair events are slightly correlated.  The 
measured $\dxf$ and $\dy$ distributions 
(Figs.~\ref{fig:fig8}--\ref{fig:fig9}) are somewhat narrower than 
what one would predict from the observed single-charm predictions
assuming no correlations. 
The $x_{F,D}$ ($y_D$) distribution 
depends on the value of $x_{F,\overline{D}}$ ($y_{\overline{D}}$),
and vice-versa (Figs.~\ref{fig:fig13}--\ref{fig:fig14}).

The single-charm $x_F$ and $y$ distributions from the three theoretical 
models do 
not agree with each other or with the measured distributions
(Fig.~\ref{fig:fig17}).
The three models predict different 
correlations between the charm and anticharm longitudinal 
momenta (Figs.~\ref{fig:fig24}--\ref{fig:fig25}) --- the 
next-to-leading-order 
calculation predicts no significant correlation; the 
{\sc Pythia/Jetset} $\ccb$ prediction indicates a slight positive 
correlation; and
the {\sc Pythia/Jetset} $\ddb$ prediction indicates a strong negative 
correlation.
The $\ddb$ data agree best with the {\sc Pythia/Jetset} $\ccb$ prediction. 
The disagreement between the models and the data might be corrected by
adjusting the non-perturbative parameters in the models, or by adding
higher order terms.

\subsection{Dependence of Yields and Longitudinal Correlations on Type
of $\ddb$ Pair}
The relative yields of the four types of charm pairs,
$\dzdb$, $\dzdm$, $\dpdb$, and $\dpdm$,
as calculated in the {\sc Pythia/Jetset} event generator,
agree
with data as far as their ordering but 
disagree with regard to number of pairs produced, predicting
%%predict 
too many
$\dzdb$
pairs and too few $\dpdm$ pairs. (See Table~\ref{tab:tab3}.)
We studied the degree to which longitudinal correlations depend on the 
type of $\ddb$ pair in data and in the {\sc Pythia/Jetset} event generator.
Although we see differences between data and the event generator
(Fig.~\ref{fig:fig29}), the statistical uncertainties on 
the measured correlations are
too large to make any conclusive statements.

\subsection{Summary and Discussion}

The charm-pair distributions presented in this paper provide an opportunity to
extend our understanding of charm production beyond what was previously
possible with single-charm and lower-statistics or partially reconstructed
charm-pair distributions.
The measured distributions and observed correlations can be compared to
the predictions of models,
testing assumptions in the models and
providing discrimination among
different values for the free parameters in the models.
Some comparisons have been
made in the paper and in the Appendix.

Before comparing the measurements to predictions,
we looked for correlations directly in the charm-pair data by
comparing the observed $\ddb$ pair distributions with
the convolution of the measured
single-charm distributions assuming no correlations.
We find that the charm-pair distributions are quite similar to the
convoluted single-charm distributions, indicating little correlation
between the two charm mesons in an event,
with the exception of the distributions for $\dphi$ and
$\pttddb$.
The $\dphi$ distribution shows clear evidence of correlations,
and the $\pttddb$ distribution is steeper
than the uncorrelated single-charm prediction.
In addition, the data are consistent with possible small correlations
in the $\dxf$ and $\dy$ pair distributions,
which are somewhat more peaked near zero than the single-charm
convolutions.

In the comparisons of the measured and predicted charm-pair distributions,
we observe less correlation between transverse momenta and different
correlations between longitudinal momenta than theoretical
models predict, for the default values of parameters in the models.
Work by other authors suggests a different
set of parameters might provide better agreement~\cite{ref:bib42}.
Both the single-charm and charm-pair distributions agree
best with the predictions for charm quark (rather than $D$ meson)
production, possibly caused by an accidental cancellation of color-dragging
and fragmentation effects.
Also, the $\dphi$ distribution
is more similar to the prediction of the NLO theory at higher
$\sptt$.

In the Appendix, we investigate the sensitivity of single-charm and
charm-pair distributions to various theoretical assumptions.
We conclude that the predictions depend not only on unknown parameters
such as the mass of the charm quark and the intrinsic transverse
momentum of the partons that collide to form the $c\overline c$ pair,
but also on the values of the
renormalization and factorization scales.
The measurements reported here, and the charm-pair measurements from
photoproduction experiments,
should allow the free parameters in the
theoretical models to be further constrained.

\section*{Acknowledgments}

We gratefully acknowledge the assistance of the staffs of Fermilab
and of all the participating institutions. This work was supported
by the Brazilian Conselho Nacional de Desenvolvimento Cient\'{\i}fico
e Tecnol\'{o}gico, CONACyT (Mexico), the Israeli Acadamy of Sciences
and Humanities, the U.~S.~Department of Energy, the U.S.--Israel
Binational Science Foundation and the U.~S.~National Science Foundation.

This work was performed at the Fermi National Accelerator Laboratory,
which is operated by the Universities Research Association, under
contract DE-AC02-76CH03000 with the U.S. Department of Energy.

%%%%%%%%%%%%%%%%%%%%%%%%%%%%%%%%%%%%%%%%%%%%%%%%%%%%%%%%%%%%%%%%%%%%%%%%%
%\input{offline_doc_290_8_r1.tex} %  \section{Appendix A - Theoretical 
%Predictions for Charm-Pair Distributions}

\appendix
\section{Theoretical Predictions for Charm-Pair Distributions}
\label{sec:append}

In Sec.~\ref{sec:theory}, we introduced the 
theoretical framework used to describe the hadroproduction of
$\ddb$ pairs.  In particular, we discussed:
\begin{itemize}
\item the leading-order perturbative QCD description of the hadroproduction 
of $\ccb$ pairs;
\item higher-order perturbative corrections to the leading-order calculation;
\item the addition of intrinsic transverse momentum to 
the hard-scattering partons that collide to form the $\ccb$ pair; and, lastly,
\item the hadronization of $\ccb$ pairs to observable $\ddb$ pairs.
\end{itemize}
Using this framework, we investigate how sensitive single-charm and
charm-pair distributions are to various theoretical assumptions.
All predictions discussed in this Appendix are for
a 500 GeV/$c$ $\pi^-$ beam  incident on a nuclear target --- the same
beam-target as the data from experiment E791 reported in this paper.

The {\sc Pythia/Jetset} event generator depends on many parameters.
Unless otherwise mentioned, we use the default settings for all 
parameters.
The next-to-leading order perturbative QCD calculation depends on the
following six parameters:
\begin{itemize}
\item the mass of the charm quark, $m_c$, 
\item the beam and target parton distribution functions, $f^{\pi}$ and
$f^N$, respectively,
\item $\Lambda_{QCD}$, the free parameter that must be 
determined experimentally, which roughly defines the mass scale below which 
quarks and gluons do not behave as independent, free partons ---
that is, below which perturbative QCD calculations are no longer valid, and 
\item the renormalization and factorization scales, $\mu_R$ and $\mu_F$.
\end{itemize}
The pairs of pion and nucleon parton distribution 
functions considered in this section, obtained
from the CERN computer library package PDFLIB~\cite{ref:bib43}, are 
listed in Table~\ref{tab:tab4}.  
Parton 
distribution functions depend on 
the fraction $x$ of the hadron momentum carried
by the hard-scattering parton, on both the
factorization and renormalization scales, and on 
$\Lambda_{QCD}$.  
In the parton distribution functions accessible from PDFLIB,
the renormalization scale is defined 
to be the same as the factorization scale.

\begin{table} [htbp]
        \caption{The pairs of pion and nucleon parton distribution
        functions considered in this Appendix, obtained from the
        CERN FORTRAN package PDFLIB.  The functions have been extracted from
        fits to data assuming a fixed value of $\Lambda_{QCD}$. The
        functions are undefined below the minimum scale $\mu_0$}.
        \label{tab:tab4}
\begin{tabular} {lclcccc} \hline \hline
Set & & Name & $\mu^2_0 [{\rm GeV^2}]$  & $\Lambda^{(4)}_{\rm{QCD}}$ [MeV] &
Order &Ref. \\  \\ \hline \hline
(1)\footnote{HVQMNR suggested default.} & $f^{\pi}$& SMRS-P2 & 5
   & 190  & NLO                   & \cite{ref:bib44} \\
    & $f^N$    & HMRS-B (4.90) & 5 & 190 & NLO & \cite{ref:bib45} \\  \\
(2) & $f^{\pi}$& GRV-P & 0.3 & 200 & NLO & \cite{ref:bib46} \\
    & $f^N$ & GRV & 0.3 & 200 & NLO &\cite{ref:bib47} \\  \\
(3) & $f^{\pi}$ & SMRS-P2 & 5      & 190 & NLO                     &
\cite{ref:bib44} \\
    & $f^N$ & HMRS-B (8.90) & 5 & 100 & NLO & \cite{ref:bib48} \\  \\
(4) & $f^{\pi}$ & ABFKW-P3 & 2 & 281 & NLO & \cite{ref:bib49} \\
    & $f^N$ & HMRS-B (8.90) & 5 & 300 & NLO & \cite{ref:bib48} \\  \\
(5)\footnote{{\sc Pythia/Jetset} default.} & $f^{\pi}$ & OW-P1 &
 4 & 200 & LO & \cite{ref:bib50} \\
    & $f^N$ & CTEQ 2L & 4 & 190 & LO & \cite{ref:bib51} \\ \hline
\end{tabular}
\end{table}

For each parton distribution function listed in Table~\ref{tab:tab4}, 
we specify the square of 
the minimum factorization scale allowed, $\mu^2_0$; 
whether the evolution equations were calculated to leading-order
(LO) or to next-to-leading order (NLO)
and the value of
$\Lambda^{(4)}_{QCD}$ used in the fit.
Querying PDFLIB for the value of a parton distribution
function at a scale below $\mu_0$ gives undefined results.
The default set of parton distribution functions 
for the {\sc Pythia/Jetset} event generator is set (5) 
in Table~\ref{tab:tab4}; the
default suggested by the authors of HVQMNR is set (1).

When possible, we choose pion and nucleon distribution functions 
that are fit assuming similar values for 
$\Lambda_{QCD}$.  
For all predictions shown below, the $\Lambda_{QCD}$ used in the 
next-to-leading order 
calculation of the partonic cross section is defined to be 
the same as the $\Lambda_{QCD}$ used to extract the
nucleon parton distribution function $f^N$.

The degree to which the charm-pair distributions are sensitive to
variations in $\mu_R$ and $\mu_F$ gives an
indication of how important higher-order corrections are; that
is, an indication of how much (or little) we can trust the $\alpha_s^3$
calculation.
In general, one tries to minimize higher-order 
contributions by choosing $\mu_R$ and $\mu_F$ 
to be of the same order
as the energy scale $Q$ of the hard-scattering process.
However, this scale cannot be defined unambiguously.
One reasonable choice is
\begin{equation}
\label{eq:q2}
Q \equiv \sqrt{m_c^2 + \frac{p^2_{t,c} + p^2_{t,\overline{c}}}{2}}.
\end{equation}
The default setting for the {\sc Pythia/Jetset} event generator is 
$\mu_R = \mu_F = Q$, leading to factorization scales as low as the
mass of the charm quark, $m_c$, which by default is set to 1.35 GeV.
The parton distribution functions used by the {\sc Pythia/Jetset} event
generator, however, are only defined for scales above 2 GeV. 
This
problem is handled by setting the parton distribution function to 
$f(x,\mu_0)$ for all factorization scales less than $\mu_0$.
%%(Fig.~\ref{fig:fig32} shows that the distributions have a
%%weak dependence on the parton distribution
%%functions, even when the minimum scale is below 2 GeV.)

The suggested default for the HVQMNR program is $\mu_R = Q$ and 
$\mu_F = 2 Q$.  Given their suggested default for the mass of the 
charm quark of $m_c = 1.5$ GeV, 
this choice ensures that the factorization scale will never go below
the minimum allowed scale, $\mu_0 = \sqrt{5}$ GeV.  

In Figs.~\ref{fig:fig30}--\ref{fig:fig35} we
show single-charm and charm-pair distributions for a wide range 
of theoretical assumptions.  
When obtaining these theoretical predictions, we only allow charm-pair
events in which both charm rapidities are greater than $-0.5$ and less than
$2.5$.  For the HVQMNR generator, which
does not hadronize the $\ccb$ pair to charmed mesons, the cut is on
the charm quark rapidities.  For the {\sc Pythia/Jetset} generator, the cut
is on the $D$ meson rapidities.
In Table~\ref{tab:tab5}, 
we show which generator (HVQMNR or {\sc Pythia/Jetset})
and what theoretical assumptions are used in each figure. 

The same set of single-charm and charm-pair distributions are shown 
in each figure.
Each charm particle in a charm-pair event 
can be described using three variables.
A common choice of
independent variables for single-charm analyses is 
$\xf$, $\ptt$, and $\phi$.

We ignore the latter variable because all theoretical predictions
give a flat $\phi$ distribution.
Although $x_F$ and $y$ are
very correlated, 
we show predictions for both distributions.  
For each single-charm variable $v$, 
we obtain predictions for two charm-pair 
distributions: $\Delta v = v_c-v_{\overline{c}}$ and 
$\Sigma v = v_c+v_{\overline{c}}$.  ($\Delta \phi$ is defined to 
be the minimum of $|\phi_c-\phi_{\overline{c}}|$ and
$360^{\circ}-|\phi_c-\phi_{\overline{c}}|$.)
As with the single-charm $\phi$ variable, we ignore the 
charm-pair $\Sigma \phi$ variable because all theoretical predictions
give a flat $\Sigma \phi$ distribution.  We do not, however, ignore the 
$\Delta \phi$ distribution which is very sensitive to theoretical 
assumptions.  
Two other commonly used charm-pair distributions that we examine are
the square of the transverse momentum of the charm-pair,
$p^2_{t,c\overline{c}} = |\vec{p}_{t,c} + \vec{p}_{t,\overline{c}}|^2$,
and the invariant mass of the charm-pair, $M_{c\overline{c}}$. 

The vertical axis of each distribution shown in 
Figs.~\ref{fig:fig30}--\ref{fig:fig35} gives the fraction of single-charm
(charm-pair) events per variable $v$ interval, 
$\frac{1}{N}\frac{{\rm d}N}{{\rm d}v}$, where $N$ is the total
number of single-charm (charm-pair) events generated.  The number
of single-charm events generated is, of course, just twice the 
number of charm-pair events generated.

\subsection{Sensitivity to Higher-Order Perturbative Corrections}
\begin{figure} %Figure 30
        \centering
        \centerline{\epsfig{file=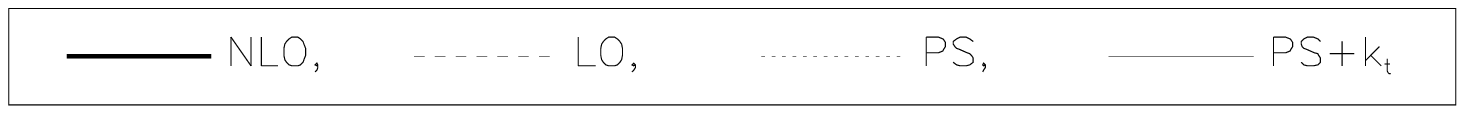,width=4.0in,angle=0}}
        \centerline{\epsfig{file=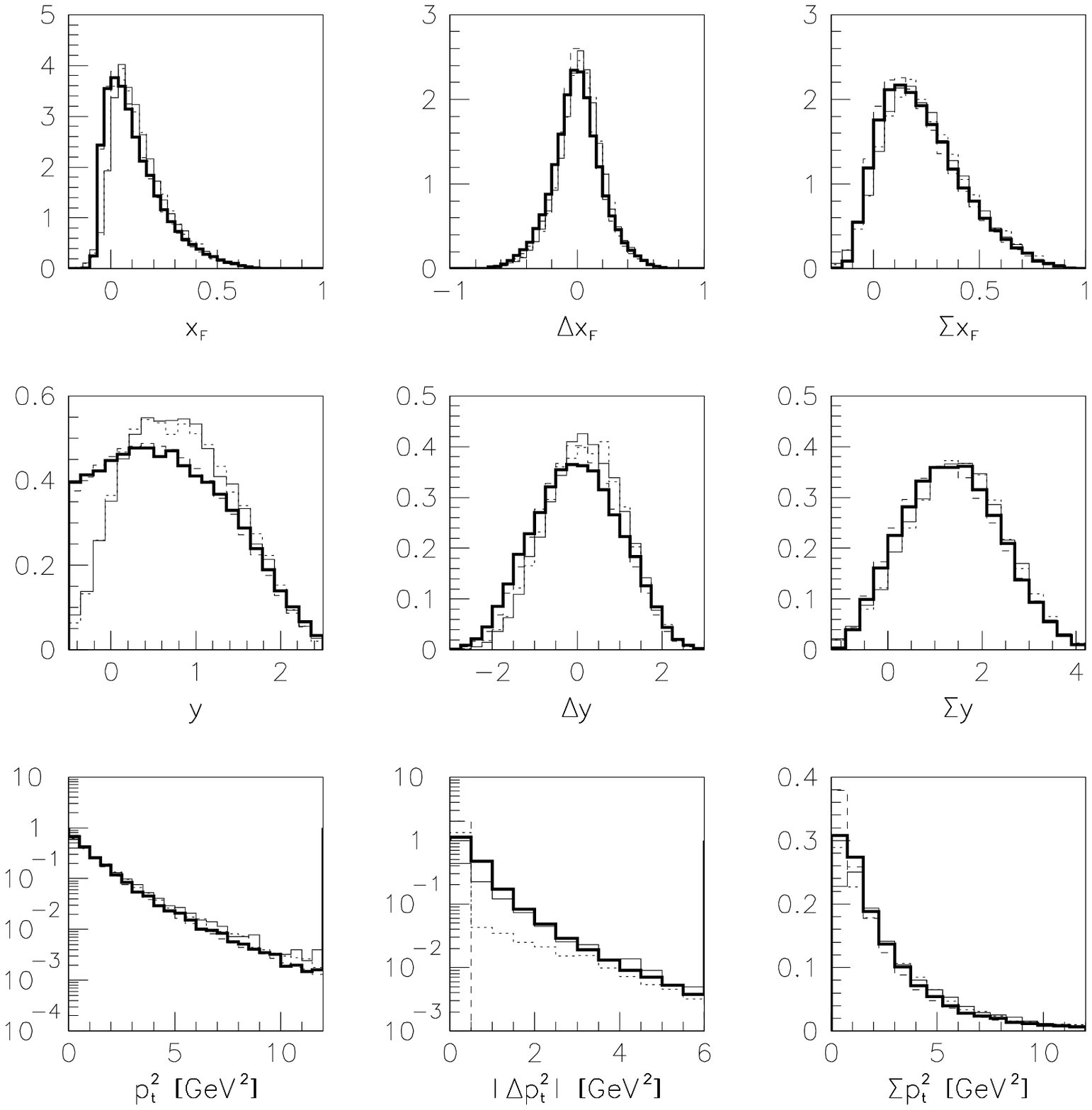,width=4.0in,angle=0}}
        \centerline{\epsfig{file=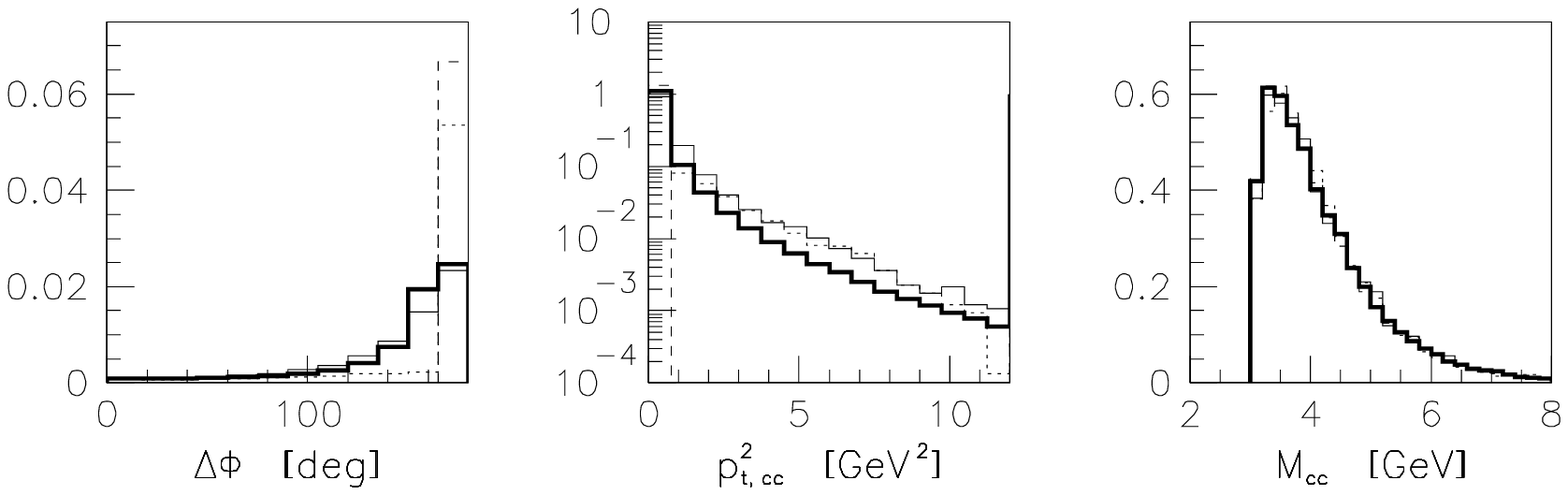,width=4.0in,angle=0}}
        \caption{Sensitivity of single-charm and charm-pair distributions
        to higher-order perturbative corrections.  The LO (dashed)
        and NLO ({\bf solid}) distributions are obtained
        from the HVQMNR generator; the parton-shower distributions, with (solid)
        and without (dotted) intrinsic transverse momentum, from the
        {\sc Pythia/Jetset} generator.
        See Table~\protect\ref{tab:tab5}.}\label{fig:fig30}
\end{figure}

In Fig.~\ref{fig:fig30}, we compare the 
complementary methods used by the HVQMNR program and the {\sc Pythia/Jetset} 
event generator to include higher-order perturbative
corrections to the leading-order partonic cross section.
As discussed in the previous section, the {\sc Pythia/Jetset} 
event generator, beginning
with leading-order matrix elements, uses parton showers to include
higher-order perturbative effects, whereas the HVQMNR program calculates 
the next-to-leading order $\ccb$ cross section.  
To more directly compare these two approaches,
we change three of the default {\sc Pythia/Jetset} settings --- 
$m_c$, $f^{\pi}$ and $f^N$ --- to match the default HVQMNR settings. (See
Table~\ref{tab:tab5}.)  We obtain {\sc Pythia/Jetset} $\ccb$ distributions
assuming no intrinsic transverse momentum for the interacting partons,
as well as assuming 
$\sigma_{k_t} = 0.44$ GeV which is the {\sc Pythia/Jetset} default.
As argued by T.~Sj\" ostrand, the intrinsic transverse momentum may, 
at least in part, be seen as a replacement for
gluon emission that is truncated in the parton shower approach due
to the introduction of an energy scale below which the parton shower 
evolution is stopped\cite{ref:bib52}.  In Fig.~\ref{fig:fig30}, 
we also show the HVQMNR leading-order distributions
to emphasize which distributions are, and which are not, 
sensitive to higher-order corrections.

\begin{table}[htbp]
        \caption{The settings used by the HVQMNR and {\sc Pythia/Jetset}
        generators to obtain the single-charm and charm-pair distributions
        shown in Figs.~\protect\ref{fig:fig30}-\protect\ref{fig:fig35}.
        The set of pion and nucleon parton distribution
        functions (PDF), labeled (1) through (5),
        are defined in
        Table~\protect\ref{tab:tab4}.
        A ``Y'' indicates
        that parton showers (PS) are included in the {\sc Pythia/Jetset} event
        generator;
        an ``N'' indicates that they are not included.
        The last column describes the histogram style corresponding
        to the settings in that row, in the figure listed in
        the second to last column.
         }
        \label{tab:tab5}
\begin{minipage}{\linewidth}
\renewcommand{\thefootnote}{\themfootnote}
\begin{tabular} {llcccccclc}  \hline \hline
          &        &         &           &           &$m_c$  &$\sigma_{k_t}$&                  &             \\
Generator &        & PDF     & $\mu_R/Q$ & $\mu_F/Q$ & [GeV] &[GeV] & Fig./PS                  &             \\ \\ \hline \hline
MNR       & NLO \footnote{NLO refers to the default next-to-leading order HVQMNR distributions.} 
                   &(1)      & 1.0       & 2.0       & 1.5   &0     & \protect\ref{fig:fig30}  & {\bf solid} \\
MNR       & LO \footnote{LO refers to the default leading-order HVQMNR distributions.} 
                   & (1)     & 1.0       &2.0        & 1.5   &0     &                          & dashed      \\
P/J       & $\ccb$ \footnote{$\ccb$ refers to the default {\sc Pythia/Jetset} $\ccb$ distributions.}
                   & (1)     & 1.0       & 1.0       & 1.5   &0     &Y                         & dotted      \\
P/J       & $\ccb$ & (1)     & 1.0       & 1.0       & 1.5   &0.44  &Y                         & solid       \\ \\
MNR       & NLO    & (1)     & 1.0       & 2.0       & 1.5   &0     &  \protect\ref{fig:fig31} & solid       \\
MNR       & NLO    & (1)     & 1.0       & 2.0       & 1.2   &0     &                          & dashed      \\
MNR       & NLO    & (1)     & 1.0       & 2.0       & 1.8   &0     &                          & dotted      \\ \\
MNR       & NLO    & (1)     & 1.0       & 2.0       & 1.5   &0     &\protect\ref{fig:fig32}   &{\bf solid}  \\
MNR       & NLO    & (2)     & 1.0       & 2.0       & 1.5   &0     &                          & dashed      \\
MNR       & NLO    & (3)     & 1.0       & 2.0       & 1.5   &0     &                          & dotted      \\
MNR       & NLO    & (4)     & 1.0       & 2.0       & 1.5   &0     &                          & solid       \\ \\
MNR       & NLO    & (2)     & 0.5       & 0.5       & 1.5   &0     &\protect\ref{fig:fig33}   & solid       \\
MNR       & NLO    & (2)     & 1.0       & 1.0       & 1.5   &0     &                          & dashed      \\
MNR       & NLO    & (2)     & 1.5       & 1.5       & 1.5   &0     &                          & dotted      \\ \\
P/J       & $\ddb$ \footnote{$\ddb$ refers to the default {\sc Pythia/Jetset} $\ddb$ distributions.} 
                   &(5)      & 1.0       & 1.0       & 1.35  &0.44  &\protect\ref{fig:fig34}/Y &{\bf solid}  \\
P/J       & $\ccb$ & (5)     & 1.0       & 1.0       & 1.35  &0     &Y                         & dashed      \\
P/J       & $\ccb$ & (5)     & 1.0       & 1.0       & 1.35  &0.44  &N                         & dotted      \\
P/J       & $\ddb$ & (5)     & 1.0       & 1.0       & 1.35  &0     &N                         & solid       \\ \\
P/J       & $\ddb$ & (5)     & 1.0       & 1.0       & 1.35  &0.44  &\protect\ref{fig:fig35}/Y &{\bf solid}  \\
P/J       & $\ddb$ & (5)     & 1.0       & 1.0       & 1.35  &0.7   &Y                         & dashed      \\
P/J       & $\ddb$ & (5)     & 1.0       & 1.0       & 1.35  &1.0   &Y                         & dotted      \\
P/J       & $\ddb$ & (5)     & 1.0       & 1.0       & 1.35  &1.5   &Y                         & solid       \\ \hline
\end{tabular}
\end{minipage}
\end{table}

Figure~\ref{fig:fig30} shows that 
higher-order perturbative corrections do not significantly 
affect the shapes of most of the single-charm and charm-pair 
distributions.  That is, the HVQMNR leading-order and next-to-leading
predictions for all distributions are very similar --- 
except for the $|\Delta p^2_t|$, 
$\Delta \phi$, and $p^2_{t,c\overline{c}}$ distributions.  
In the leading-order calculation,
these latter distributions are delta functions  ---
at 0 (GeV/$c)^2$, $180^\circ$, and 0 (GeV/$c)^2$, 
respectively --- because
the leading-order charm and anticharm quark are 
back-to-back in the plane transverse to the beam axis.

The next-to-leading order predictions and the parton shower 
prediction are also quite similar.
The $|\Delta p^2_t|$ and $\Delta \phi$ {\sc Pythia/Jetset} 
distributions with no intrinsic transverse momentum included,
indicate that the parton shower evolution is playing a very small role.
The $\Delta \phi$ parton shower distribution, in particular, is closer to the
leading-order delta-function prediction than to the next-to-leading 
order prediction.  Adding intrinsic transverse momentum, with 
$\sigma_{k_t} = 0.44$ GeV, brings the {\sc Pythia/Jetset} prediction 
very close to the next-to-leading order HVQMNR prediction.  
  
\subsection{Sensitivity to the Mass of the Charm Quark}
\begin{figure} %Figure 31
        \centering
        \centerline{\epsfig{file=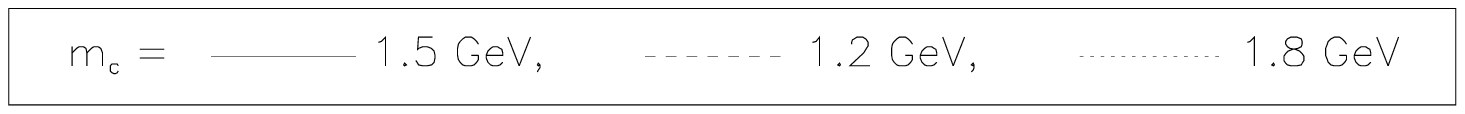,width=4.0in,angle=0}}
        \centerline{\epsfig{file=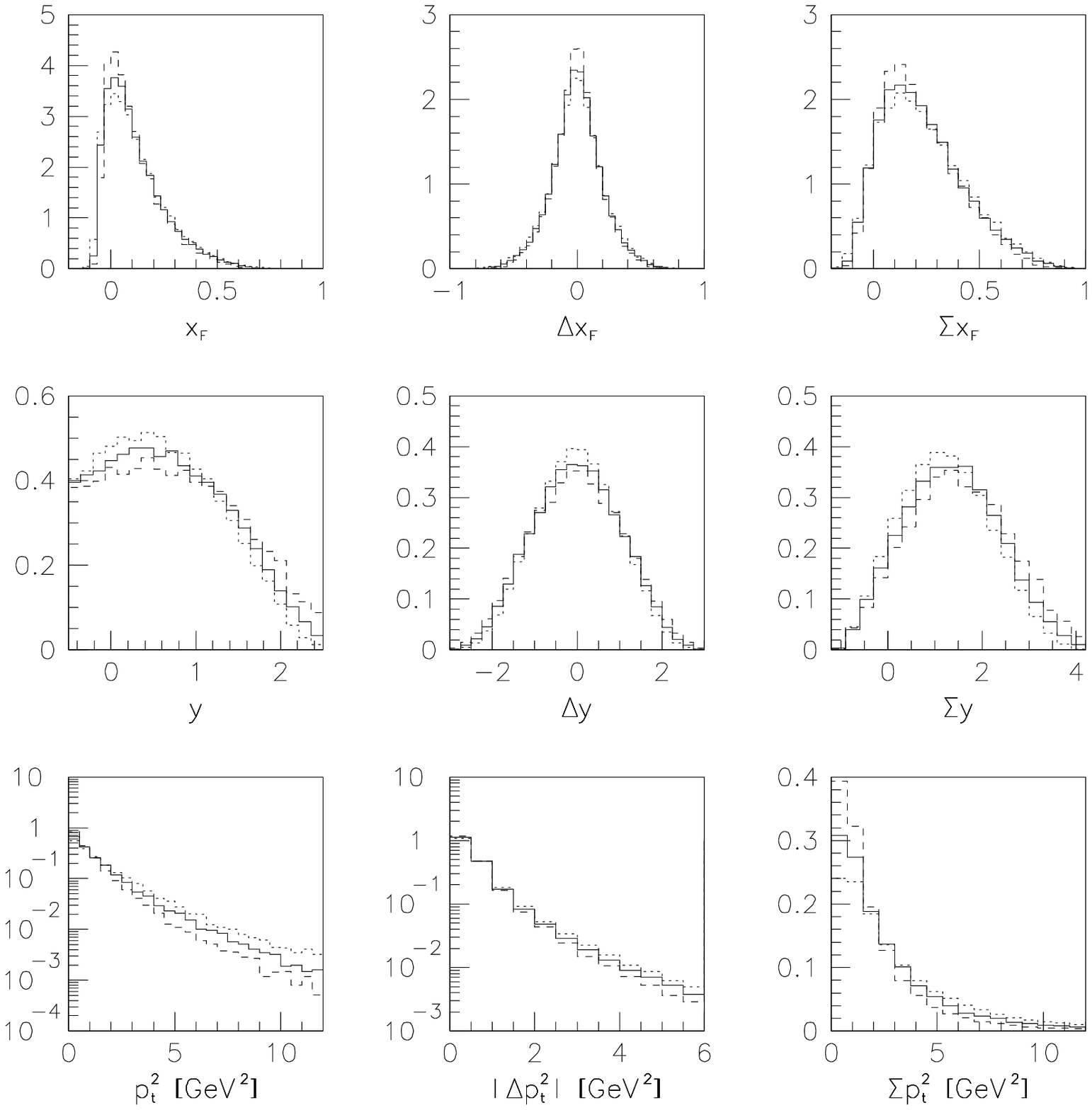,width=4.0in,angle=0}}
        \centerline{\epsfig{file=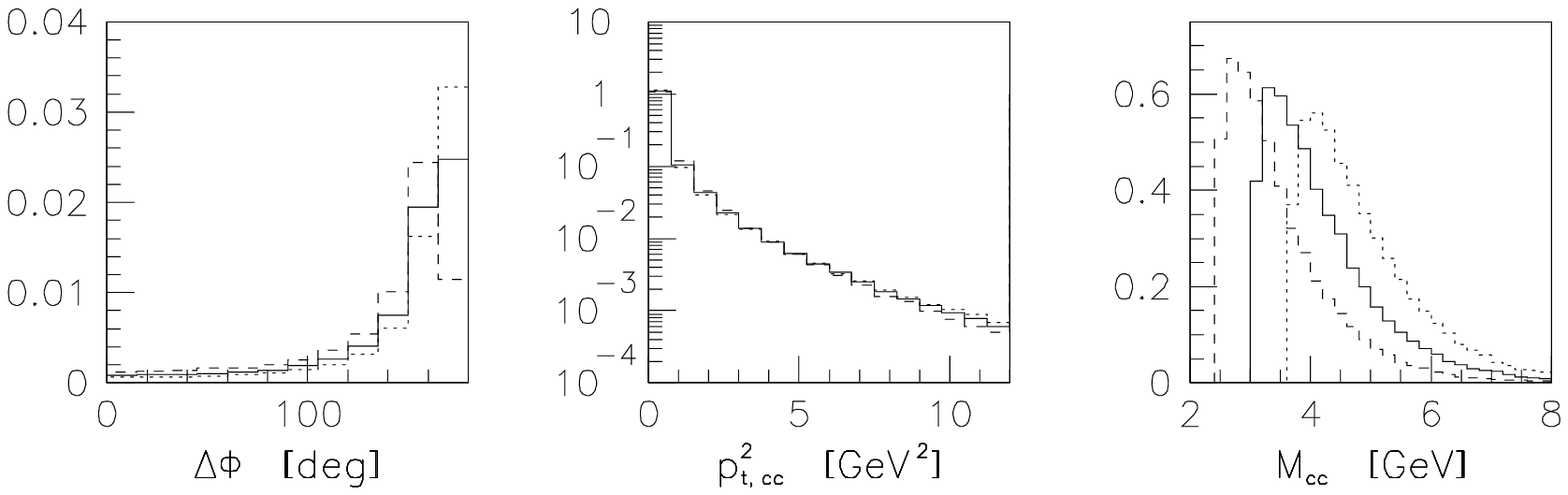,width=4.0in,angle=0}}
        \caption{Sensitivity of single-charm and charm-pair distributions
        to variations in the mass of the charm quark.  All distributions are
        obtained from HVQMNR NLO calculations with
        the default values for all parameters except $m_c$.  The values of
        $m_c$ used are 1.5 GeV (solid), 1.2 GeV (dashed) and 1.8 GeV (dotted).
        See Table~\protect\ref{tab:tab5}.}\label{fig:fig31}
\end{figure}

In Fig.~\ref{fig:fig31}, we investigate 
the degree to which the
single-charm and charm-pair distributions are sensitive to
variations in
the mass of the charm quark.  All distributions are
obtained from HVQMNR NLO 
calculations using the default values
for all parameters --- except for $m_c$.  
Higher-order effects play a larger role as the charm-quark mass decreases
because
the ratio $Q/\Lambda_{QCD}$ decreases, where $Q$ is the energy scale of
the interaction (Eq.~\ref{eq:q2}).
For the lightest charm-quark mass
($m_c = 1.2$ GeV/$c^{2}$), the single-charm $x_F$ and $p^2_t$ distributions 
are steepest because the outgoing charm quark can more easily 
radiate gluons; the single-charm $y$ distribution is less peaked near $y=0$;
and the invariant mass of the charm-pair is significantly steeper than the
higher mass predictions.  The increase in higher-order effects for 
smaller $m_c$ is also evident in the $\Delta \phi$ distribution, which
is flattest for $m_c = 1.2$ GeV/$c^{2}$.

\subsection{Sensitivity to Parton Distribution Functions}

\begin{figure} %Figure 32
        \centering
        \centerline{\epsfig{file=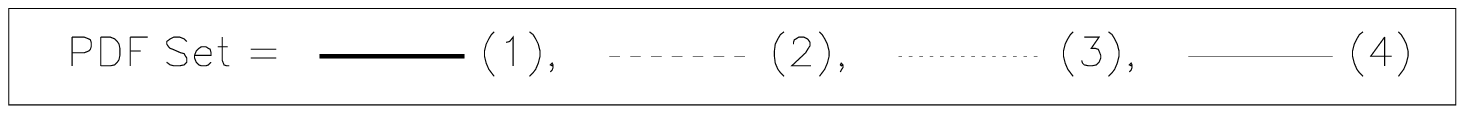,width=4.0in,angle=0}}
        \centerline{\epsfig{file=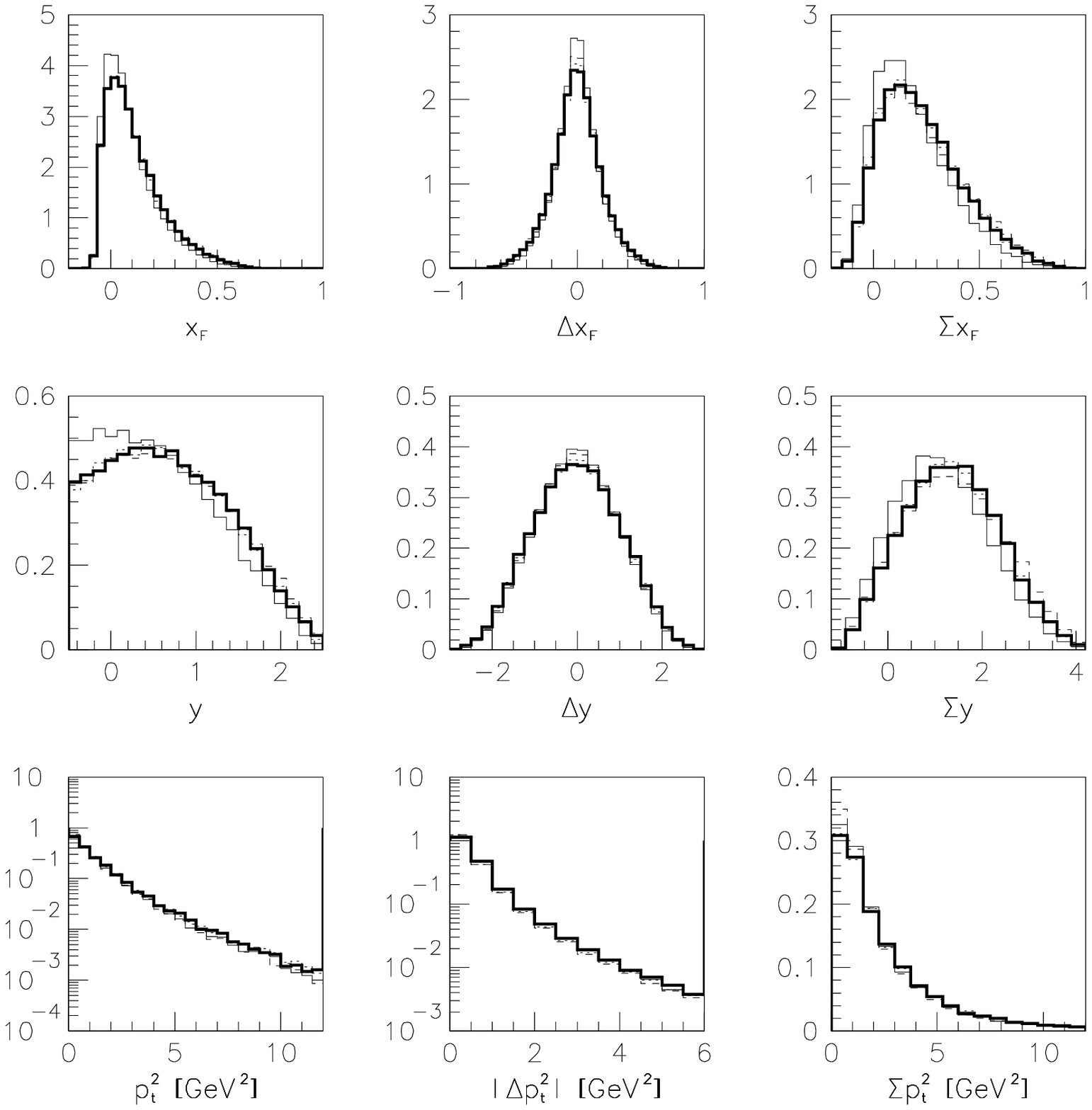,width=4.0in,angle=0}}
        \centerline{\epsfig{file=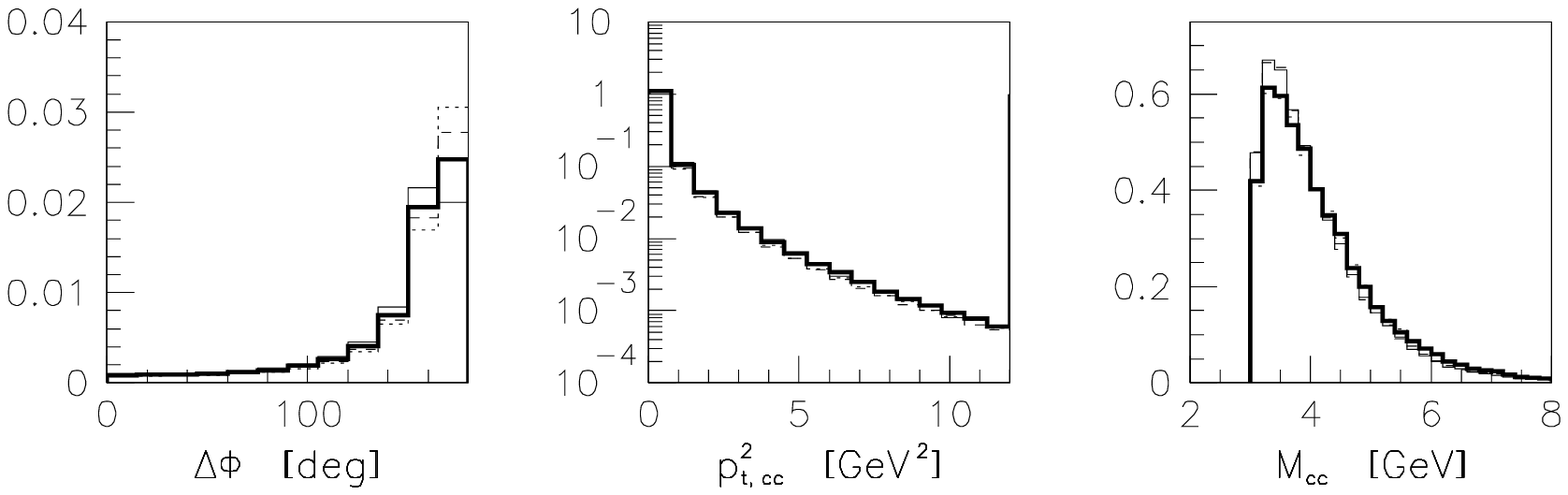,width=4.0in,angle=0}}
        \caption{Sensitivity of single-charm and charm-pair distributions
        to variations in the parton distribution functions.  Sets (1)
        ({\bf solid}),
        (2) (dashed), (3) (dotted) and (4) (solid) are defined in
        Table~\protect\ref{tab:tab4}.  All distributions are
        obtained from HVQMNR NLO
        calculations with the default values
        for all parameters except the parton distribution
        functions. See Table~\protect\ref{tab:tab5}.}
        \label{fig:fig32}
\end{figure}

In Fig.~\ref{fig:fig32}, we investigate 
the degree to which the
single-charm and charm-pair distributions are sensitive to
variations in
the parton distribution functions and $\Lambda_{QCD}$.  
All distributions are
obtained from HVQMNR NLO
calculations using the default values
for all parameters --- except for the parton distribution functions.
We examine predictions for four pairs of pion and nucleon 
parton distribution functions, sets (1) through (4) defined 
in Table~\ref{tab:tab4}.

\begin{figure} %Figure 36
        \centering
        \centerline{\epsfig{file=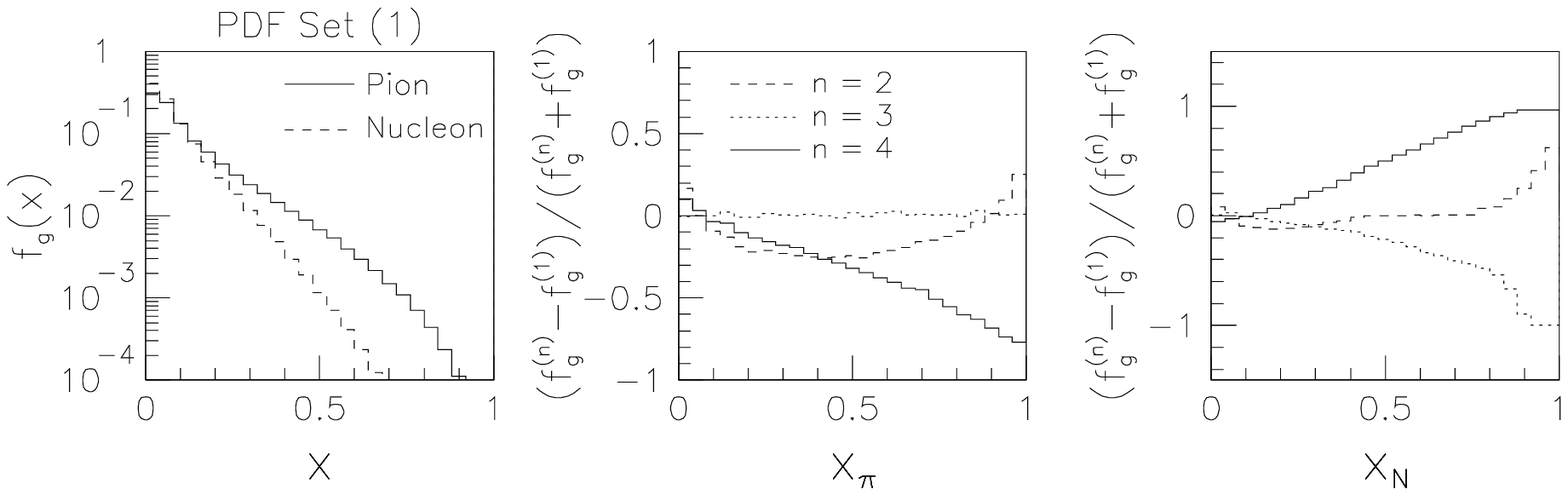,width=4.0in,angle=0}}
        \caption{Comparison of the gluon distribution functions
        for sets (1) through (4), defined in Table~\protect\ref{tab:tab4}.
        The pion and nucleon functions for each set are obtained from PDFLIB
        after imposing the constraint
        $\protect\sqrt{x_{\pi} x_N} \geq \frac{2 m_c}{\sqrt{s}}$, where
        $\sqrt{s}$
        is the center-of-mass energy of the colliding hadrons.  The left plot
        shows the pion and nucleon gluon distribution functions for set (1).
        The middle and right plots show, for the pion and nucleon, respectively,
        the asymmetries between the set (1) gluon distribution function and each
        of the other three (n=2-4) gluon distribution functions.}
        \label{fig:fig36}
\end{figure}

At fixed-target
energies, the dominant contribution to the $\ccb$ cross section
is from gluon fusion.  In Fig.~\ref{fig:fig36}, we compare the 
gluon distribution functions $f_g$ for sets (1) through (4).  
By energy conservation, the energy of the two colliding partons must
be at least twice the mass of a charm quark to produce a $\ccb$ \space pair;
that is, $\sqrt{x_{\pi} x_N} \geq \frac{2 m_c}{\sqrt{s}}$
where $\sqrt{s}=30.6$ GeV is the center-of-mass energy of the colliding hadrons.
Hence, for each set, the pion and nucleon functions are obtained 
after imposing the constraint $x_{\pi} x_N \geq \frac{4 m^2_c}{s}$.  
We impose this constraint because we want to investigate how the 
four sets compare in the region of $x$ that we explore, not in the 
very low $x$ region where the functions are largest.

Although the four sets of parton distribution functions 
differ significantly, the single-charm and 
charm-pair distributions shown in Fig.~\ref{fig:fig32} 
are not very sensitive to these differences.
The sensitivity of the $\Delta \phi$ distribution is due to the 
variation in $\Lambda_{QCD}$ in sets (1) through (4) 
(see Table~\ref{tab:tab4}).
As the value of $\Lambda_{QCD}$ increases, the ratio $Q/\Lambda_{QCD}$ 
decreases,
where $Q$ is the energy scale of
the interaction (Eq.~\ref{eq:q2}),
causing higher-order effects to play a larger role.
Hence, the flattest $\dphi$ distribution results from using 
Set (4) ($\Lambda^{(4)}_{QCD} = 300$ GeV); the steepest $\dphi$ 
distribution, from using set (3) ($\Lambda^{(4)}_{QCD} = 100$ GeV).

\subsection{Sensitivity to Factorization and Renormalization Scales}
\begin{figure} %Figure 33
        \centering
        \centerline{\epsfig{file=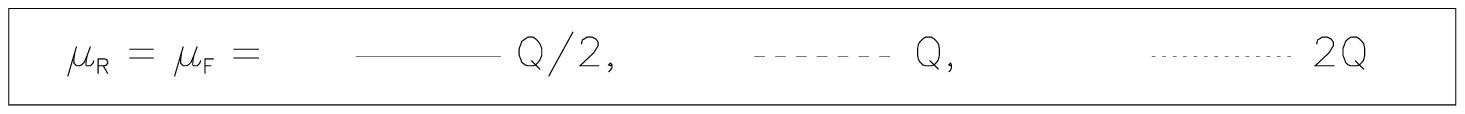,width=4.0in,angle=0}}
        \centerline{\epsfig{file=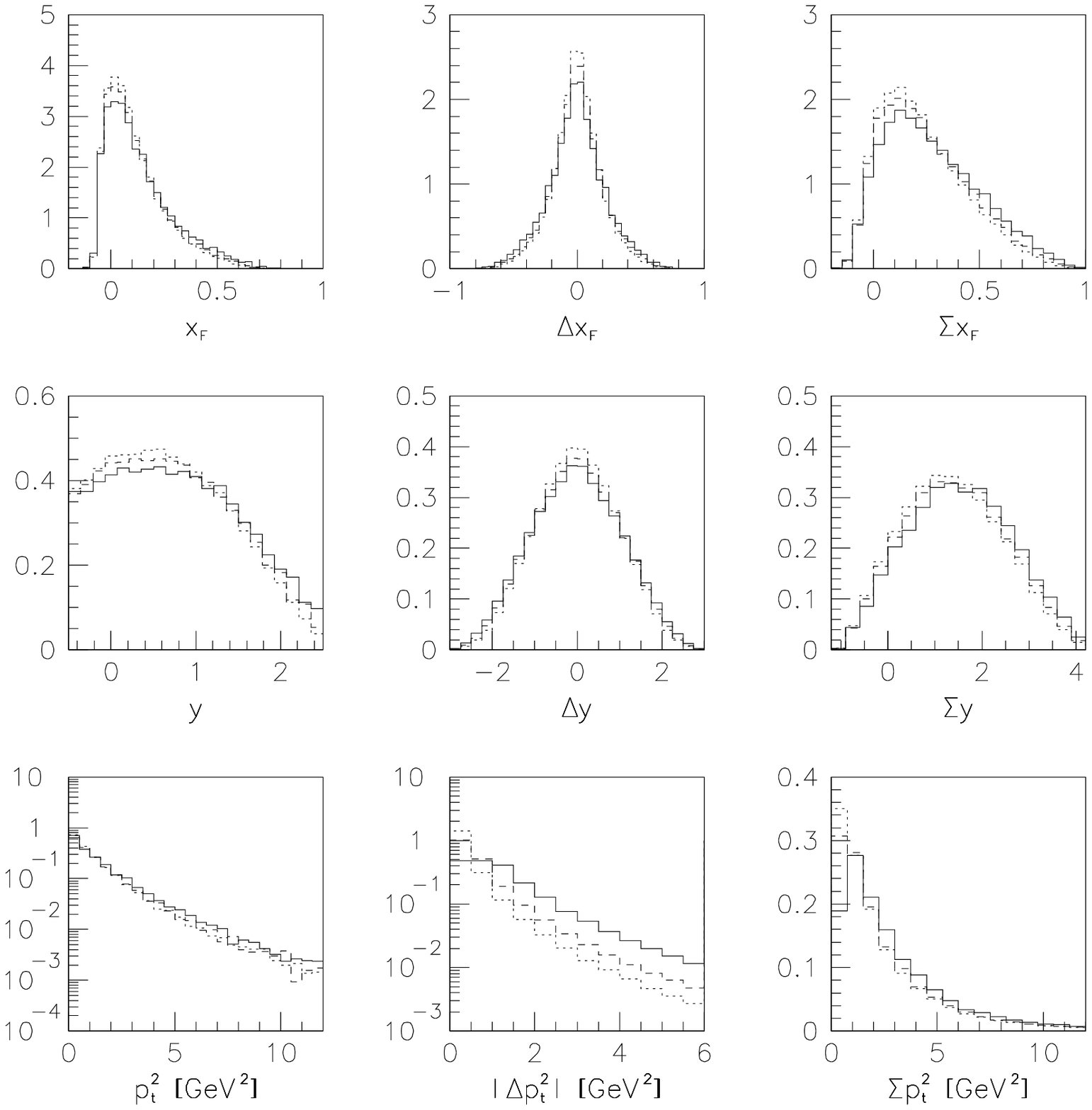,width=4.0in,angle=0}}
        \centerline{\epsfig{file=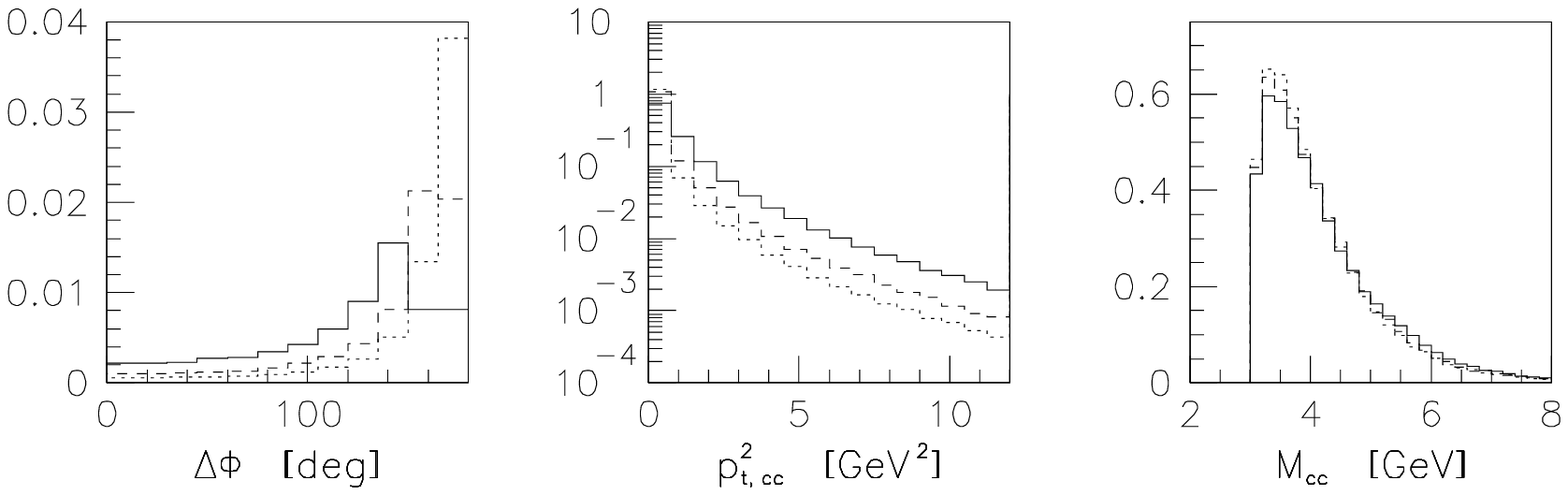,width=4.0in,angle=0}}
        \caption{Sensitivity of single-charm and charm-pair distributions
        to variations in the factorization and renormalization scales.
        All distributions are obtained from HVQMNR NLO
        calculations.
        Rather than using the default set of parton distribution functions,
        we use the GRV functions,
        which are evolved down to $\mu_0^2$ = 0.3 ${\rm GeV^2}$.
        The values of $\mu_0/Q$ used are 0.5 (solid),
        1.0 (dashed) and 1.5 (dotted). The energy scale $Q$ is defined in
        Equation~\protect\ref{eq:q2}.
        See Tables~\protect\ref{tab:tab4} and~\protect\ref{tab:tab5}.}
        \label{fig:fig33}
\end{figure}

In Fig.~\ref{fig:fig33}, we investigate 
the degree to which the
single-charm and charm-pair distributions are sensitive to
variations in
the renormalization and factorization scales.
All distributions are
obtained using the HVQMNR NLO calculation.
We set the two arbitrary scales equal to each
other, $\mu \equiv \mu_F = \mu_R$, and obtain distributions
for $\mu = Q/2$, $Q$, and $2Q$, where $Q$ gives the energy scale of
the interaction (Eq.~\ref{eq:q2}).
We use 
the GRV parton distribution functions for both the pion and the 
nucleon (set (2) in Table~\ref{tab:tab4}),
which have been evolved down
to $\mu_0^2$ = 0.3 ${\rm GeV^2}$.  With this choice, the factorization
scale $\mu$ can go 
as low as $m_c/2$ without going below $\mu_0$.
As mentioned, the degree to which the distributions are sensitive to
variations in the renormalization and factorization scales
gives an indication of how much (or little) we can trust the $\alpha_s^3$
calculation.
As expected, the distributions that are most sensitive to variations
in $\mu$ are those distributions that are trivial at leading-order:
$|\Delta p^2_t|$, $\Delta \phi$, and $p^2_{t,c\overline{c}}$.  The 
smaller the factorization and renormalization scales, the broader
these distributions are.  That is, the higher-order $\alpha^3$ 
terms play a larger role, compared to the leading-order $\alpha^2$ terms,
as renormalization and factorization scales decrease.
The sensitivity of these scales indicates that the model may
have several ways of obtaining accurate predictions, both
by adjusting the model's parameters, and by adding higher order terms.

\subsection{Sensitivity to Higher-Order Nonperturbative Effects}
\label{ssec:npeffects}
\begin{figure} %Figure 34
        \centering
        \centerline{\epsfig{file=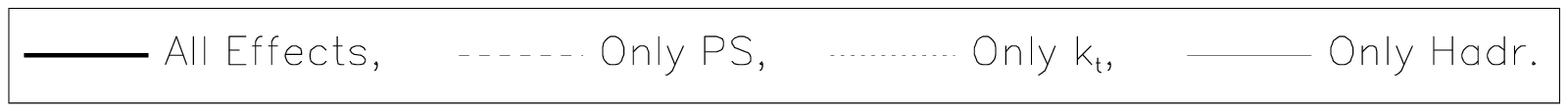,width=4.0in,angle=0}}
        \centerline{\epsfig{file=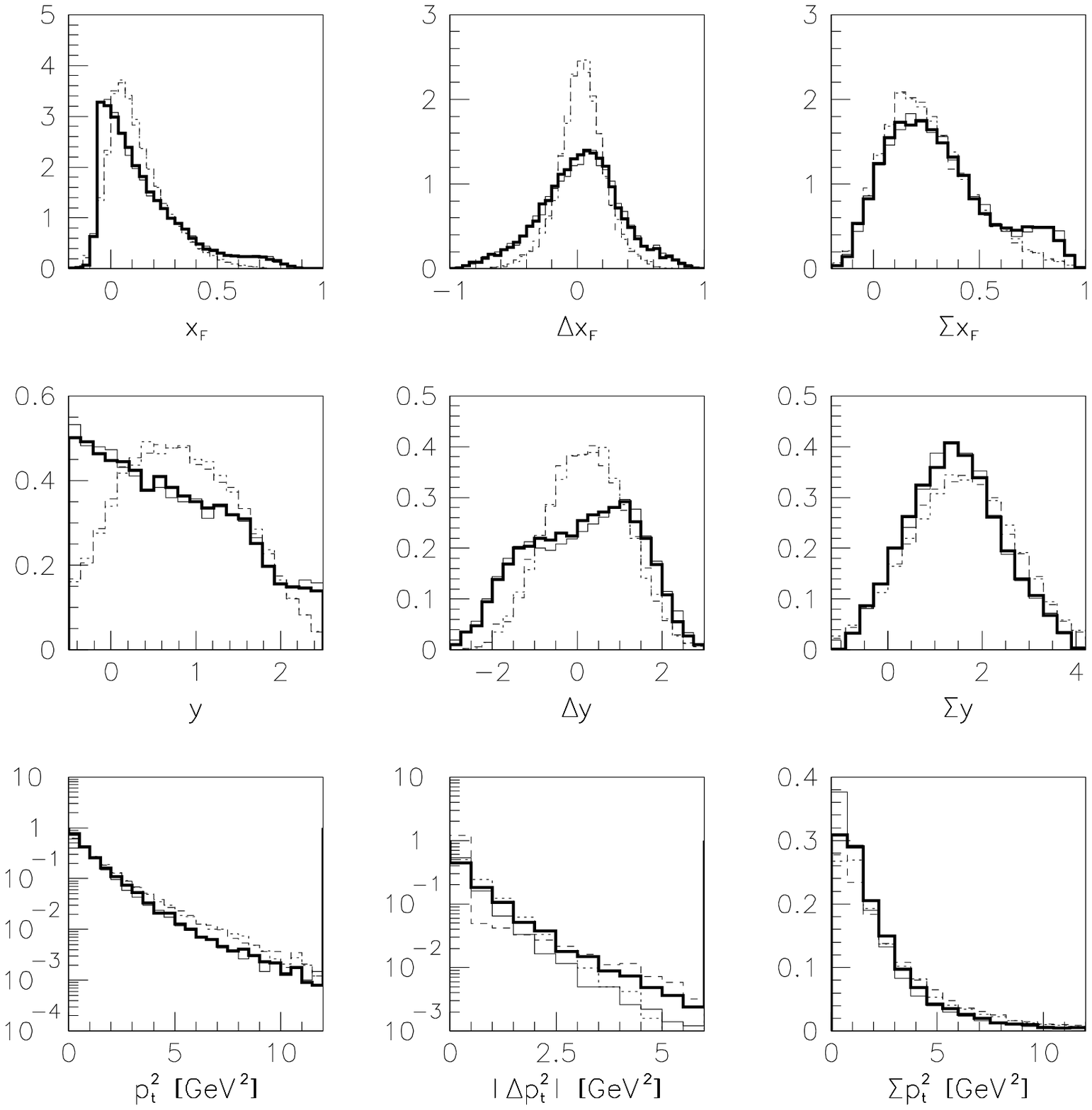,width=4.0in,angle=0}}
        \centerline{\epsfig{file=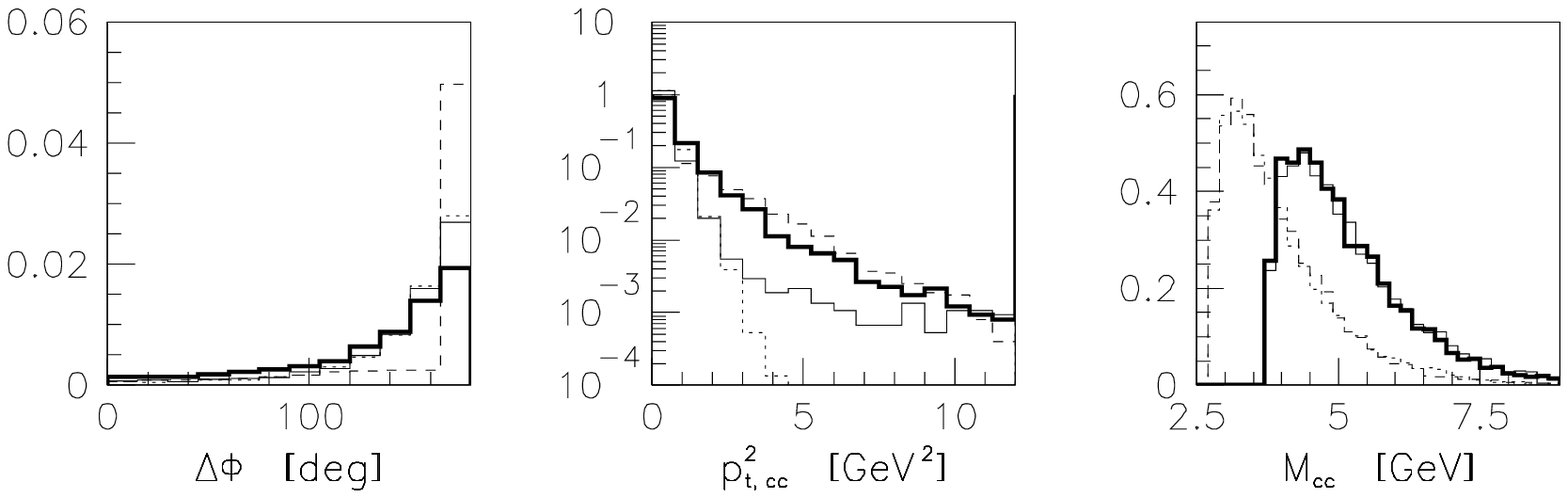,width=4.0in,angle=0}}
        \caption{Sensitivity of the single-charm and charm-pair distributions
        to the parton shower evolution (PS), the addition of
        intrinsic transverse momentum, and the hadronization process.
        All distribution are obtained from the {\sc Pythia/Jetset} event
        generator.  The {\bf solid} distributions include all three effects;
        the dashed distributions include {\it only} the parton shower
        evolution; the
        dotted distributions include {\it only} intrinsic transverse momentum;
        the solid distributions include {\it only} hadronization.
        See Table~\protect\ref{tab:tab5}.}
        \label{fig:fig34}
\end{figure}

In Fig.~\ref{fig:fig34}, we look separately at the effects of parton
showers, intrinsic transverse momentum, and hadronization.
All distributions are obtained using the {\sc Pythia/Jetset} event
generator.  The distributions obtained using the default settings 
({\bf solid}) include all three effects.  We compare these default
distributions to three sets of distributions that are obtained by including:
\begin{itemize}
\item {\it only} hadronization, but no parton shower evolution 
or intrinsic transverse momentum (solid);
\item {\it only} the parton shower evolution, but no intrinsic transverse 
momentum or hadronization (dashed);
\item {\it only} intrinsic transverse momentum, but no hadronization
or parton shower evolution (dotted).
\end{itemize}
For the longitudinal momentum distributions ($x_F$, $\Sigma x_F$, 
$\Delta x_F$, $y$, $\Sigma y$, $\Delta y$), the most important factor
is whether or not hadronization is included.  The two sets of 
distributions that include hadronization effects are quite similar;
the two sets of distributions that do not
include hadronization effects are similar; but the latter two sets
of distributions differ significantly from the former two sets.
In the {\sc Pythia/Jetset} hadronization model, the broadening of the 
longitudinal momentum distribution is the result of color-connecting
the charm quark to a valence antiquark (or diquark) 
from one of the colliding hadrons 
and the anticharm quark to a valence quark from the other colliding hadron.

All three higher-order effects broaden the leading-order 
delta function prediction for the $\Delta \phi$ distribution.
The broadening due to the parton shower evolution, however, is
significantly smaller than the broadening due to either the hadronization 
process or the addition of intrinsic transverse momentum
($\sigma_{k_t} = 0.44$ GeV$/c$).  The latter
two effects broaden the $\Delta \phi$ distribution by roughly the same 
amount.

All three higher-order effects also broaden the leading-order
delta function prediction for the $p^2_{t,c\overline{c}}$ distribution.
In this case, however, the broadening due to the 
parton shower evolution is larger than the broadening due to either 
hadronization effects or the addition of intrinsic 
transverse momentum ($\sigma_{k_t} = 0.44$ GeV$/c$).  

\subsection{Sensitivity to Intrinsic Transverse Momentum}
\label{ssec:itm}
\begin{figure} %Figure 35
        \centering
        \centerline{\epsfig{file=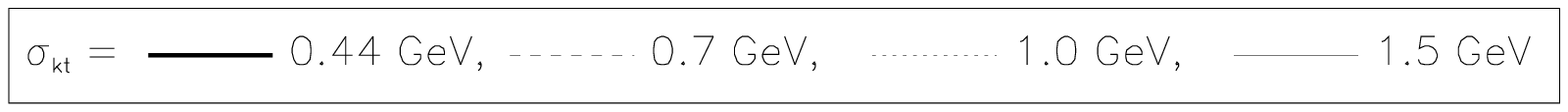,width=4.0in,angle=0}}
        \centerline{\epsfig{file=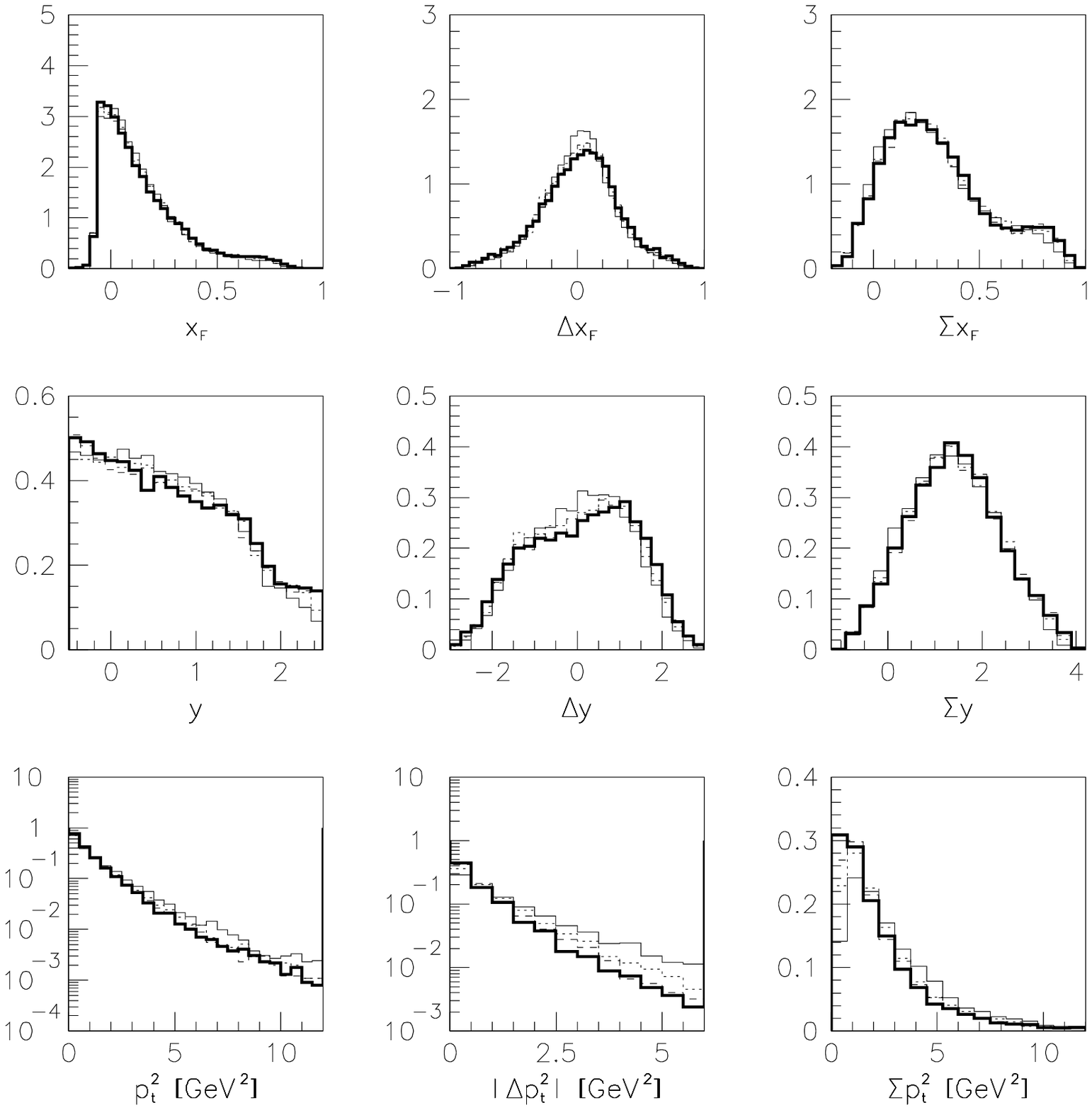,width=4.0in,angle=0}}
        \centerline{\epsfig{file=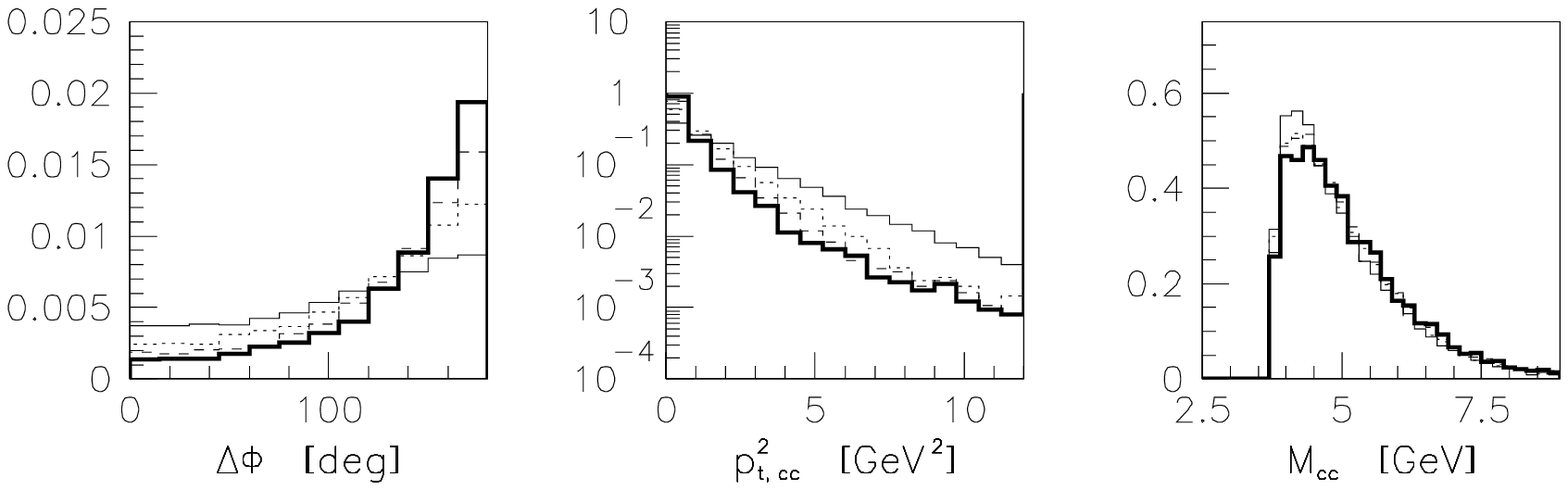,width=4.0in,angle=0}}
        \caption{Sensitivity of single-charm and charm-pair distributions
        to variations in the amount of intrinsic transverse momentum added to
        the hard-scattering partons.  All distributions are
        obtained from the {\sc Pythia/Jetset} event generator,
        with default settings for all parameters except $\sigma_{k_t}$.
        The values of $\sigma_{k_t}$ used are 0.44 GeV/$c$ ~({\bf solid}),
        0.7 GeV/$c$ ~(dashed), 1.0 GeV/$c$ ~(dotted) and 1.5 GeV/$c$ ~(solid).
        See Table~\protect\ref{tab:tab5}.}
        \label{fig:fig35}
\end{figure}

In Fig.~\ref{fig:fig35}, we investigate the degree to which the
single-charm and charm-pair distributions are sensitive to variations in
the amount of intrinsic transverse momentum added to the hard-scattering
partons that collide to form a $\ccb$ pair.
All distributions are
obtained from the {\sc Pythia/Jetset} event generator, with default settings
for all parameters except for the width of the Gaussian 
intrinsic transverse momentum distribution, $\sigma_{k_t}$.

When intrinsic transverse momentum is included, the 
hard-scattering partons from the colliding hadrons are no longer
necessarily moving parallel to the colliding hadrons.  The 
plane that is transverse to the axis of the parton-parton collision ---
which cannot be determined experimentally --- is no longer the same
as the plane that is transverse to the beam axis.
Hence, including intrinsic transverse momentum smears
the leading-order prediction 
$\vec{p}_{t,c} = -\vec{p}_{t,\overline{c}}$.
Not surprisingly, the distributions that are most sensitive to variations
in $\sigma_{k_t}$ are those transverse 
distributions that are trivial at leading-order:
$|\dptt|$, $\dphi$, and $p^2_{t,c\overline{c}}$. 
As the width $\sigma_{k_t}$ increases, these distributions become
flatter.

\subsection{Summary}
In this section, we briefly summarize the results
of the comparisons shown in Figs.~\ref{fig:fig30}--\ref{fig:fig35}.

The longitudinal momentum distributions --- $x_F$, $\Sigma x_F$,
$\Delta x_F$, $y$, $\Sigma y$, and $\Delta y$ ---
are relatively insensitive to all variations considered above, 
except for inclusion or omission of the {\sc Pythia/Jetset} hadroproduction 
hadronization (Fig.~\ref{fig:fig34}).  The steepness of the invariant mass 
distribution is also sensitive to whether or not hadronization 
is included, as well as to the mass of the charm quark (Fig.~\ref{fig:fig31}).
Therefore, the measured distributions for these physics variables, 
discussed in Sec.~\ref{sec:results}, provide a test of the 
{\sc Pythia/Jetset} hadronization model --- in particular, a test
of the string topology scheme that color-connects the charm quark
to a valence quark from one of the colliding hadrons
and the anticharm quark to a valence quark from the other colliding hadron.

The transverse distributions $|\dptt|$, $\sptt$,
$\dphi$, and $p^2_{t,c\overline{c}}$ are sensitive to 
almost all variations considered above because they 
are sensitive to the degree of correlation between the charm and anticharm
transverse momenta.
Varying $m_c$ (Fig.~\ref{fig:fig31}), 
$\Lambda_{QCD}$ (Fig.~\ref{fig:fig32}),
or $\mu_R$ (Fig.~\ref{fig:fig33}) in the 
next-to-leading order calculation changes the definition of
the running coupling constant $\alpha_s$, which 
is approximately proportional to
$1/\ln(\mu_R/\Lambda_{QCD})$.  As the coupling
constant increases --- that is, as $m_c$ decreases, $\Lambda_{QCD}$ increases,
or $\mu_R$ decreases --- higher-order effects play a larger role,
and consequently the charm and anticharm transverse momenta 
become less correlated.  The other methods we discussed 
for including higher-order 
effects were parton showers, 
intrinsic transverse momentum, and hadronization.

In Sec.~\ref{sec:results}, 
we quantify the degree of correlation
between the transverse momenta of the 
$D$ and $\overline{D}$ mesons from 
our $\ddb$ data sample.  
The sensitivity of the NLO predictions to the
arbitrary renormalization and factorization 
scales (Fig.~\ref{fig:fig33}) indicates that higher-order perturbative
corrections are important.  
In 
principle,
one could determine the sets of
theoretical parameters that generate
predictions that are in good agreement with the full range of 
experimental results.  
The set of fit values chosen, however,
may not be unique.
The fit values of 
({\it e.g.}, the mass of the charm quark)
would depend on the values of the 
renormalization and factorization 
scales.  For example, if a renormalization scale of $Q/2$, rather 
than $Q$, is assumed, then a smaller value for 
$\sigma_{k_t}$ or a larger value for $m_c$ could
each be used to fit the data.

%%%%%%%%%%%%%%%%%%%%%%%%%%%%%%%%%%%%%%%%%%%%%%%%%%%%%%%%%%%%%%%%%%%%%%
%\input{offline_doc_290_9_r1.tex} %   {references}
%%\begin{references}

%%\end{references}

\end{document}